\pdfoutput=1
\documentclass[11pt]{article}

\usepackage{wrapfig}
\usepackage{graphicx}

\def\la{\mathrel{\mathchoice {\vcenter{\offinterlineskip\halign{\hfil
$\displaystyle##$\hfil\cr<\cr\sim\cr}}}
{\vcenter{\offinterlineskip\halign{\hfil$\textstyle##$\hfil\cr
<\cr\sim\cr}}}
{\vcenter{\offinterlineskip\halign{\hfil$\scriptstyle##$\hfil\cr
<\cr\sim\cr}}}
{\vcenter{\offinterlineskip\halign{\hfil$\scriptscriptstyle##$\hfil\cr
<\cr\sim\cr}}}}}
\def\ga{\mathrel{\mathchoice {\vcenter{\offinterlineskip\halign{\hfil
$\displaystyle##$\hfil\cr>\cr\sim\cr}}}
{\vcenter{\offinterlineskip\halign{\hfil$\textstyle##$\hfil\cr
>\cr\sim\cr}}}
{\vcenter{\offinterlineskip\halign{\hfil$\scriptstyle##$\hfil\cr
>\cr\sim\cr}}}
{\vcenter{\offinterlineskip\halign{\hfil$\scriptscriptstyle##$\hfil\cr
>\cr\sim\cr}}}}}
\def\HII{H{\sc ii}}
\newcommand{\gtrsim}{\mathrel{\hbox{\rlap{\hbox{\lower4pt\hbox{$\sim$}}}\hbox{$>
$}}}}
\newcommand{\lesssim}{\mathrel{\hbox{\rlap{\hbox{\lower4pt\hbox{$\sim$}}}\hbox{$<
$}}}}
\def\Dfrac{$D_{\rm frac}$}

\newcommand{\xco}{$X_{\rm CO}$}
\newcommand{\htwo}{H$_2$}
\newcommand{\jj}{\textit{J}}
\newcommand{\lsun}{$L_\odot$}
\newcommand{\hcoplus}{{HCO$^+$}}
\newcommand{\hcn}{HCN(1--0)}
\newcommand{\aap}{A\&A}
\newcommand{\apj} {ApJ}
\newcommand{\araa} {ARA\&A}
\newcommand{\mnras}{MNRAS}
\newcommand{\apjl} {ApJL}
\newcommand{\nat}{Nature}
\newcommand{\apjs} {ApJS}
\newcommand\blfootnote[1]{%
  \begingroup
  \renewcommand\thefootnote{}\footnote{#1}%
  \addtocounter{footnote}{-1}%
  \endgroup
}

\begin{document}

\begin{figure}[tbh]
\vspace{-1cm}
\includegraphics[angle=0,scale=0.25]{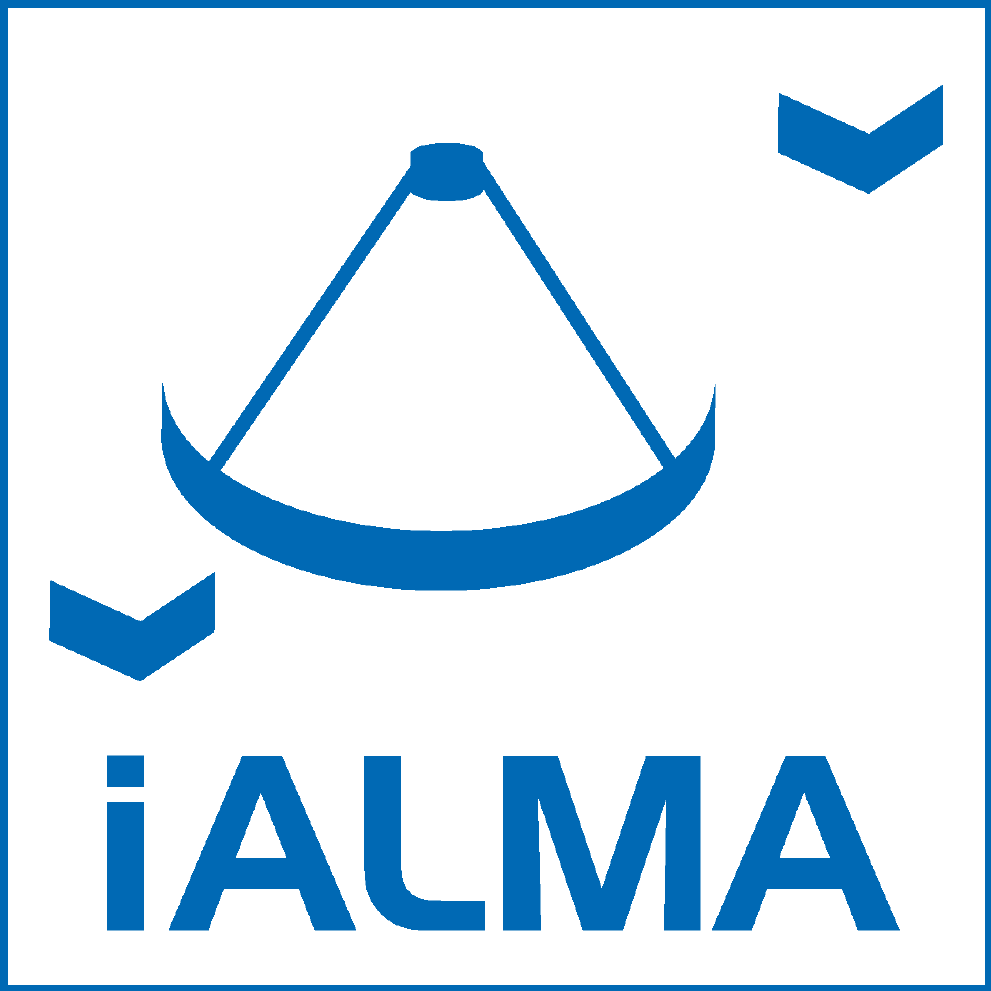}
\end{figure}

\begin{center}
{\LARGE{\bf Italian Science Case for ALMA Band 2+3$^\star$\blfootnote{$^\star$This white book 
expands and partly overlaps the European Science Case for ALMA Band 2 (Fuller et al.,
 in preparation). The scientific cases here presented are focused on the
research interests of the members of the iALMA Premiale Project.}}}
\end{center}

\begin{center}

{\large{M.\ T.\ Beltr\'an\footnote{INAF-Osservatorio Astrofisico di Arcetri, Largo
E.\ Fermi 5, 50125, Firenze, Italy}, 
E. Bianchi$^1$, J.\ Brand\footnote{INAF-Istituto di
 Radioastronomia, via Gobetti 101, 40129 Bologna, Italy}, V. Casasola$^1$, R.\ Cesaroni$^1$,  C.\ Codella$^1$,
 F.\ Fontani$^1$, L. Gregorini\footnote{Dipartimento di Fisica e Astronomia,
 Universit\`a di Bologna, Viale Berti Pichat 6/2, 40127 Bologna, Italy},  G. Guidi$^1$, L.\ Hunt$^1$, E.\ Liuzzo$^2$, 
 A.\ Marconi\footnote{Dipartimento di Fisica e Astronomia, Universit\`a di Firenze, Largo E. Fermi 2, I-50125 Firenze, Italy}, M. Massardi$^2$, L. Moscadelli$^1$, R. Paladino$^2$, 
 L. Podio$^1$, I. Prandoni$^2$, V.\ Rivilla$^1$, K.\ L.\ J.\ Rygl$^2$,  L.\ Testi$^{1,
 }$\footnote{ESO, Karl-Schwarzschild-Strasse 2, 85748 Garching bei M\"unchen, Germany}}}

\end{center}





\section{Introduction}

The Premiale Project ``Science and Technology in Italy for the upgraded ALMA
Observatory -- iALMA" has the goal of strengthening the scientific, technological and
industrial Italian contribution to  the Atacama Large Millimeter/submillimeter Array
(ALMA),  the largest ground based international infrastructure for the study of the
Universe in the microwave. One of the main objectives of the Science Working Group
(SWG) inside iALMA, the Work Package 1, is to develop the Italian contribution to the
Science Case for the ALMA Band 2 or Band 2+3 receiver. ALMA Band 2 reveiver spans from
$\sim$\,67~GHz (bounded by an opaque line complex of ozone lines) up to 90~GHz which
overlaps with the lower frequency end of ALMA Band 3.  Receiver technology has advanced
since the original definition of the ALMA frequency bands. It is now feasible to
produce a single receiver which could cover the whole frequency range from 67\,GHz to
116\,GHz, encompassing Band 2 and Band 3 in a single receiver cartridge, a so called
Band 2+3 system. In addition, upgrades of the ALMA system are now foreseen that should
double the bandwidth to 16~GHz. The
science drivers discussed below therefore also discuss the advantages of these two
enhancements over the originally foreseen Band 2 system.

\section{Galactic Science}

\subsection{\underline{High-mass star formation}}

High-mass stars are commonly defined as those stars whose mass exceeds
$\sim$8~$M_\odot$, a limit that may vary depending on the border conditions,
such as, e.g., the accretion rate. What makes these stars different from their
low-mass siblings, is that they reach the zero-age main sequence still
undergoing heavy accretion. While the theoretical details of the high-mass
star formation process are still a matter of debate, a solid observational
finding is that young OB-type stars are deeply embedded inside their parental
molecular cores -- the so-called hot molecular cores (HMCs). At the same time,
massive stars are known to form in rich stellar clusters and are located at
typical distances of a few kpc. All these facts make it very difficult to
study the cirumstellar environment, due to the large extinction and small
separation from the nearby cluster members. These constraints call for high
angular resolution ($\ll$1$''$) observations at long wavelengths, where the
dust opacity is sufficiently low. In practice, interferometric observations
at (sub)mm wavelengths are the ideal tool to investigate these regions.

Another important characteristic of young massive stars is that they
are surrounded by dense gas, rich in molecular species. These are believed
to form on grains and then be released to the gas phase when the grain mantles
are evaporated by the photons of the newly born star. As a matter of fact,
the rotational transitions of rare, complex species can be effectively
used to derive the physical parameters of the circumstellar gas and study its
distribution and velocity field. Moreover, detailed census of many molecular
species may allow one to derive the chemical composition of the gas and thus
decide what reactions may explain the formation of the observed molecules,
which in turn may shed light on the genesis of the star itself.

With all this in mind, there is little doubt that ALMA is the ideal
instrument for studies of high-mass star formation. The unprecedented angular
resolution in the (sub)mm regime coupled with the high sensitivity, as well
as the broad, simultaneous frequency coverage, are bound to satisfy all the
requirements discussed above. In particular, the dust opacity in molecular
clouds is known to increase with frequency, which makes it convenient to
observe at (sufficiently) long wavelengths. However, one must also keep in
mind that all rotational transitions of molecules lie in the (sub)mm range,
which sets a lower limit to $\lambda$. Therefore, as a rule of thumb,
observations of high-mass star forming regions are best performed between a
few~mm and a few 100~$\mu$m.

In the following, we identify three major contributions that ALMA Band 2+3
may give to the study of high-mass star formation.

\subsubsection{The chemical content of HMCs}

As discussed in more detail in other parts of this report, the study of the
chemical content of HMCs is expected to help us to set tight constraints
on the star formation process. But what is the best frequency range for
this type of studies? To answer this question,
it is important to stress that the number of rotational transitions is
increasing with decreasing wavelength and their excitation energies are
on average higher. If the ``quantity'' of the information obtained
from broad band observations increases dramatically with the observing
frequency, the corresponding ``quality'' is not necessarily improving.
In fact, one sees that the overwhelming number of lines detected towards
HMCs often hinders precise line identifications due to multiple overlaps
of adjacent transitions. Moreover, the high excitation energies make such
lines weaker and their emitting regions significantly smaller than those
traced by lower excitation (and lower frequency) transitions. These problems
make it very difficult and sometimes even impossible to resolve both in
space and velocity the emission from different species. For HMC studies
it may thus be worthy to perform observations at longer wavelengths. In this
context the ALMA Band~2+3 project will play a crucial role.

Another aspect that also favours the choice of low frequencies for studies of
HMC chemistry, is the dust opacity. The dust optical depth in the
(sub)mm domain increases with frequency as $\tau\propto N_{\rm
H_2}\nu^\beta$, where $\beta\simeq1$--2 and $N_{\rm H_2}$ is the H$_2$
column density along the line of sight through the HMC. Clearly, this implies
that at sufficiently large frequencies the core will become optically thick
and the line photons will be absorbed by the dust, thus reducing the observed
line intensities dramatically. This effect is bound to affect especially the
transitions of the typical HMC tracers, as opposed to those molecules that
are present (also) in the surrounding, lower density envelope. For the same
reason, the effect will be more prominent in the innermost parts of the HMCs,
because the density is believed to increase towards the core center as a
power law, which boosts the value of the dust opacity. While it is difficult
to establish the frequency and radius at which the dust opacity will overcome
the line emission, it seems likely that in the ALMA Band~2+3 such an effect
will be negligible: in fact even for a column density as high as
$N_{\rm H_2}=10^{25}$~cm$^{-2}$, the dust opacity at, e.g., 85~GHz is only
$\tau\simeq 0.01\ll 1$.

\subsubsection{Infall studies}

One of the important issues concerning the formation of an OB-type star is 
the time at which accretion stops. Some authors (see Keto 2002) have proposed
that accretion onto the star may initially quench the formation of an \HII\
region, until the stellar mass has increased to the point that the Lyman
continuum luminosity overcomes the mass accretion rate. This model predicts
also that even after the formation of an ionized region around the star,
acrretion should go on through the \HII\ region itself. It is thus interesting
to establish observationally whether infall can be detected towards the
youngest \HII\ regions. In particular, hypercompact \HII\ regions (see
Kurtz 2005) should be the ideal targets to detect red-shifted absorption in
suitable molecular lines.

To search for red-shifted absorption towards these objects, one needs a good
tracer of the infalling gas and a bright \HII\ region. Unfortunately, these
two requirements are to some extent conflictual, as the brightness
temperature of the free-free emission decreases with frequency, wehereas the
number of molecular lines increases (as explained above). It is thus
necessary to find a compromise and select the longest possible wavelength
where a sufficient number of rotational transitions can be found. Cesaroni
(2008) demonstrates that employing observations at frequencies below
$\sim$100~GHz even infall towards hypercompact \HII\ regions around stars as
late as B3 can be detected.

One should keep in mind that infall studies with molecular lines may be
nicely complemented by observations of recombination lines, which trace the
ionized gas of the \HII\ region. Given the broad ALMA frequency coverage,
it will be possible to perform observations of both suitable molecular lines
to detect infall {\it towards} the \HII\ region, and a recombination line
to trace the infall {\it inside} the \HII\ region.

\subsubsection{Circumstellar disks}

It is known that circumstellar accretion disks around B- and even more O-type
stars appear to be elusive (see Cesaroni et al.~2007 for a review). Recently,
the existence of such disks around B-type stars has found further support from
observations with ALMA (see S\'anchez-Monge et al. 2013, 2014; Beltr\'an et
al.~2014) and similar results are expected also for O-type stars when the
longest array baselines will be available. Clearly, angular resolution is
an important issue in this type of studies, which calls for high-frequency
observations. However, one should not forget that also low-frequency lines
may be worthy. As previously explained, most high-frequency transitions arise
from high-energy levels and hence trace the innermost regions of the disk.
Therefore, the advantage of a better angular resolution is basically compensated
by the smaller emitting region. One may roughly conclude that the ratio
between source angular size and instrumental beam is approximately
independent of the observing frequency.

Despite this consideration, the study of disks around O-type stars is likely
best performed at high frequencies, given the expected size of the disks (a
few 1000~au) and the large distances of the objects (several kpc), but
observations at ALMA Band~2+3 may play an important role to investigate the
structure of the nearest (2--3~kpc) disks around B-type stars. Beside the
advantages already discussed above (limited line overlap, negligible dust
absorption), observations at low frequencies are less affected by phase
instabilities and allow to target objects at lower elevations. The linear
resolution achievable at 3~mm ($\sim$50~mas) will suffice to resolve the disk
plane and establish for example the presence of spiral arms.  The targets for
this type of observations will be known luminous ($\sim$$10^4~L_\odot$)
(proto)stars for which evidence of circumstellar disks has been previously
found with Plateau de Bure and ALMA observations.

\subsection{\underline{High-mass star-formation under the magnifying glass}}

\subsubsection{High-mass star forming regions with methanol masers}

As already mentioned, some of the biggest challenges in high-mass star formation
(HMSF) studies is the identification of high-mass star-forming objects and
understanding how mass accretion continues beyond a stellar mass of 8\,$M_\odot$
(Palla \& Stahler 1993). With the advance of mm-interferometry, many massive cores
have been found to fragment to lower mass proto-stellar objects, and multiplicity
in star formation is often difficult to exclude, casting doubts on masses of
single star-forming objects. The search for accretion discs, predicted by recent
2-D models of HMSF (e.g., Kuiper et al. 2010) and considered essential for
continuing accretion despite the stellar radiation pressure, is only possible
through high angular resolution kinematic studies of the gas surrounding the
protostar, searching for disc signatures and/or jets emitted along the disc axis.

The methanol maser emission at 6.7\,GHz is one of the strongest and most
widespread interstellar masers (Menten et al. 1991). It can be used both for
pinpointing high-mass star-forming regions (HMSFRs) and for determining  the
structure and kinematics of the gas at high resolution. As the excitation of this
maser transition requires high densities and strong infrared emission and/or
shocks (Sobolev \& Deguchi 1994, Cragg et al 2005), methanol masers are found
exclusively toward HMSFRs (Minier et al. 2003). Their brightness  and compactness
make the methanol masers an excellent tool for astrometry (Rygl et al. 2010) and
structure and 3-D velocity studies at distances of 10--1000\,AU from the newly
born high-mass star (e.g., Moscadelli et al. 2011). VLBI imaging of 6.7\,GHz
methanol masers showed that in $\sim$30\% of the sources they are distributed in a
ring-like shape (with radii of 20--200\,mas) around the central protostar, with
kinematics suggesting that these masers originate at the interface between the
protostellar outflow and a disc or torus (Bartkiewicz et al. 2009). Though the
6.7\,GHz masers environments have been identified only in a few high-mass
protostars (e.g., Sanna et al. 2010, 2014, Moscadelli et al. 2011), the 6.7 GHz
maser emission was found to trace both disc rotation and expansion, the latter
likely induced by interaction with protostellar wind/jet, suggesting that
ring-shaped methanol masers are the ideal high-mass disc-candidates. 

\subsubsection{Justification for ALMA Band 2+3}

\paragraph{Ring-like methanol maser sources in continuum emission with ALMA}

A high-angular resolution continuum and molecular line study with ALMA in Band 2+3
of the ring-shaped 6.7\,GHz methanol masers could find conclusive evidence of
discs, outflows, and constrain the evolutionary stage of the high-mass protostar.
Two emission mechanisms can contribute to the spectral energy distribution of disc
candidates: free-free emission from the ionized gas and thermal black body
emission from the dusty envelope. In its turn, the free-free emission associated
with these maser sources can originate both from hyper- or ultra-compact H{\sc ii}
regions and ionized, wide-angle or collimated, stellar winds. Following the
paradigm of the disc/jet system, one would expect to observe a flaring dusty
envelope, oriented perpendicular to the ionized jet. 

Establishing the nature and structure of the protostellar continuum requires
sensitive (rms noise $\lesssim$10\,$\mu$Jy), high angular resolution
($\theta_\mathrm{FWHM}$$<$100\,mas) imaging in the mm wavelength range.  The ALMA
Band 2+3 frequency range, from 67 to 116\,GHz, is crucial to visualize the
distribution of the continuum emission and separate the  free-free emission from
the thermal dust component by means of the spectral index, $\alpha$, analysis. In
fact, while the free-free emission is expected to be optically thin
($\alpha$$\sim$0) in this frequency range, the dust emission should rapidly
increase with frequency ($\alpha$$>$1). Previous observations show that above a
detection threshold of a few 100\,$\mu$Jy only a small fraction of  ring shaped
methanol masers (10\%) is associated with cm continuum emission, which motivated
the interpretation that these sources harbor young H{\sc ii} regions with a high
turnover frequency (Bartkiewicz et al. 2009). ALMA will achieve an extraordinary
sensitivity at 3-5\,mm, rms noise $<$10\,$\mu$Jy in less than one hour, allowing
one to image much weaker free-free emission than in previous cm studies, and
possibly finding dusty flattened disc structures (Sanna et al. 2014). ALMA's long
baseline configuration of $\sim$15\,km offers $\sim$ 70 and 40\,mas resolution at
5 and 3\,mm, respectively, sufficient to resolve structures that have similar
angular scales as the observed maser rings.

\paragraph{Ring-like methanol maser sources in spectral line emission with
ALMA}

Simultaneously with the continuum observations, ALMA can perform high-resolution
imaging of the disc/jet structure in various molecular lines available in Band
2+3, such as the low-$J$ HCO$^+$ and SiO transitions, and various deuterated
species (see other contributions in this white paper) that can be compared to the
structure of the maser ring. If the envelope is truly flattened, then one might
observe an inner gap at radii smaller than the methanol maser ring, since
temperature and turbulence should be too high for these molecular species to
survive.

In summary, ALMA Band 2+3 observations of ring shaped methanol masers, will offer
a wealth of continuum and molecular data that could provide crucial evidence for
disc structures around high-mass star-forming objects. 

\subsection{\underline{Low excitation lines of deuterated molecules}}

There are no doubts that the two most important chemical processes occuring in the
cold (T $\leq 20$ K) and dense ($\geq 10^4$ cm$^{-3}$) pre--stellar cores are:
{\it (i)} the
enhancement of deuterated molecules, and {\it (ii)} the freeze-out of neutral atoms and
molecules on the surface of dust grains,  where the depletion of both C-bearing
and N-bearing molecules is found  to be high in low- and high-mass pre--stellar
cores.  The first one, initiated mainly from the proton-deuteron exchange reaction
$H_3^+ + HD \rightarrow H_2D^+ + H_2 + \Delta E$, is favoured at low temperature 
because of the endothermicity of the backward chemical reaction.  The second one,
occurring on the surfaces of dust grains, is favoured  by the combination of low
temperature and high density, by which atoms  and molecules stick on the surface
of a dust grain and remain frozen on it (see Caselli \& Ceccarelli for a review).
The two processes are believed to be correlated in the cold gas: in fact,  depletion
of CO boosts the deuterium fractionation process because other  neutrals can
combine with the abundant H$_2$D$^+$ and  form D-bearing molecules, and indeed
correlation between the amount of  CO depletion and the deuterium fractionation
(\Dfrac , defined as the abundance ratio between a D-bearing molecule and its
hydrogenated counterpart)  was found in low-mass dense cores (e.g., Crapsi et
al.~2005). Freeze-out of CO and other neutrals has also another crucial
implication: frozen on grain mantles, atoms and molecules can undergo
hydrogenation (and deuteration), due to the high mobility of the light H (and D)
atoms.  In particular, from hydrogenation of CO, the most abundant neutral
molecule after H$_2$, one forms sequentially HCO, H$_2$CO and CH$_3$OH, which is
thus the species that is formed last. Therefore, as time proceeds, the formation
of methanol and its deuterated forms (CH$_2$DOH, CH$_3$OD, CHD$_2$OH, etc.) is
boosted, until the energy released by the nascent protostellar object in the form
of radiation increases the temperature of its environment, causing the evaporation
of the grain mantles and the release of these molecules into the gas. As the
temperature increases, the deuterated species are expected to get gradually
destroyed due to the higher efficiency of the backward endothermic reactions
(Caselli \& Ceccarelli 2012). Therefore, high \Dfrac\ of species formed in the
gas, like N$_2$D$^+$ and DCN, are good tracers of the pre--stellar phase, while
high \Dfrac\  of methanol are expected to better trace the short-living
evolutionary stage in between the pre-stellar and the protostellar phase, at
which  the evaporation/sputtering of the grain mantles is efficient, and  the warm
gas-phase reactions have no time to change significantly  the chemical composition
of the gas. Recent observations of these  molecules in dense high-mass
star-forming cores belonging to  different evolutionary stages (Fontani et
al.~2015) indicate the  essential validity of this scenario.

In this framework, observations of deuterated molecules are crucial to derive the
kinematics and other chemical/physical properties of pristine  dense cores. This
information is mandatory to understand the initial phases of the star formation
process and put constraints on current theories. In particular, because of the low
temperature, the ground rotational  transition (1--0) is certainly the most
sensitive probe because it is the easiest line to excite even at very low
temperature. However, unfortunately the (1--0) rest frequency of many deuterated
isotopologues of common, abundant interstellar molecules is below 80~GHz, not
accessible by most of the current millimeter single-dish telescopes and
interferometers. For this reason, the large majority of the studies  performed so
far have used the higher excitation transitions, which are not fully
representative of the whole core but just of the densest inner parts. Moreover,
the deuterated fractions are often derived by comparing  the column density
computed from the fundamental transition of the hydrogenated  molecule to that
derived from higher excitation lines of the deuterated counterpart, which make the
ratio dependent on the possibly different excitation conditions. To solve this
problem, ALMA Band 2 will be perfectly suitable, because the (1--0) lines of many
abundant deuterated  molecules, including DCO$^+$ and N$_2$D$^+$, are  unique to
this band. A list of these lines is reported in Table~1. Also, the excitation
temperature of the (1--0) lines can be derived directly from their hyperfine
structure, assuming the line to be optically thick. This method in principle is
valid also for higher excitation lines, but it is more uncertain because they are
expected to be fainter and  optically thinner. Finally, in the same single
spectral setup one can observe simultaneously   many lines of the {\it
ortho}-NH$_2$D, including the (1$_{1,1}$--1$_{0,1}$) line at $\sim 85$~GHz, as
well as several transitions of HDCO, which can help, all together, to obtain a
complete and consistent picture of deuterated molecules in the target cores.

\begin{table}
\caption[] {Ground rotational transitions of deuterated molecules in ALMA Band 2}
\begin{center}
\begin{tabular}{lc}
\hline \hline
molecule & rest freq. (GHz) \\
\hline
DCO$^+$ (1--0) & 72.03931 \\
DCN (1--0) &  72.41469 \\
CCD (1--0) & 72.09843 -- 72.20019 \\
DNC (1--0) &  76.30570 \\
N$_2$D$^+$ (1--0) & 77.10924 \\
{\it ortho-}NH$_2$D ($J_{K1K2}=1_{1,1}-1_{0,1}$) & 85.92772 \\
\hline
\end{tabular}
\end{center}
\end{table}

\begin{figure}
\begin{center}
\includegraphics[angle=-90,width=0.75\textwidth]{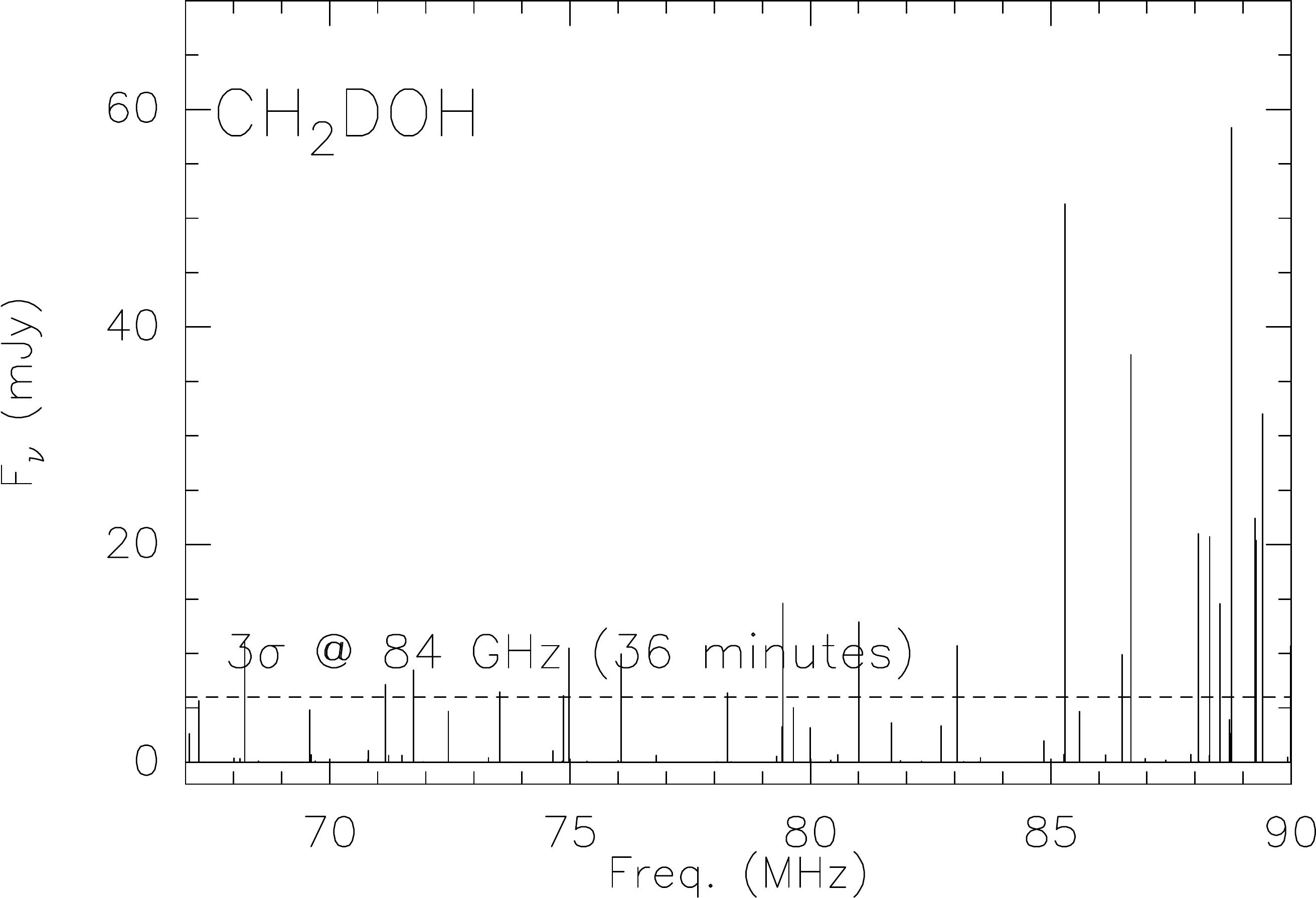}
\caption{\it Synthetic spectrum of CH$_2$DOH modeled with WEEDS
in the range $\sim 67-90$~GHz, assuming $T_{\rm ex}=20$ K, 
$N$(CH$_2$DOH)=$5\times 10^{15}$ cm$^{-2}$, source size = 1", 
and line FWHM = 1 km s$^{-1}$. The dashed line represents the
expected 3$\sigma$ level in the spectrum that can be achieved after
36 minutes of integration on source with ALMA (see text for details).}
\label{fig_FF_band2}
\end{center}
\end{figure}

Other important lines that can be observed in Band 2 are those of  deuterated
methanol, which are important to trace the short-living evolutionary stage in
between the pre- and the protostellar phase (see above).  Observations of
deuterated methanol are possible in several bands,  but their identification is
difficult at wavelengths shorter than 3~mm  because they can be easily overwhelmed
by nearby stronger emission lines of lighter and more abundant molecules.
Therefore, their detection must be checked carefully with synthetic spectra, which
is not an easy task usually (see e.g. Fig.~1 in Fontani et al.~2015).  This
problem can be bypassed in band 2, because several transitions of CH$_2$DOH can be
observed in the range 67--90~GHz without the contamination of lines of more
abundant molecules,  as can be seen in Fig.~\ref{fig_FF_band2}. The combination
of high angular resolution  and high sensitivity offered by ALMA in Band 2 will be
eminently  suitable to map these lines, which are expected to arise from  very
compact regions. As an example, we can estimate the time required  to detected
some of the lines shown in the synthetic spectrum in  Fig.~\ref{fig_FF_band2}. The
spectrum includes all transitions of CH$_2$DOH in the  spectroscopic band $\sim
67-90$~GHz, as modeled by WEEDS  assuming the following parameters: $T_{\rm
ex}=20$ K,  $N$(CH$_2$DOH)=$5\times 10^{15}$ cm$^{-2}$, source size = 1",  line
FWHM = 1 km s$^{-1}$. The column density assumed is a  mean source-averaged value
measured towards massive  protostellar cores (Fontani et al.~2015). Based on the
OT time exposure calculator, at a representative frequency of 84~GHz (the lower
frequency currently observed by ALMA),  assuming $T_{\rm sys}=$60 K, 36 dishes of
12~m, 2 polarisations,  a velocity resolution of $\sim 1$ km s$^{-1}$, the
1$\sigma$~rms noise in the spectrum after just 36~minutes of integration on source
is 2~mJy. Fig.~\ref{fig_FF_band2} tells us that more than 10 lines have intensity 
peak well above 3$\sigma$. Therefore, this will allow us not only to just detect
the presence of the molecule in the same spectral setup where the (1--0) lines of
the other species will be detected, but also to derive  estimates of some
important physical parameters from methanol (e.g., gas temperature and column
density from the rotation diagrams).

\subsection{\underline{Understanding the formation of astrobiological molecules}}

The increasing number of detections of complex organic molecules around young
stellar objects strongly suggests that they are part of the material of which
planetary systems are made. These molecules play a central role in interstellar
prebiotic chemistry and may be directly linked to the origin of life.  While the
detection of the simplest amino acids such as glycine still remains elusive, the
search of two other families of prebiotic molecules has been more successful:
aldoses and polyols. The monosaccharide sugar glycolaldehyde (CH$_2$OHCHO,
hereafter GA) is the simplest representative of the aldoses. This molecule can
react with propenal to form ribose, a central constituent of RNA (Collins and
Ferrier 1995; Weber et al.\ 1998).  The simplest representative of the polyols is
the reduced alcohol of GA, ethylene glycol (CH$_2$OH)$_2$, hereafter EG).

Theoretical chemical models have been developed in the last years to understand
how these complex molecules can be formed in the interstelar medium. Woods et al.\
(2012, 2013) tested six different mechanisms of GA synthesis proposed in the
literature, both in the gas phase and on the surface of grains, and concluded that
the most likely pathways are 3 grain-surface formation routes involving the formyl
radical (HCO):

\vspace{2mm}

\begin{small}

1) CH$_3$OH + HCO $\longrightarrow$ CH$_2$OHCHO + H;

2) H$_2$CO + HCO + H $\longrightarrow$ CH$_2$OHCHO;

3) 2HCO $\longrightarrow$ CO + H$_2$CO $\longrightarrow$ HOCCOH;  \\
\hspace*{.9cm} HOCCOH + H $\longrightarrow$ CH$_2$OCHO; CH$_2$OCHO + H $\longrightarrow$ CH$_2$OHCHO.

\end{small}

\vspace{2mm}

\begin{figure}[tbh]
\centering
\includegraphics[angle=0,scale=0.4]{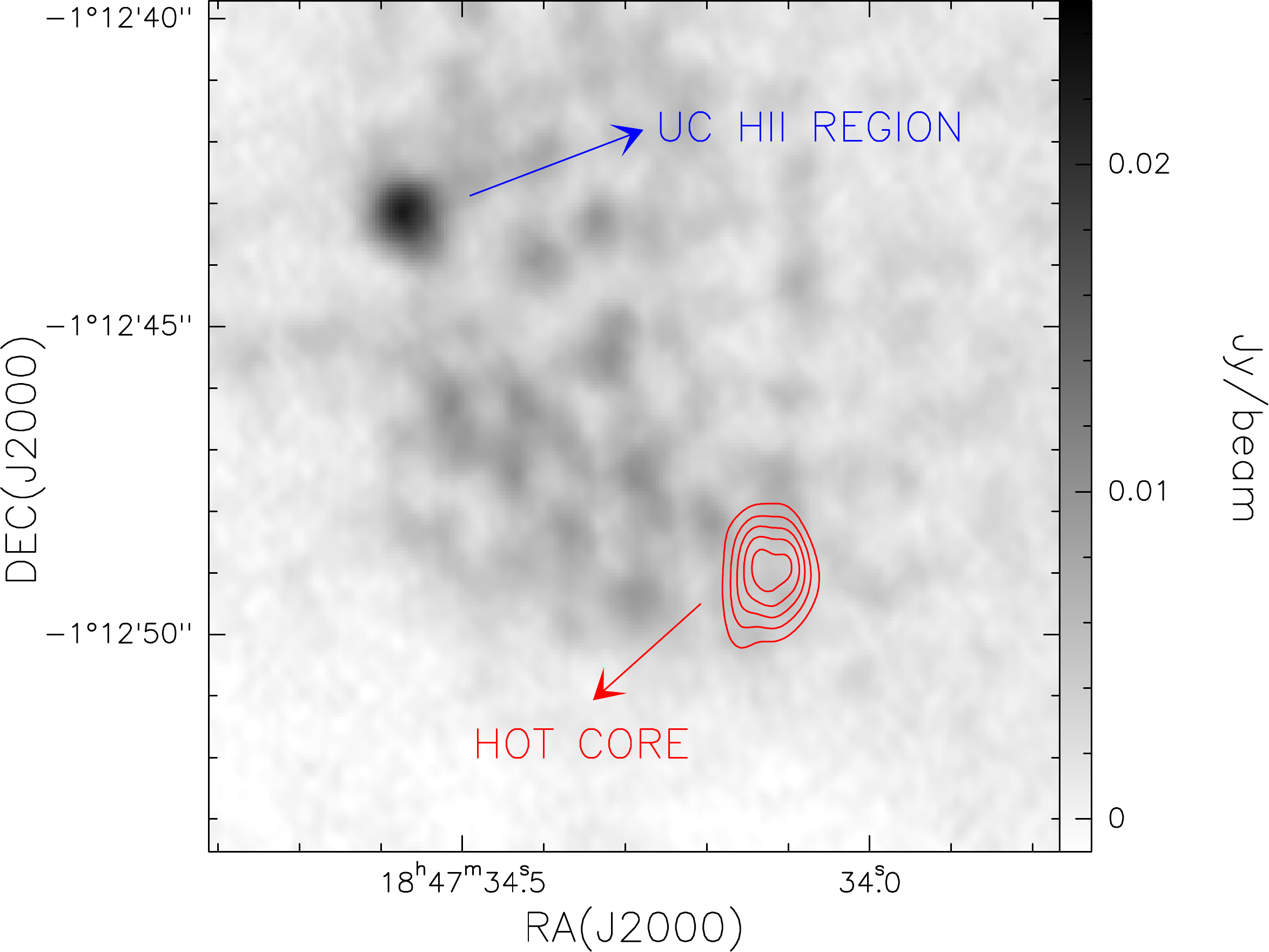}
\caption{\it Integrated intensity of the CH$_2$OHCHO (GA) line at 1.4 mm
(220.46 GHz) detected with PdBI towards the HMC G31.41+0.31 (red contours;
Beltr\'an et al.\ 2009), overplotted on the VLA map of the 1.3 cm continuum
emission (grey scale; Cesaroni et al.\ 1998). The positions of the HMC and the
nearby UC HII region are indicated.}
\label{cesa-glyco}
\end{figure}

\begin{figure}[tbh]
\centering
\includegraphics[scale=0.425]{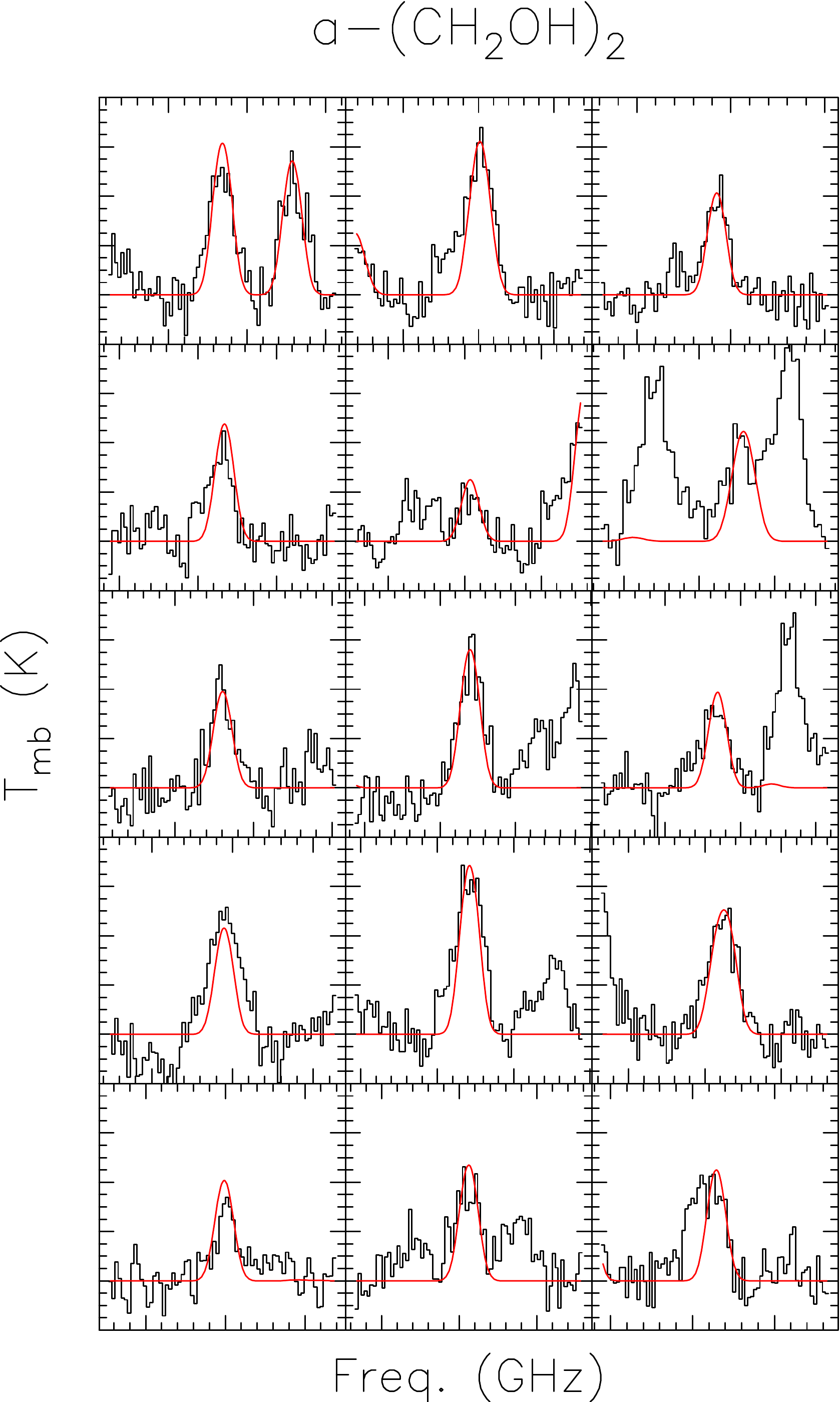}
\includegraphics[scale=0.425]{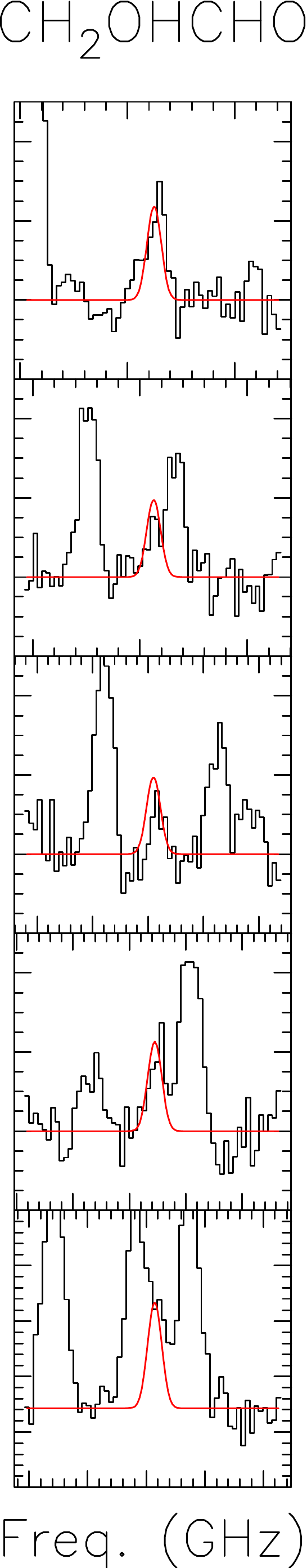}
\includegraphics[scale=0.425]{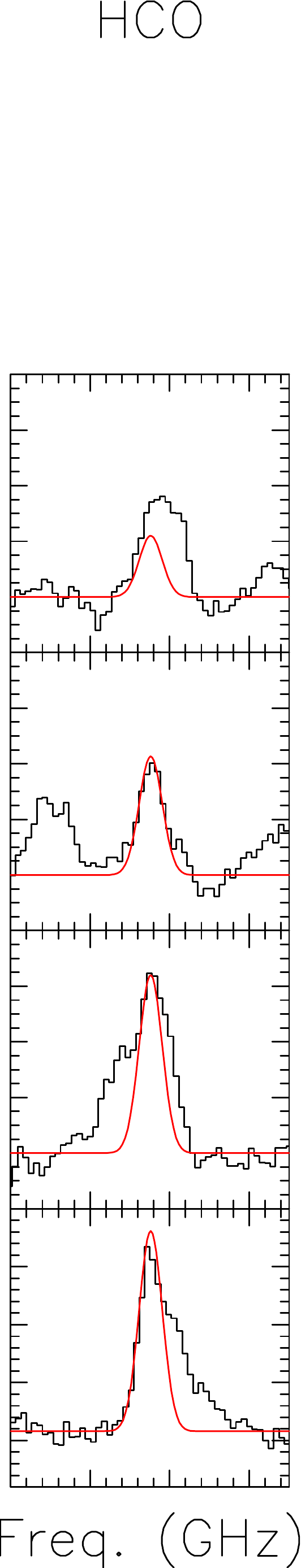}
\caption{\it EG, GA and HCO transitions detected towards G31 HMC with the IRAM
30m telescope (Rivilla et al., in preparation). The red line is the spectrum
simulated assuming LTE conditions. The extra emission in the HCO transition in the
upper right panel is due to contamination from an unidentified molecule.}
\label{HCO-30m}
\end{figure}

From a chemical perspective, these proposed formation routes look promising. It is
known that HCO is formed on the surface of interstellar grains by the
hydrogenation of CO (Woon et al.\ 2002). Since CO is the secondmost abundant
molecule in insterstellar ices, it is expected that the abundance of HCO in the
grains is high enough to produce efficiently GA and EG.  The third mechanism
proposed has been recently supported by the laboratory experiments by Fedoseev et
al.\ (2015). This work also shows that this mechanism is also very efficient in
forming EG through sequential hydrogenation by two H atoms: {\scriptsize
CH$_2$OHCHO + H + H $\longrightarrow$  (CH$_2$OH)$_2$.} 

Interstellar GA and EG were first detected towards the Galactic Center in the
SgrB2 region (Hollis et al.\ 2000, 2002; Halfen et al.\ 2006). Beltr\'an et
al.\ (2009) reported the first detection of GA outside the Galactic Center towards
the Hot Molecular Core (HMC) G31.41+0.31 (hereafter G31), using the IRAM PdBI
(Fig.~\ref{cesa-glyco}). Recently, the a-conformer of EG, a-(CH$_2$OH)$_2$), has
been detected towards G31 using the IRAM 30m telescope (Fig.~\ref{HCO-30m};
Rivilla et al., in preparation). Calcutt et al.\ (2014) also confirmed the
detection of GA towards 5 additional HMCs. Interestingly, GA and EG have also been
recently reported towards two low-mass protostellar systems, the binary IRAS
16293$-$2422 and NGC 1333 IRAS2A (J{\o}rgensen et al.\ 2012; Maury et al.\ 2014;
Coutens et al.\ 2015), and towards an intermediate-mass protostar NGC7129 (Fuente
et al.\ 2014). The HCO molecule was first detected in the interstellar medium by
Snyder et al.~(1976) towards a sample of molecular clouds associated with HMCs.

In agreement with theoretical predictions, very recent single-dish observations
suggest a link between HCO, GA and EG. Rivilla et al., (in preparation) has
detected for the first time HCO towards G31 using the IRAM 30m telescope. Figure
\ref{HCO-30m} shows the 3 mm spectrum towards G31, including the HCO, GA and EG
transitions. Moreover, the spectral line survey by Armijos-Abenda\~no et al.
(2015) towards two Galactic Center molecular complexes has shown that GA is
detected only if HCO is also present. These findings support the hypothesis of HCO
as a precursor of GA and EG.

\subsubsection{Justification for ALMA Band 2+3}

\begin{figure}[tbh]
\centering
\includegraphics[scale=0.55]{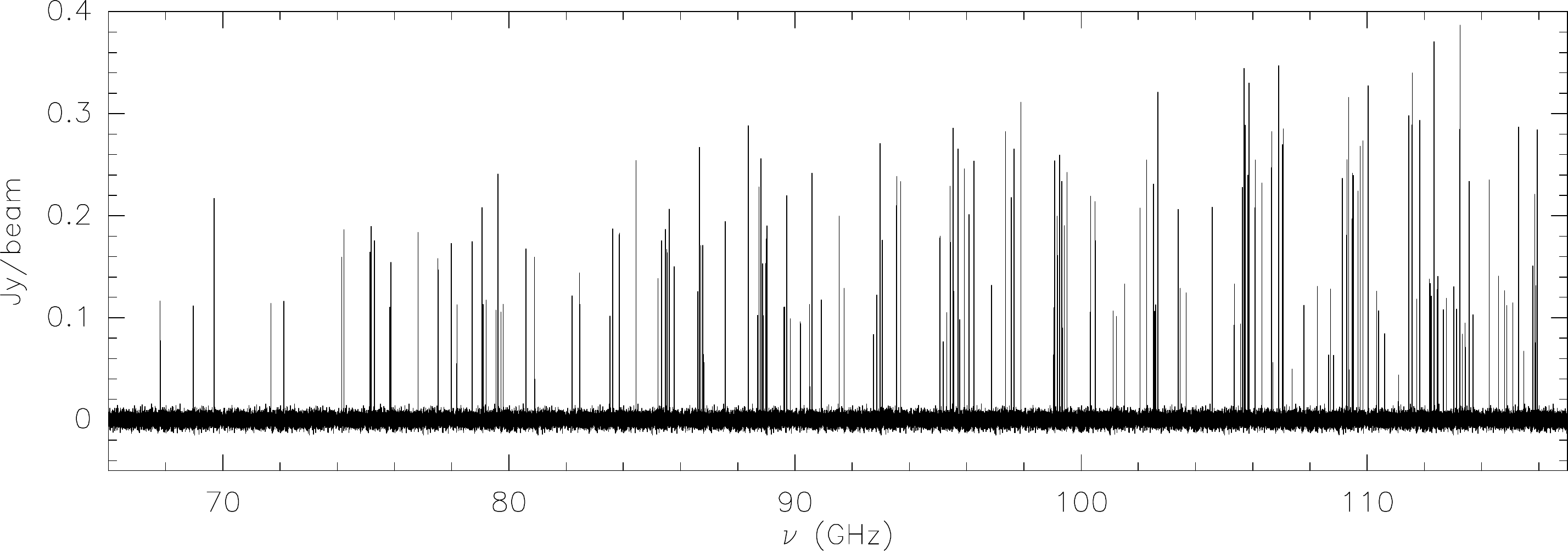}
\caption{\it Simulated spectrum of GA, EG and HCO in the spectral range covered by
ALMA Band 2+3 (67-116 GHz), considering the size of the G31 HMC  ($\sim1"$), a
temperature of 100 K and the molecular column densities estimated from the
single-dish detections (Rivilla et al., in preparation). The noise provided by ALMA
in 30 min hour on-source within beam of 1$"$ is 4.35 mJy/beam,
and has  been added to the theoretical spectrum.}
\label{simulation}
\end{figure}

The advent of ALMA offers an unprecedented opportunity to study the formation of
complex molecules of astrobiological interest in star forming regions. Fig.
\ref{simulation} shows the simulated spectrum of HCO, GA and EG in the ALMA Band
2+3, assuming Local Thermodynamic Equilibrium (LTE) conditions, a temperature of
100 K and the column densities derived from the IRAM 30m detections (Rivilla et
al., in preparation). The Band 2+3 will be particularly suitable for the study of
complex organic molecules due to several reasons:

\begin{itemize} 

\item The number of molecular rotational lines excited at these frequencies
(67--116 GHz) is significantly lower than at higher frequencies, and hence the
transitions of the complex molecules suffer less from line blending with other
species.

\item The broad frecuency coverage of Band 2+3 will allow us to detect multiple
transitions of the molecular species with different excitation energies, which is
needed to confirm robustly the detections and to derive the physical parameters
(column densities $N$ and temperatures $T_{\rm rot}$).

\item  The high spatial resolution of ALMA contributes to reduce the line
confusion with respect to single-dish observations since different molecules are
expected to arise from different regions. If all the transitions attributed to a
molecule exhibit the same spatial distribution, the identification is further
supported.

\item  Since the abundance of complex organic molecules is significantly lower
than that of simpler molecules, the high-sensitivity of ALMA opens up the
possibility to detect a large number of complex molecules that have remained
undetected so far.

\end{itemize}

\subsubsection{Comparison between observations and chemical models} 

\begin{figure}[tbh]
\centering
\includegraphics[scale=0.45]{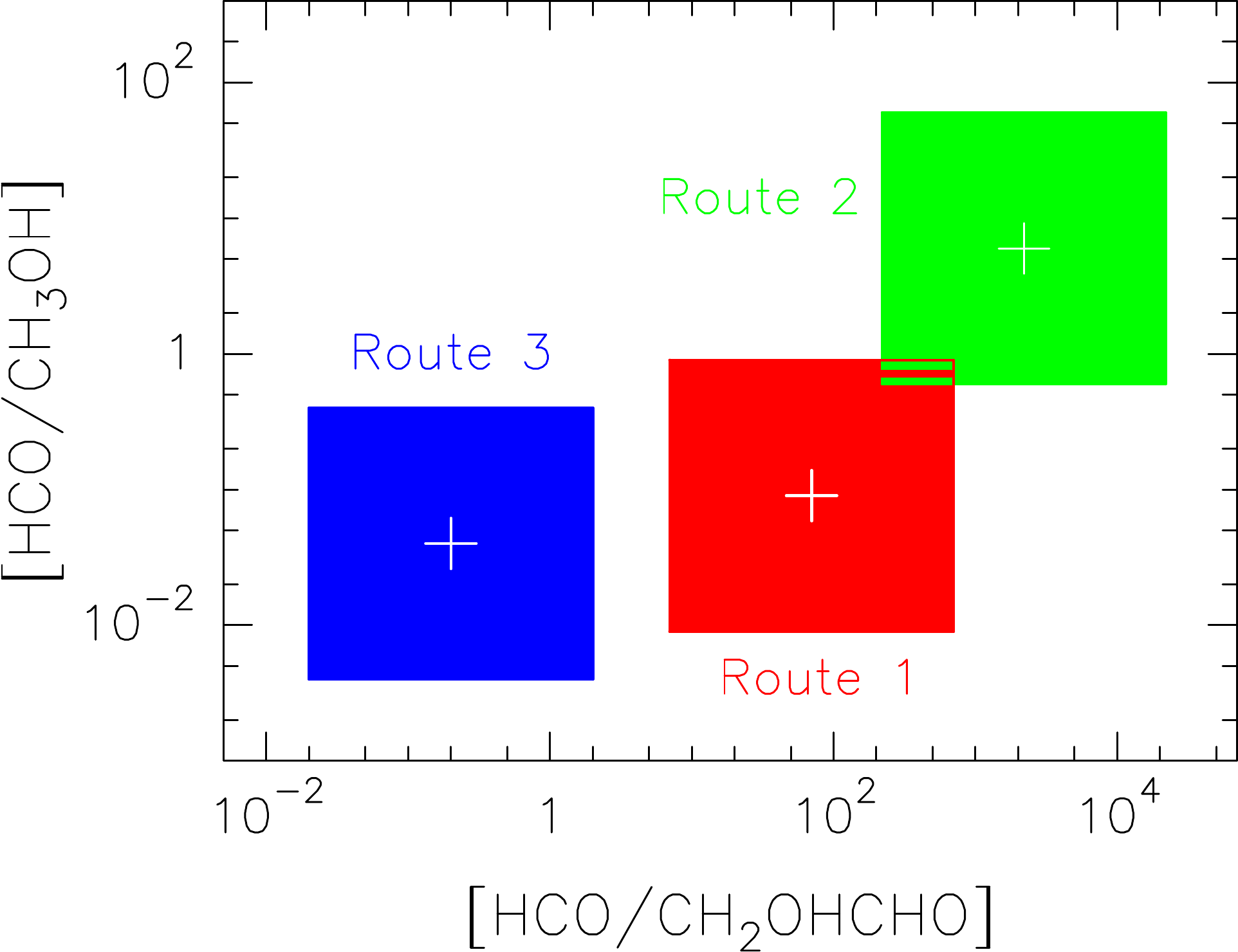}
\caption{\it Theoretical predictions of the HCO, CH$_3$OH and GA abundances
from the chemical models by Woods et al.\ (2012) at the end the surface-chemistry
phase, when the complex organic molecules (GA and EG) are formed. A source with
the physical conditions of G31 was considered. The colored boxes consider an
uncertainty of $\pm$1 order of magnitude.}
\label{models}
\end{figure}

Once the molecular species are clearly identified and its abundances well
determined, the comparison with chemical models will shed light about their
formation processes. Figure~\ref{models} shows the predictions of the suface
chemistry on the grains considering the 3 different pathways proposed by (Woods et
al.\ 2012). The different routes predict well differentiated values of the
relative surface abundances. Since gas-phase formation of GA and EG is unlikely
(Woods et al.\ 2012), the present gas abundances are expected to be directly
linked with the original surface abundances. In the case of HCO, it is also
expected that the observed gas abundance is mainly inherited from the abundance in
the grains mantles after evaporation (Gerin et al.\ 2009; Cernicharo et al.\
2012). Therefore, the comparison of the observed gas abundances with the chemical
models would reveal which of the proposed chemical routes is more likely.

\subsection{\underline{Complex organic molecules in protostellar environments in
the ALMA era}}

\subsubsection{The molecular complexity in a Sun-like forming system}

Molecular complexity builds up at each step of the Sun-like star formation
process, starting from simple molecules and  ending up in large polyatomic
species. Understanding how our own solar system, Earth and life, have come to be
is one  of the most important and exciting topics in science. Complex organic
molecules (COMs; such as methyl formate, HCOOCH$_3$,  dymethyl ether,
CH$_3$OCH$_3$, or glycolaldehyde, HCOCH$_2$OH) have been found in all the
components of the star formation  recipe (prestellar cores, hot-corinos,
circumstellar disks, shocks induced by fast jets). These species are thought  to
be mostly formed in solid state chemistry of grain mantles and then released in
the gas phase due to ice  grain mantle sublimation or sputtering, and in some
cases enhanced by gas-phase reactions. Intriguingly,  the simplest, non-chiral
amino acid, glycine (C$_2$H$_5$NO$_2$), has been found in comet and meteorite
samples  from our own Solar System, but its detection in star-forming regions is
still a matter of intense debate. The level of chemical  processing and complexity
of ices in low-mass star forming regions and protoplanetary disks and their 
relationship with the material delivered on planetary surfaces is still very
uncertain, in spite of the  astrobiological implications. The combination of
theoretical, observational and laboratory studies has  the potential of advancing
this field considerably, especially in the ALMA era and with the prospect of
expanding the  ALMA capabilities to the Band 2 (or 2+3) range.

\subsubsection{COMs: from prestellar cores to planetary systems}

\begin{figure}
\includegraphics[angle=0,width=0.9\textwidth]{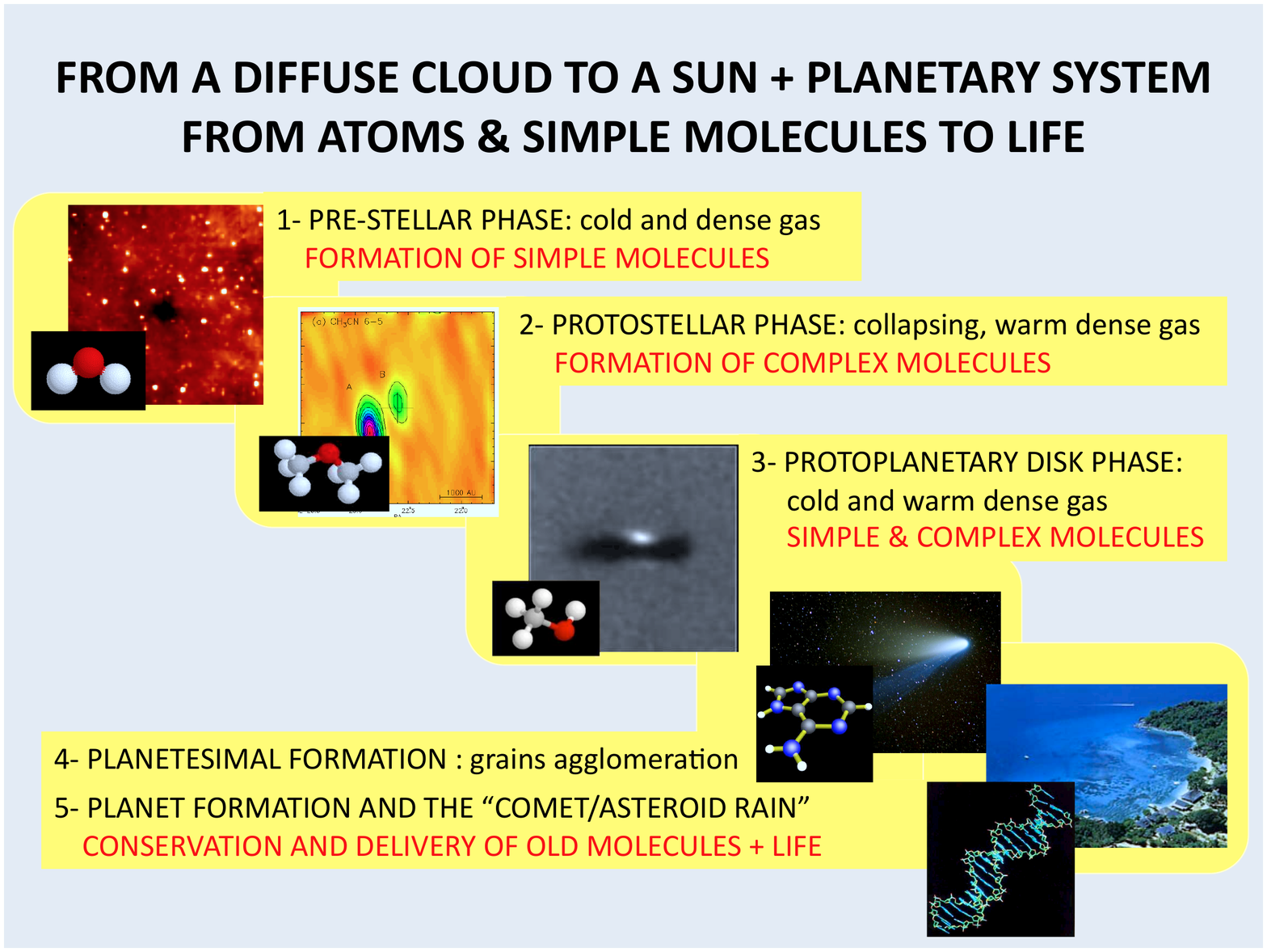}
\caption{\it Star formation and chemical complexity
(from Caselli \& Ceccarelli 2012). The formation of a star and a planetary
system, like the Solar System, passes through different phases, marked in the sketch.}
\label{life}
\end{figure}

The formation of Sun-like stars and the chemical complexity of the molecular gas
involved in the process can  be summarised as follows (see Fig.~\ref{life}, from Caselli \&
Ceccarelli 2012): (1) matter slowly accumulates toward the center  of a molecular
cloud. The central density increases while the temperature decrease. Atoms and
molecules in the gas phase  freeze-out onto the cold surfaces of the dust grains,
forming the grain mantles. Hydrogenation of atoms and  molecules takes place,
forming molecules such as water (H$_2$O), and formaldehyde (H$_2$CO).   The recent
detection of COMs in cold cores (e.g., Bacmann et al.\ 2012) support that in these
regions  the formation of new molecules in icy mantles is also caused by the
effects of UV photons and low-energy cosmic rays;  (2) the collapse starts, the
gravitational energy is converted into radiation and the envelope around  the
central object warms up. The molecules frozen on the mantles acquire mobility and
form new, more complex species.  When the temperature reaches about 100 K the
mantle sublimates, given origin to the so-called hot corino phase,  which has a
very small size ($\leq$ 0.01 pc). Molecules in the mantle are injected in the gas,
where they react  and form new, more complex, molecules.  The abundance of COMs
(such as methyl formate, HCOOCH$_3$, or  dymethyl ether, CH$_3$OCH$_3$) 
dramatically increases. A classical example is provided by IRAS16293--2422 (e.g.,
Ceccarelli et al. 2007),  where recently also glycolaldehyde (HCOCH$_2$OH), a
crucial molecule for the formation of metabolic molecules,  has been detected
(J\"orgensen et al.\ 2012; see also Taquet et al.\ 2015);  (3) simultaneously to
the collapse, a newborn protostar generates a fast and well collimated jet, 
possibly surrounded by a wider angle wind. In turn, the ejected material drives
shocks travelling through the surrounding high-density medium. Shocks heat the gas
up to thousands of K and trigger several processes such as endothermic chemical
reactions  and ice grain mantle sublimation or sputtering (e.g., van Dishoeck \&
Blake 1998). Several molecular species  undergo significant enhancements in their
abundances. The prototypical chemical rich shock in a molecular  outflow is
L1157-B1 (Bachiller et al.\ 2001). Toward this source, not only relatively simple
complex molecules,  like methanol, have been detected, but also molecules
considered hot corinos tracers, like methyl formate (HCOOCH$_3$),  ethanol
(C$_2$H$_5$OH), formic acid (HCOOH) and methyl cyanide (CH$_3$CN) (Arce et al.\
2008; Codella et al.\ 2009).   The emission of these species is concentrated in a
small ($<$1000 AU) region associated with the violent  shocks at the head
of the outflowing material. The presence of COMs in molecular outflows strongly
suggests  that these species were part of the sputtered icy mantles, because the
time elapsed since the shock is  too short for any gas-phase route to build up
COMs; (4) the envelope dissipates with time and eventually  only a circumstellar
disk remains, which is also called protoplanetary disk. In the hot regions, close
to the  central forming star, new complex molecules are synthesized by reactions
between the species formed in the protostellar phase.  In the cold regions of the
disk, where the vast majority of matter resides, the molecules formed in the
protostellar phase  freeze-out onto the grain mantles again.  Dust grains then
coagulate into larger planetesimals, the bricks of  future planets, comets, and
asteroids.

\subsubsection{The IRAM--ALMA synergy}

In the last years several large programs started collecting  unbiased spectral
surveys with high frequency resolutions and unprecedented sensitivities of
different targets considered among the typical laboratories of different stages of
the low-mass star forming process (as e.g., prestellar cores, hot-corinos,
protostellar shocks, more evolved Class I objects). One of the main goals is
indeed the detection of complex and rare molecular species in the interstellar
space through emission due to their ro-vibrational transitions. In particular,
these efforts have been recently carried out in the 80--300 GHz range using the
IRAM 30-m ground based observatory  (ASAI: Astrochemical Surveys At IRAM;
http://www.oan.es/asai) and between 500 GHz and 2000 GHz using the ESO Herschel
Space Observatory  (CHESS: Chemical HErschel Surveys of Star forming regions;
http://www-laog.obs.ujf-grenoble.fr/heberges/chess/). 

A huge effort is carried out also using interferometers to provide high-spatial
resolution images. The IRAM Plateau de Bure interferometer (PdBI) large program
CALYPSO (Continuum and Lines from Young ProtoStellar Objects;
http://irfu.cea.fr/Projects/Calypso) is providing the  first sub-arcsecond
statistical study, in the 80--300 GHz window, of the inner  environments of the
low-luminosity Class 0 sources. Also in this case spectacular forests of lines are
observed, showing an amazing large number of lines at high excitation due to COMs
such as glycolaldehyde and ethylene glycol ($aGa^{'}$--(CH$_2$OH)$_2$), 
originating from a region of radius 40--100 AU, centered on the protostar. The
collected spectra are very rich in COMs showing amazing forest of lines, 
reflecting the chemical complexity of all the components involved in the low-mass
star formation.  A particular case is represented by formamide (NH$_2$CHO), the
simplest possible amide and a central molecule in the synthesis of metabolic and
genetic molecules, which has been revealed towards both hot-corinos and
protostellar shocks  (e.g., Mendoza et al.\ 2014; L\'opez-Sepulcre et al.\ 2015).
Finally, the IRAM NOEMA large program SOLIS (Seed Of Life In Space) started its
operations with the goal to provide the first systematic COMs study towards a
large sample covering the first phases of Solar-type star formation.  The
high-resolution images will pin down where COMs are located, their abundances, and
how they are influenced by evolutional and environmental factors, shedding light 
on COMs formation and destruction routes. 

In conclusion, the synergy between spectral surveys using single-dishes and  
interferometric observations is of paramount importance to detect and analyse the COM emission.
On the one hand, the unbiased spectral surveys provide the largest possible
frequency range, and thus the most complete census of COMs; a multiline approach
is also needed to safely identify the emission spectrum of complex species. 
On the other hand, interferometric images provide the size and morphology of the emitting
regions, overcome the filling factor limitation of single-dish spectra, and allow one
to propely derive physical conditions such as density, temperature, and abundance.
In addition, the linewidths of COMs are typically reduced in interferometric observations, 
which allows a better identification of the molecular lines since it reduces line blending. 

\subsubsection{Amino acids in Solar-system precursors}

The increasing performance of millimeter instrumentation in the past years has
opened up  the possibility to carry out high-sensitivity molecular line surveys
even toward the earliest  (and coldest) stages of low-mass star formation. These
initial conditions are represented  by cold dark cores and in particular, by
pre-stellar cores, i.e., dense and cold  condensations on the verge of
gravitational collapse (central H$_2$ densities of some 10$^4$~cm$^{-3}$  and
temperatures $\leq$10 K). These observations have shown  that COMs are also
present  in the cold gas of these cores, unexpectedly revealing a high chemical
complexity in these objects  (e.g., Bacmann et al.\ 2012). Among these complex
organics, amino acids are of high interest in Astrochemistry and Astrobiology 
because of their role in the synthesis of proteins in living organisms. It is
currently believed  that their formation may have occurred in the interstellar
medium (ISM) since amino acids, including  glycine and alanine, have been found in
meteorites (e.g., Pizzarello et al.\ 1991;  Glavin et al.\ 2006) and comets (as in
Wild 2; Elsila et al.\ 2009). However,  the detection of amino acids in the ISM
remains to be reported.

Recently, it has been proposed that pre-stellar cores may be well suited for the
detection of  amino acids in the ISM (Jim\'enez-Serra et al.\ 2014). The gas
temperatures in pre-stellar cores  are $\leq$ 10 K, which yields a low level of
line confusion since the number of molecular  transitions excited at these
temperatures is small. The linewidths of the molecular line emission  in these
cores are $\leq$ 0.5 $\,$km$\,$s$^{-1}$, which allows accurate identifications of
the  observed molecular lines because they suffer less from line blending. In
addition, water vapour  has recently been found toward the central few thousand AU
of one pre-stellar core, L1544, which  indicates that a small fraction of the ices
($\sim$0.5\% of the total water abundance in the mantles)  has been released into
the gas phase (Caselli et al.\ 2012). These authors have proposed that water 
vapour in L1544 is produced by the partial photo-desorption of ices by secondary,
cosmic ray induced  UV-photons. Since COMs form in the outer layers of the
mantles,  these molecules are also expected to be photo-desorbed together with
water in pre-stellar cores, making  these objects excellent candidates to test the
detectability of amino acids in Solar-system precursors. The detection of glycine,
and of its COM precursors, in these Solar-system precursors will represent a major
milestone in Astrochemistry and Astrobiology, providing a unique opportunity to
link the pre-biotic chemistry in the ISM to their subsequent delivery onto
planetary systems.

\begin{figure}
\centering
\includegraphics[angle=0,width=0.9\textwidth]{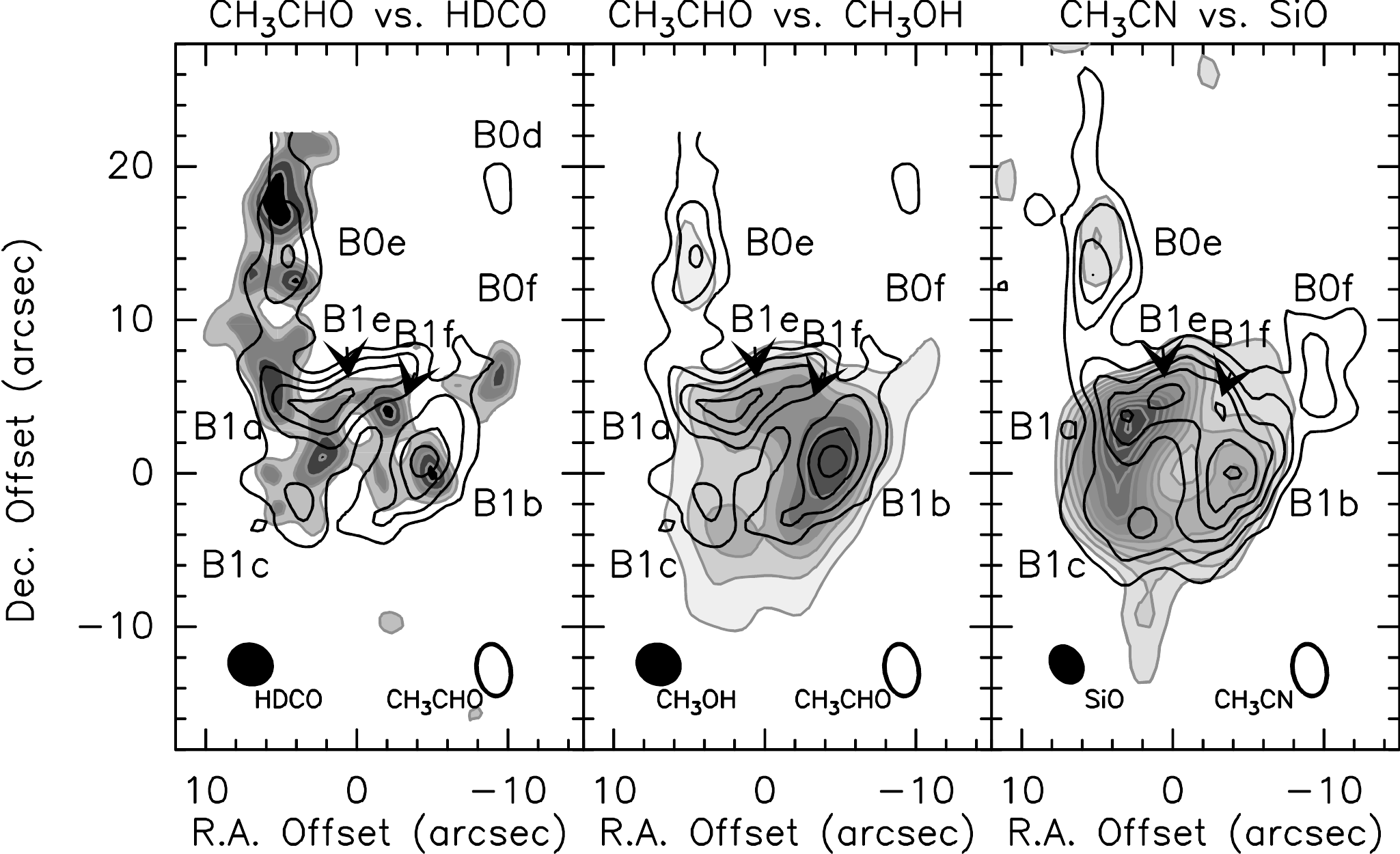}
\caption{\it Chemical differentiation in L1157-B1: the maps are centred at
$\Delta\alpha$ = +21$^{o}$ and $\Delta\delta$ = --64$^{o}$
from the driving protostar L1157-mm. 
Overlay of the integrated intensity IRAM PdBI map at 2 mm of the 
CH$_3$CHO~(7$_{0,7}$--6$_{0,6}$)AE (contours, from Codella et al.\ 2015)
on the HDCO~(2$_{1,1}$--1$_{0,1}$)  
(left panel, from Fontani et al.\ 2014) and CH$_3$OH~(3$_{K}$--2$_{K}$) 
emission (middle panel; from Benedettini et al. 2013).
Righ panel compares the CH$_3$CN~(8$_{K}$--7$_{K}$) and SiO~(2--1)
spatial distributions (Gueth et al.\ 1998, Codella et al.\ 2009).
Labels are for the different L1157-B1 molecular clumps well
imaged in the CH$_3$CN map. 
The synthesised beams are shown in the bottom part of the panels.}
\label{L1157}
\end{figure}

\subsubsection{Hot protostellar shocks versus hot-corinos}

The L1157 region at 250 pc hosts a Class 0 protostar (L1157-mm) driving a
spectacular chemically rich bipolar outflow associated with   cavities seen in the
IR H$_2$ and CO lines (e.g., Neufeld et al.\ 2009; Gueth et al.\ 1996). The bow
shocks, when mapped with the IRAM PdB and VLA interferometers, are well traced by
typical products of both grain mantle sputtering (such as  NH$_3$, CH$_3$OH, and
H$_2$CO) and refractory core disruption (SiO).  In particular, the bright and
young ($\sim$2000 yr) L1157-B1 shock offers   an exceptional opportunity to
investigate in details the chemical composition of the grain ice mantles as well
as how the gas phase  is chemically enriched after the shock occurrence.

Thanks to PdBI observations (see Fig.~\ref{L1157}), three COMs have been so far imaged in
L1157-B1: CH$_3$OH, acetaldehyde (CH$_3$CHO), and CH$_3$CN (Benedettini et al.\
2013; Codella et al.\ 2009, 2015). The CH$_3$CN image shows a clumpy structure
superimposed to the classical B1 arch-like shape,  displaying a unique continuous
structure tracing the propagation of a large bow shock. Recently, also CH$_3$CHO
has been mapped at PdBI (Codella et al.\ 2015), and also in this case there is a
good agreement with the CH$_3$OH (and CH$_3$CN) spatial distribution, confirming
COMs are emitting where mantles have been recently sputtered. This is further
indicated by the excellent spatial correpondance with HDCO emission (Fontani et
al.\ 2014) tracing deuterated formaldehyde formed on grains and then injected in
the shocked gas. For acetaldehyde, the abundance ratio with respect to methanol is
quite high ($\sim$ 10$^{-2}$) and in any case similar to that observed towards a
typical hot-corino such as IRAS16293--2422 (Bisschop et al.\ 2008).  These findings
support for CH$_3$CHO either a direct formation on grain mantles or a quick ($\le$
2000 yr) formation in a chemically enriched gas phase.  

These first results show the importance of studying shocked regions as
laboratories where (i) to investigate the COMs formation routes  and (ii) to
verify whether the study of hot-corinos chemistry can be affected by COM emission
arising from shocked envelope material at the base of the inner (unresolved) jet. 
Interferometric observations are instructive to minimise   beam dilution effects.
In addition, the multiline approach  is needed to carefully sample the excitation
conditions and correctly  derive the abundances of COMs. 

\begin{figure}
\centering
\includegraphics[width=0.8\textwidth]{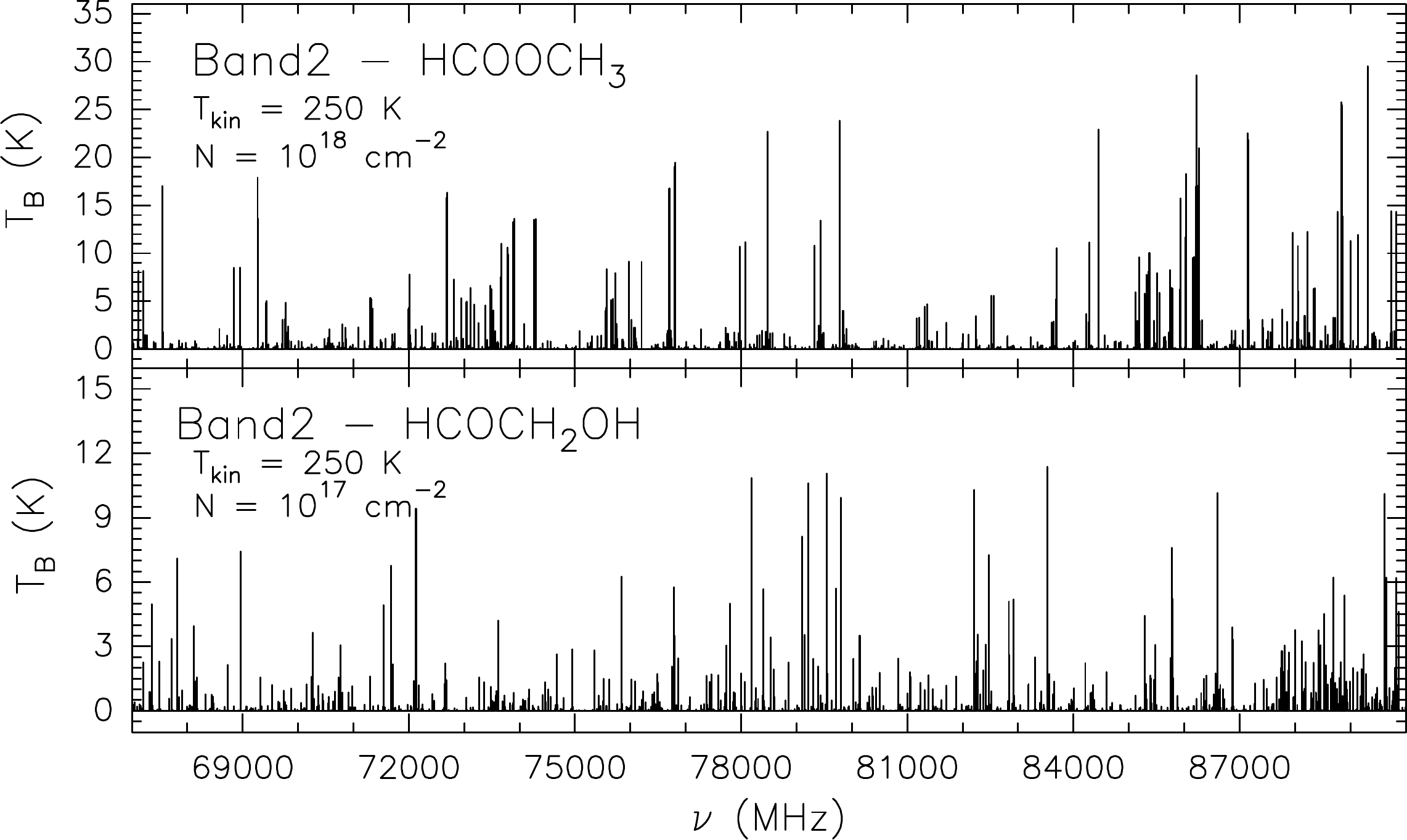}
\caption{\it Simulations of the spectrum of HCOOCH$_3$ and HCOCH$_2$OH as observed by
ALMA Band 2 modeled with GILDAS-Weeds
(Maret et al.\ 2011) in the range $\sim$ 67--90~GHz, assuming LTE
conditions and physical parameters expected for methyl formate and
glycolaldehyde in hot-corinos
(see e.g., J\"orgensen et al.\ 2012): $T_{\rm kin} = 250$ K,
$N$(HCOOCH$_3$) = $10^{18}$ cm$^{-2}$,
$N$(HCOCH$_2$OH) = $10^{17}$ cm$^{-2}$, source size = 1",
and line FWHM = 4 km s$^{-1}$.}
\label{glyco}
\end{figure}

\subsubsection{Justification for ALMA Band 2+3}

The numerous detections of high-excitation COM lines at frequencies larger than 80
GHz   call for observations at lower wavelengths, where heavy species are expected
to emit (JPL, http://spec.jpl.nasa.gov, and CDMS,
http://www.astro.uni-koeln.de/cdms). Even more important, the millimeter frequency
bands are often so full of lines that it is paradoxically difficult to identify
species through a large number of emission lines simply due to confusion. On the
other hand, lower frequency bands are relatively clear, given that low energy
transitions of light molecules fall at much higher frequencies. The completion of
the results obtained at mm-wavelengths with  spectral surveys at lower
frequencies  will allow one to have the possibility to have, for different
species,  a large number of transitions, which is needed to reliably detect the
largest COMs for which the population is distributed over many energy states,
having large partition functions. Figure~\ref{glyco} shows as an example the simulation of
the  HCOOCH$_3$ and HCOCH$_2$OH spectrum as expected for  ALMA Band 2 modeled with
GILDAS-Weeds (Maret et al.\ 2011) in the range $\sim$ 67--90 GHz, assuming LTE
conditions and physical parameters expected in hot-corinos (see e.g., J\"orgensen
et al.\ 2012).  In particular, if we consider bright lines (S$\mu^2$ $\geq$ 1
D$^2$) at low excitation (E$_u$ $\le$ 20 K), then the ALMA Band 2+3 band contains
a considerable number of COMs transitions. As an example, the line densities for
bright, low-excitation transitions of   methyl formate and glycoaldehyde in the
67--116 GHz range is $\sim$ 0.4--0.8 line/GHz, ten times higher than in the
116--350 GHz spectral window.

\begin{figure}
\centering
\includegraphics[angle=-90,width=.99\textwidth]{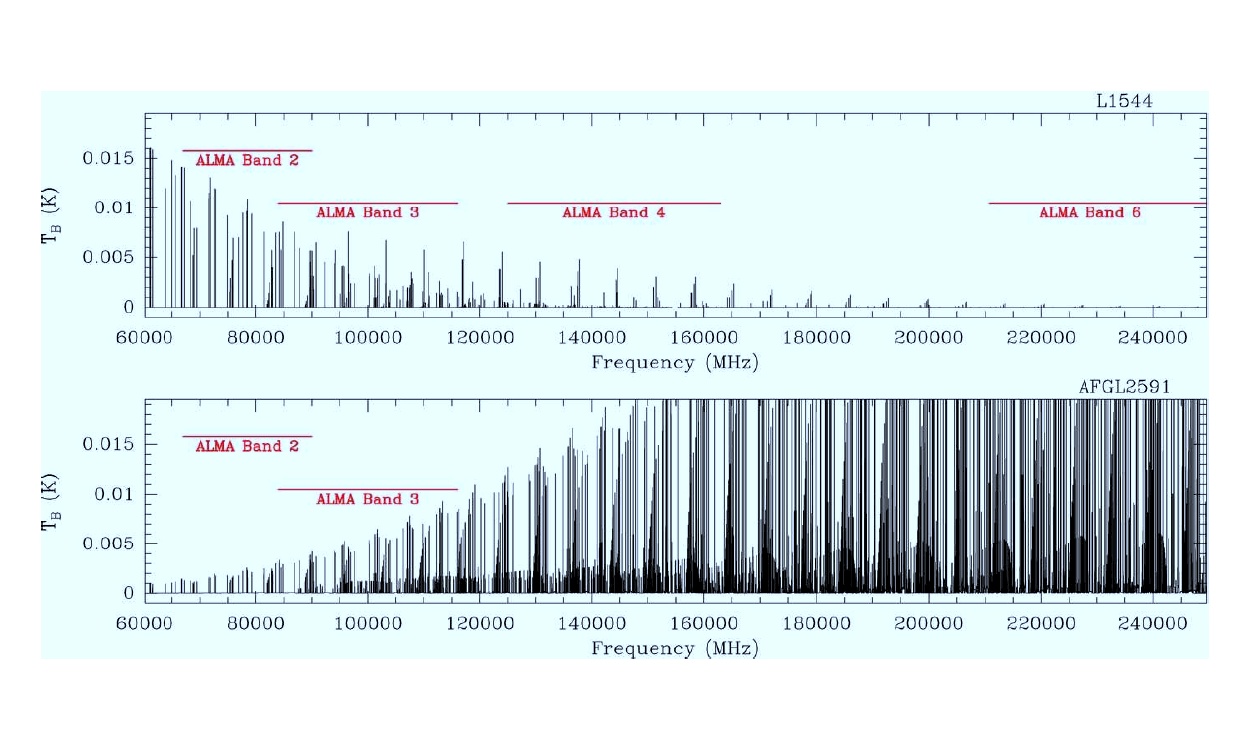}
\vspace{-.8cm}
\caption{\it Upper panel: Simulations of the spectrum of glycine (conformer I) obtained
between 60 GHz and 250 GHz, considering the physical structure
of the L1544 pre-stellar core (Jim\'enez-Serra et al.\ 2012), a solid glycine abundance of a few 10$^{-4}$
with respect to water in ices, and LTE conditions. Horizontal red lines indicate the frequency coverage
of ALMA Bands 2, 3, 4, and 6. The glycine lines with frequencies 60--80 GHz show
peak intensities $\geq$ 8--10 mK,
detectable with the Band 2 receivers of ALMA.
Lower panel: LTE spectrum of glycine predicted for the same frequency range
(60--250 GHz) but for the AFGL2591 hot molecular core.
The glycine lines at 1 mm are brighter than in the pre-stellar core case. However, line confusion and
line blending is a major issue in the detection of glycine in hot cores.}
\label{glycine}
\end{figure}

The COM emission is expected to come from regions at sub-arcsec scale (hot
corinos, jets, shocks). With a largest baseline of about 16~km, ALMA will provide
in Band 2+3 a well suited spatial resolution  of $\simeq$ 50 mas. As a comparison,
the IRAM NOEMA interferometer currently reaches   a synthetisized beam $\sim$ 1
arcsec at 3~mm (the array does not cover frequencies lower than 80 GHz;
http://www.iram.fr). In addition, the effective area offered by ALMA will be
definitely larger than that available using the IRAM--NOEMA, even in its final 12
antennas configuration ($\sim$ 2100~m$^2$). Therefore the combination of spatial
resolution and sensitivity offered by ALMA  is fundamental to resolve and image
the emitting region. In this way, (i) we will have bright line emission, and (ii)
we will correctly evaluate COMs abundances. 

A special case is represented by glycine.  Simple radiative transfer calculations 
towards the L1544 pre-stellar core show that several glycine lines could reach
detectable levels (peak line intensities $\geq$10~mK) in the frequency range
between 60 and 80 GHz (Jim\'enez Serra et al.\ 2014). These calculations assume a
solid abundance of glycine in ices of a few 10$^{-4}$ with respect to water,
similar to those synthesized in laboratory experiments of UV photon- and
ion-irradiated interstellar ice analogs (e.g., Holtom et al.\ 2005). This solid
abundance translates into a maximum gas-phase abundance of glycine of $\sim$
8$\times$10$^{-11}$ after ice photo-desorption. In Fig.~\ref{glycine},
Jim\'enez-Serra et al.\ (2012) report the predicted intensities of glycine for the
L1544 core, assuming that the gas-phase abundance of glycine remains constant
across the core. The detection of glycine, and of its precursors, in these
Solar-system precursors will represent a major milestone in Astrochemistry and
Astrobiology, providing a unique opportunity to link the pre-biotic chemistry in
the ISM to their subsequent delivery onto planetary systems.

\subsection{\underline{Chemistry of protoplanetary disks with ALMA Band 2}}
 
\subsubsection{Protoplanetary disks and the origin of water and organic molecules on
Earth and Earth-like planets}

Protoplanetary disks are the birthplaces of planets; thus, the study of their
physical and chemical structure is fundamental to comprehending the formation of
our own solar system as well as that of extrasolar planetary systems. According to
our current understanding the primordial disk is composed by $\mu$m-sized dust
grains which grow and settle towards the midplane forming planetesimal, i.e.
bodies from meters to kilometer sizes which are the building blocks of planets,
asteroids, and comets (see Testi et al.\ 2014 for a review on grain growth
processes). In the outer cold disk midplane molecules freeze-out onto dust grains
covering them with icy mantles. This process is thought to be the first step
towards the formation of Complex Organic Molecules (COMs) and prebiotic molecules,
which are then inhereted by the forming planetesimal and planets.

One of the main scientific goals of iALMA is to understand the chemistry of COMs
and especially their role as building blocks of pre-biotic molecules in
exoplanetary systems. There is strong evidence that water on Earth has been
delivered after its formation by icy bodies from the outer region of the Solar
System pushed inside by the rearrangement of the orbits of the four major planets
(Jupiter, Saturn, Uranus and Neptune). These small icy bodies are supposed to have
delivered water and other volatiles on the inner rocky planets. As part of this
process, complex organic species locked in the ices may have been delivered to
Earth as impurities, together with water. Indeed, complex molecules and even
simple amino acids, such as Glycine, are known to be present in comets and
meteorites, well mixed with water and carbon monoxide ice (e.g.,
Pizzarello et al.\ 1991; Glavin et al.\ 2006). 

COMs are thought to form in the interstellar ices around young stars as well as in
the ices in circumstellar disks as a consequence of solid-state chemistry induced
by energetic radiation from the forming stars (see Caselli \& Ceccarelli 2012 for a
review). This scenario has been confirmed by laboratory experiments of ice
irradiation by energetic particles(e.g., Modica et al.\ 2010). To date complex
organic molecules have been observed at all stages of the star formation process,
i.e. in prestellar cores (e.g., Bacmann et al.\ 2012), in the hot corinos around
Class 0 protostars (e.g., Cazaux et al.\ 2003; Bottinelli et al.\ 2007), and in
the shocks induced by protostellar jets (e.g.,  Arce et al.\ 2008; Codella et al.\
2009, 2015a). However, the search for  COMs in protoplanetary disks has only
recently started thanks to the high angular resolution and sensitivity offered by
millimeter interferometers such as the Plateau de Bure Interferometer and ALMA.

\subsubsection{Chemical complexity in protoplanetary disks}
\label{sect:disk_chem}

Sub-millimeter and millimeter surveys of disk continuum emission allowed deriving
their main physical properties, e.g. the inclination, the size, the dust
properties and mass. On the other hand, the disk chemical composition is still
poorly known as so far only emission from a few simple molecules has been
detected, such as CO, SO, CS, CN, HD, OH, CH$^{+}$, HCN, H$_2$O, CO$_2$,
HCO$^{+}$, N$_2$H$^{+}$, C$_2$H$_2$, H$_2$CO, HC$_3$N (e.g., Bergin et al.\ 2013,
Carr et al.\ 2008; Chapillon et al.\ 2012a, 2012b; Dutrey et al.\ 1997, 2011;
Guilloteau et al.\ 2013; Oberg et al.\ 2010, 2011; Pontoppidan et al.\ 2010a; Thi
et al.\ 2004, 2011). The chemical inventory of protoplanetary disks has been very
recently enlarged by ALMA observations with the first detection of c-C$_3$H$_2$
and CH$_3$CN (Qi et al.\ 2013b, Oberg et al.\ 2015; see Dutrey et al.\ 2014 for a
review). 

\begin{figure}
\centering
\includegraphics[angle=-90,width=0.8\textwidth]{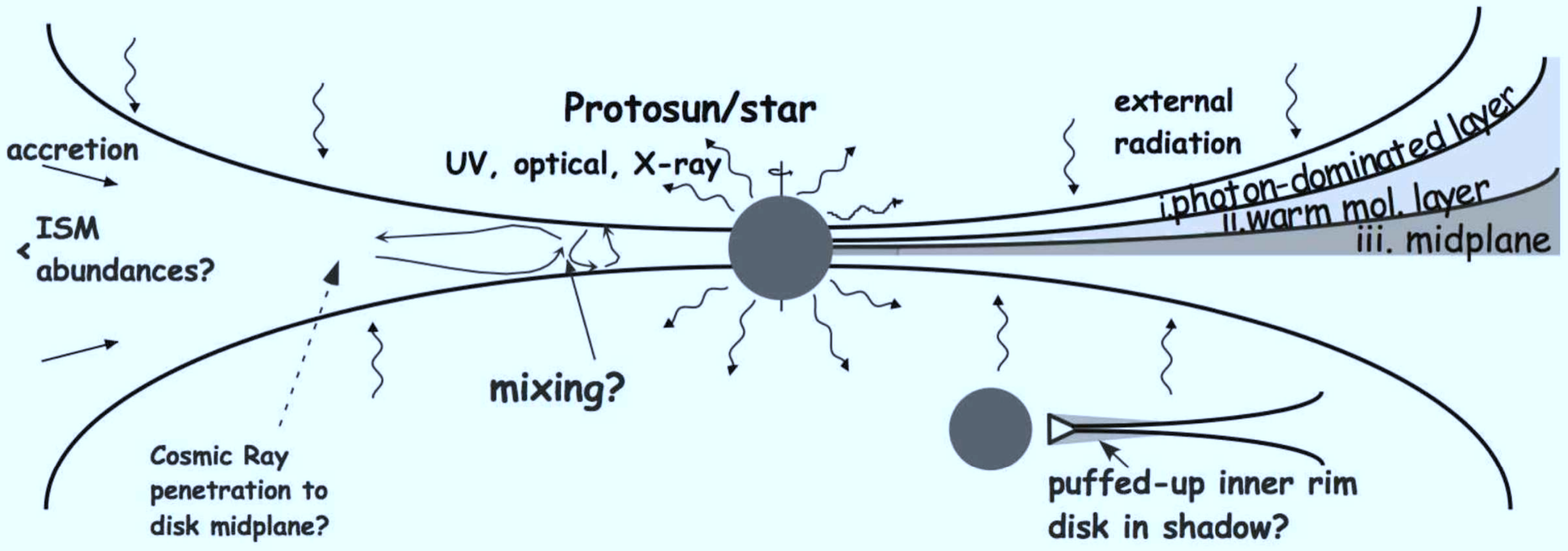}
\caption{\it Chemical structure of protoplanetary disks. Vertically the disk is 
schematically divided into three zones: a photon-dominated layer,  a warm
molecular layer, and a midplane freeze-out layer. Various non-thermal inputs,
cosmic ray, UV, and X-ray drive chemical reactions. From Bergin et al.\ (2007).}
\label{fig:disk_chem}
\end{figure}

Protoplanetary disks have a stratified structure both vertically and radially with
strong gradients in temperature and density as illustrated in
Fig.~\ref{fig:disk_chem} (e.g., Bergin et al.\ 2007; Dullemond et al.\ 2007;
Dutrey et al.\ 2014). In the disk surface layers molecules are destroyed due to
photodissociation by ultraviolet (UV) and X-rays radiation from the central
protostar. In the outer disk and close to the disk midplane the temperature
quickly drops to $\le$100 K, therefore molecules freeze out onto dust grains at
temperatures determined by their binding energies. Hydrogenated molecules, such as
CH$_3$OH and H$_2$CO, are thought to be efficiently formed on the grain mantles
due to hydrogenation of simple molecules and atoms (e.g., CO, N, C). Then
reactions between radicals induced by UV photons or cosmic rays (CR) gives rise to
complex organic molecules such as methyl formate (CH$_3$OCHO) and dimethyl ether
(CH$_3$OCH$_3$). According to recent protoplanetary disk models, COMs can reach
ice-phase abundances of 10$^{-6} - 10^{-4}$. However, only a small percentange of
those molecules are released in gas-phase by non-thermal processes such as photo-
and CR- desoprtion ($\sim 10^{-12} - 10^{-7}$) (Walsh et al.\ 2014). Hence, the
chemical complexity of protoplanetary disks remains basically hidden in their ices
making it difficult to observe it.

The recent detection of water (Caselli et al.\ 2012) and complex molecules
(Bacmann et al.\ 2012) in prestellar cores seems to confirm that COMs can be
produced also in cold environments by cold dust grains chemistry and then released
by the icy mantles by non-thermal processes. The high densities and low
temperature conditions in the cold disk midplane resemble those of pre-stellar
cores calling for a deep search of those molecules in protoplanetary disks.

\subsubsection{Justification for ALMA Band 2+3}

Protoplanetary disks have sizes of a few tens to a few hundreds of AU, i.e.
comparable or larger than our solar system. Hence, even for the nearest star
forming regions ($\sim 100$ pc), disks extend on scales $\le 1''$, therefore high
angular resolution is needed to resolve them and to avoid beam dilution. Moreover,
as explained above, the abundance of complex molecules in disks is expected to be
low ($\sim 10^{-12} - 10^{-7}$) asking for high sensitivity. The ALMA
interferometer is the ideal instrument to search for molecular complexity in
protoplanetary disks as the full ALMA array will offer a resolution down to $\sim
6$ milliarcseconds (which will allow resolving the disk structure down to AU
scales for the nearest sources) coupled to very high sensitivity.

In particular, ALMA Band 2+3 will allow us to observe at low frequencies (from
67 to 116 GHz) and to address a number of scientific goals as listed below:

\begin{itemize}

\item[-] {\bf The search for complex organic and pre-biotic molecules in
protoplanetary disks: } ALMA already proved to be sensitive enough to search for
complex molecules in disks, as it recently allowed obtaining the first resolved
map of H$_2$CO in the transitional disk around Oph IRS~48 (van der Marel et al.\
2014), the first detection of c-C$_3$H$_2$ in the disk around HD 163296 (Qi et
al.\ 2013b), and the first detection of the complex cyanides CH$_3$CN (and
HC$_3$N) in the disk around MWC 480 (Oberg et al.\ 2015). As mentioned in
Sect.~\ref{sect:disk_chem} complex molecules are thought to be formed in large
abundances on the icy mantles of dust grains in the cold disk midplane, then
partially released in gas-phase due to non-thermal processes, e.g. cosmic rays or
UV desorption. In these cool conditions, the rotational spectra of complex
molecules is significantly skewed towards low frequencies, as a consequence the
detection with ALMA is limited to the lowest bands, e.g.\ Band 2+3 covering the
$67-116$ GHz range. 

In particular ALMA Band 2+3 would allow searching for simple amino acids  like
Glycine (NH$_2$CH$_2$COOH) desorbed from icy mantles in disks. Glycine has been
found in meteorites and comets  of our own Solar System (Oizzarello et al.\ 1991;
Glavin et al.\ 2006) and laboratory experiments reported the formation of simple
amino acids, including Glycine, via ultraviolet and ion photolysis of interstellar
ice analogs (e.g., Mu\~noz-Caro et al.\ 2002). The recent detection of water
vapour in the prestellar core L1544 shows that even in the cold interior of cores
cosmic rays and secondary high energy radiation can desorb a measurable quantity
of molecules from the ices (Caselli et al.\ 2012). Jim\'enez-Serra et al.\ (2014)
have shown that also pre-biotic molecules (like Glycine) may be detectable if
desorbed together with the water molecules. In particular, the brightest emission
lines can be observed with ALMA Band 2+3 (see Fig.~\ref{glycine} for the
predictions of Glycine emission in the prestellar core L1544). We expect that this
result can be extended to the cold midplanes of protoplanetary disks.

The direct detection in the gas phase and measurements of the abundance with
respect to water of complex organic molecules would be a significant milestone in
studying our cosmic heritage and the ability of ISM and disks chemistry to produce
the raw material required for the development of life on exoplanets.


\item[-] {\bf The CO snowline:} The snow lines in protoplanetary disks are the condensation fronts where
volatiles (e.g. CO, H$_2$O) freeze-out onto the icy mantles of dust grains. The
snow lines are believed to play a crucial role in the formation of planets as
the planet formation efficiency and composition are intimately linked to their
locations in the disk (see, e.g., Qi et al.\ 2013a; Pontoppidan et al.\ 2014). In fact, icy
grains located outside the snow line have higher mass surface densities and
stickiness compared to bare grains, which favors dust coagulation and particle
growth, thus enhancing the efficiency of planet formation. Moreover, the pile-up
of dust grains in pressure traps just inside the snow line could allow the
grains to grow beyond the radial drift and fragmentation barrier up to
planetesimal size. The unprecedented angular resolution offered
by ALMA allows us to locate the snow lines in protoplanetary disks around young
solar-analogs by observing continuum emission at different wavelengths (e.g.,
Guidi et al. in preparation) or by means of the so-called chemical rings. For
example, the CO snowline can be traced by molecular ion which are present in large
abundance where CO is frozen out, e.g. DCO$^{+}$ (Mathews et al.\ 2013) and
N$_2$H$^{+}$ (Qi et al.\ 2013a).  Interestingly the lowest transition of
N$_2$H$^{+}$ (the 1--0 transition) is at 93.2 GHz; therefore ALMA Band 3
observations allow a statistical study of the CO snowline in a large sample of
protoplanetary disks.

\begin{figure}
\centering
\includegraphics[angle=0,width=0.45\textwidth]{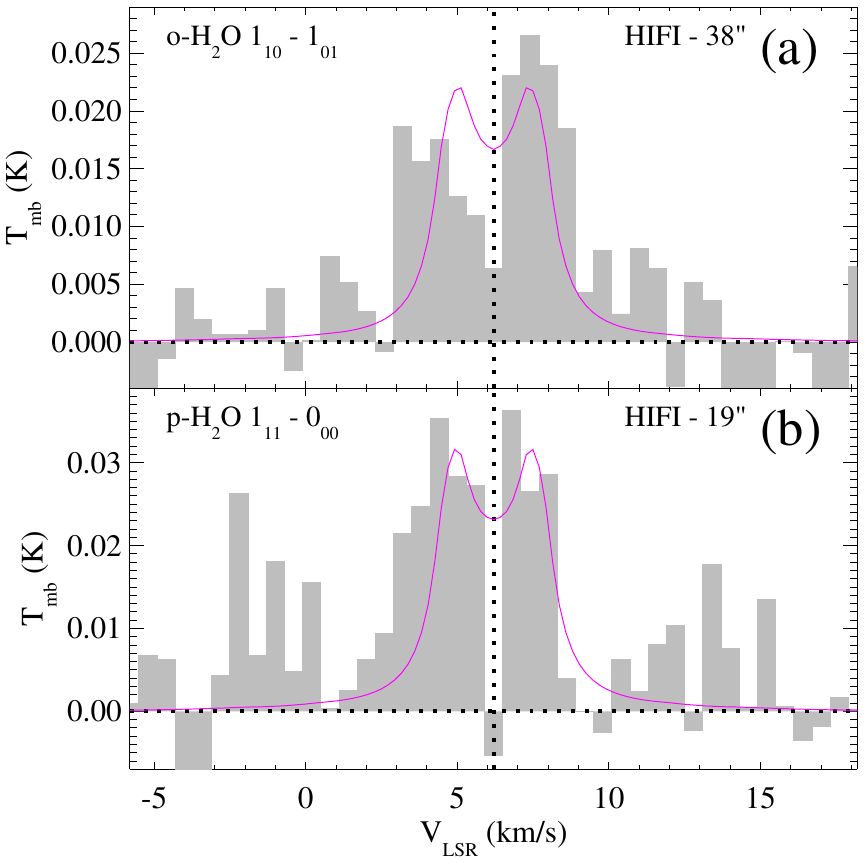}
\includegraphics[angle=0,width=0.45\textwidth]{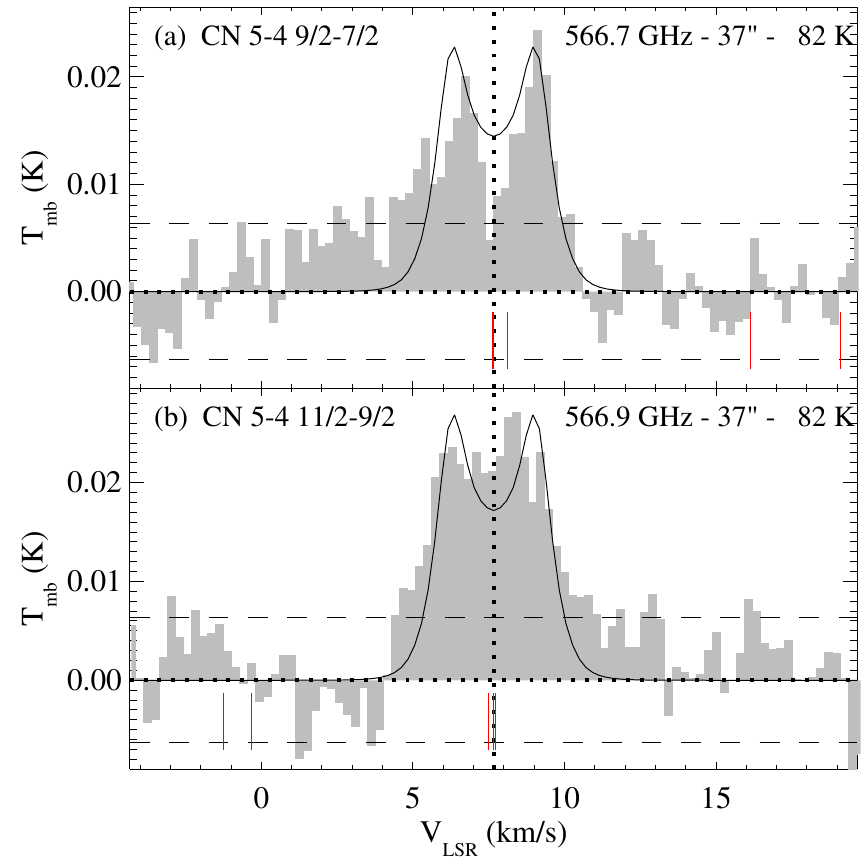}
\caption{{\it Left panel:} Water emission from the outer region of the disk of DG
Tau. The detected water vapour implies a water reservoir on ices of thousands to
hundred thousands Earth Oceans (from Podio et al.\ 2013). {\it Right panel:} The
embedded disk of T Tau N traced by high-excitation CN~(5--4) lines (from
Podio et al.\ 2014a).}
\label{fig:disk_obs}
\end{figure}

\item[-] {\bf Water distribution and deuteration ratio in protoplanetary disks:}
Water distribution and deuteration ratio (HDO/H$_2$O) in protoplanetary disks is
key to understand its origin on Earth and to test the scenario of water delivery
on Earth by icy bodies from the outer regions of the Solar System, such as
asteroids and comets (e.g., Matsui et al.\ 1986; Morbidelli et al.\ 2000;
Pontoppidan et al.\ 2014; van Dishoeck et al.\ 2014). Recently {\it Herschel}
allowed to detect for the first time cold water vapour in the outer regions of
protoplanetary disks around young solar-analogs, implying a water reservoir
trapped on the icy mantles of dust grains of thousands to hundred thousands Earth
Oceans (Hogerheijde et al.\ 2011; Podio et al.\ 2013) (see
Fig.~\ref{fig:disk_obs}). The observations obtained with {\it Herschel},
however, are spatially unresolved. To obtain spatially resolved images in the HDO
and H$_2^{18}$O lines is key to validate the scenario of water delivery, as it
allows to (i) investigate the radial distribution of water in disks, (ii) to
measure the D/H ratio of water in the disk, i.e. the HDO/H$_2$O abundance ratio.
The D/H ratio of Earth Ocean's water ($1.5576 \times 10^{-4}$ 'Vienna Standard
Mean Ocean Water', VSMOW or SMOW) is at least a factor of 6 higher than the
protosolar nebula value ($2.5 \times 10^{-5}$: Robert et al.\ 2000), i.e. of the
D/H ratio of the gas out of which our solar system formed. Thus, there should have
been an enhancement of the D/H ratio in water at some stages during the solar
system evolution. Indeed, the Earth Ocean's HDO/H$_2$O ratio is similar to what
measured in carbonaceous chondrites from the outer asteroid belt (Dauphas et al.\
2000) and recently in Jupiter-family comets (Hartogh et al.\ 2011), while it is a
factor two lower than in Oort Cloud comets (Mumma et al.\ 2011). This suggests
that both asteroids and comets may have contributed to delivering water to our own
Earth during the Late Bombardment phase(Morbidelli et al.\ 2000; Hartogh et al.\
2011). On the other hand, the HDO/H$_2$O ratio measured in our solar system
(Earth, asteroids, and comets) is much lower than what measured in the cold outer
regions of protostellar envelopes (HDO/H$_2$O $\sim 10^{-2}-10^{-3}$, e.g.
(Coutens et al.\ 2012, 2013) but similar to the ratio in the warm inner envelope
(HDO/H$_2$O $\sim 3-5 \times 10^{-4}$, e.g. J\"orgensen et al.\ 2010; Persson et
al.\ 2014). Therefore, it is still not clear whether the D/H ratio in the solar
system water was already set by ices in the early (pre-)collapse phase and
transported largely unaltered to the comet-forming zone, or whether further
alteration of D/H took place in the solar nebula disk.  ALMA Band 2 should allow
us to answer the open questions about the origin of water on Earth as it will
allow observing at high resolution and sensitivity one of the lowest transitions
of HDO, the $1_{1,0} - 1_{1,1}$ at 80.6 GHz (Note that this line is not covered by
the IRAM telescope).

\end{itemize}

Besides the listed scientific cases, the ALMA Band 2+3 also covers other
molecular transitions which can help our understanding of protoplanetary disks
chemistry, e.g.:

\begin{itemize}

\item[-] the lowest transition of CN (the 1--0 line at 113 GHz, E$_{\rm up} \sim
5$ K). CN has proved to be an efficient tracer of protoplanetary disks (e.g.,
Chapillon et al.\ 2012a; Guilloteau et al.\ 2013; Podio et al.\ 2014a) (see
Figure~\ref{fig:disk_obs});

\item[-] the SO $2_2 - 1_1$, $3_2 - 2_1$, $4_5 - 4_4$, $2_3 - 1_2$ lines falling
between 86 and 109 GHz (E$_{\rm up} \sim 9-21$ K).  SO is a key molecule to
understand the chemistry of sulfur-bearing molecules but is hard to detect in
protoplanetary disk (Fuente et al.\ 2010; Dutrey et al.\ 2011; Guilloteau et al.\
2013). On the other hand, recent studies suggest that SO can probe the accretion
shock occuring at the interface between the envelope and the disk in Class 0
protostars (Lee et al.\ 2014; Sakai et al.\ 2014; Podio et al. 2015).

\end{itemize}

While to detect forming disks around young protostars (Class 0, I) requires to
observe high excitation transitions of those molecules to avoid confusion with the
emission from the surrounding envelope and/or jets (e.g.,
Podio et al.\ 2014a, 2015; Sakai et al.\ 2014), the lowest transitions of CN and
SO can be efficiently used to image the evolved disks around Class II and III
sources (e.g., Chapillon et al.\ 2012a; Guilloteau et al.\ 2013).

\subsection{\underline{Constraining the nature of flares from young stars in the millimeter regime}}

\subsubsection{Flare events from young stars}

High energy processes during the first evolutionary stages of star formation are
responsible for both centimeter/millimeter and X-ray emission (Feigelson \&
Montmerle 1999). Low-mass pre-main sequence (PMS) stars are well known strong
X-ray emitters. Their enhanced magnetic activity with respect to more evolved
stars produces violent reconnection events in the corona of the stars, where the
plasma heated to high temperatures strongly emits variable X-ray emission. Our
understanding of the X-ray emission from young stars has dramatically increased in
the recent years due to Chandra and XMM-Newton (Getman et al.\ 2008, Arzner et
al.\ 2007). X-ray observations have revealed thousands of PMS stars in tens of
stellar clusters, resulting in good constraints on their X-ray properties such as
plasma temperatures, levels of variability, luminosities and X-ray flare rate
(e.g., Wolk et al.\ 2005).

In contrast, the physics associated with the cm/mm events (nature and origin of
the emission, variability, timescales, flaring rate) are still poorly constrained.
Drake \& Linsky (1989) proposed that these flares might be produced by the same
coronal activity that is responsible for bright X-ray emission (see review by
G\"udel 2002). It would be expected then that electrons spiraling in the magnetic
field of the corona produce non-thermal and highly variable gyrosynchrotron
radiation. Moreover, ionized material in the vicinity of stars, in circumstellar
disks or envelopes or at the base of bipolar outflows, also produce thermal
free-free (bremsstrahlung) radiation.

\begin{figure}[tbh]
\centering
\includegraphics[angle=0,scale=0.5]{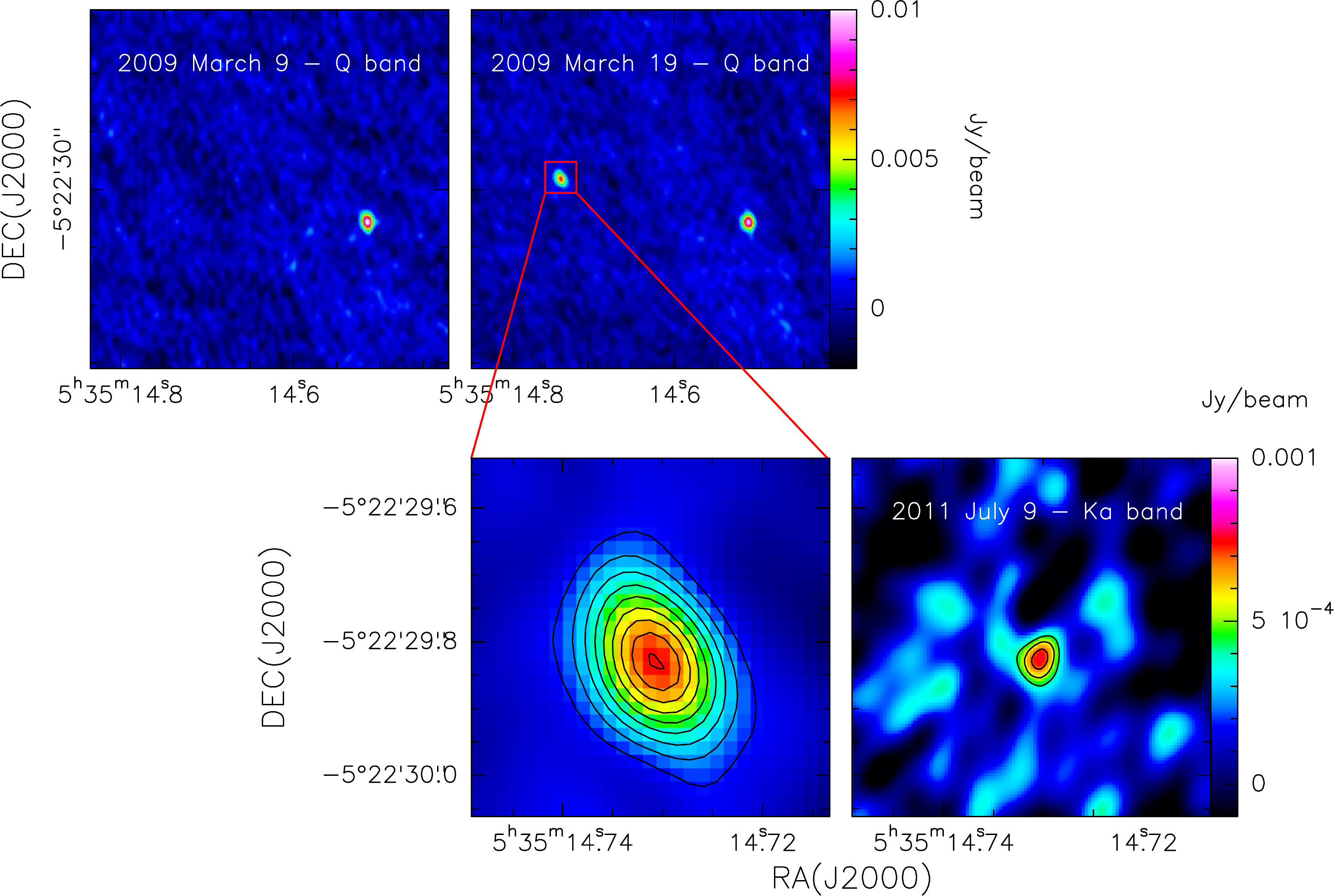}
\caption{\it (Upper panels): Very Large Array (VLA) observations of the
Orion Hot Core region at 7 mm from 2009 March 9 (left) and 2009 March 19 (right),
from Rivilla et al. (2015). The well known massive source I is detected
in both images, while a new centimeter flaring source OHC-E appeared 3.3$''$
toward the northeast on 2009 March 19. (Lower left
panel): Zoom-in view of the detection of the source OHC-E from the 2009 March 19
observation. (Lower right panel): Second detection of OHC-E, from a 9 mm
observation carried out in 2011 July 9.}
\label{OHCE}
\end{figure}

\begin{figure}[tbh]
\centering
\includegraphics[angle=0,scale=0.7]{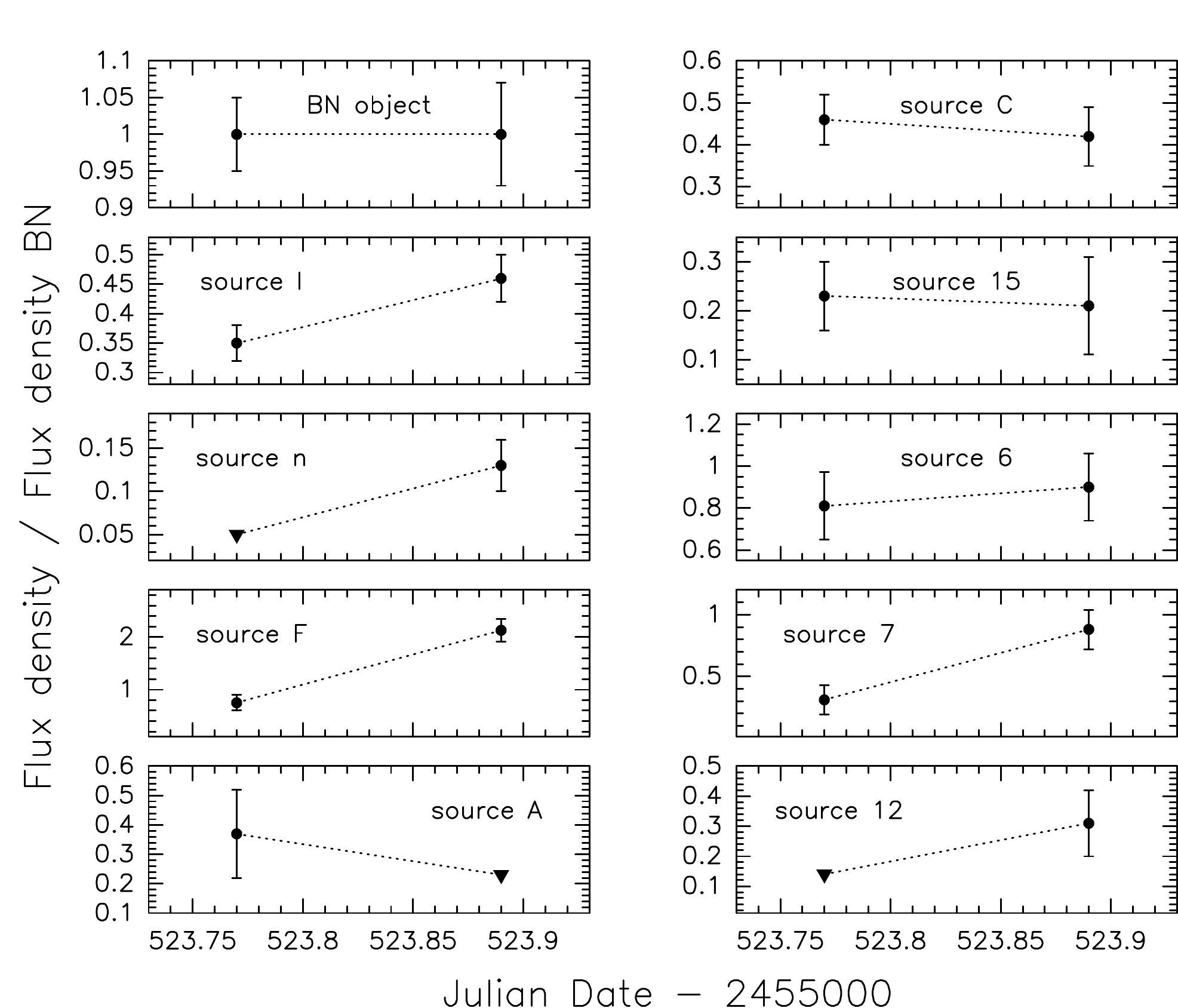}
\caption{\it 9 mm flux density variations (normalized by the flux density of the
constant massive source BN) on hour timescales for sources of the Orion Nebula
Cluster, from Rivilla et al. (2015). The two observations are separated
by only 3 hours.}
\label{hours-variability}
\end{figure}

\begin{figure}[tbh]
\centering
\includegraphics[angle=0,scale=0.28]{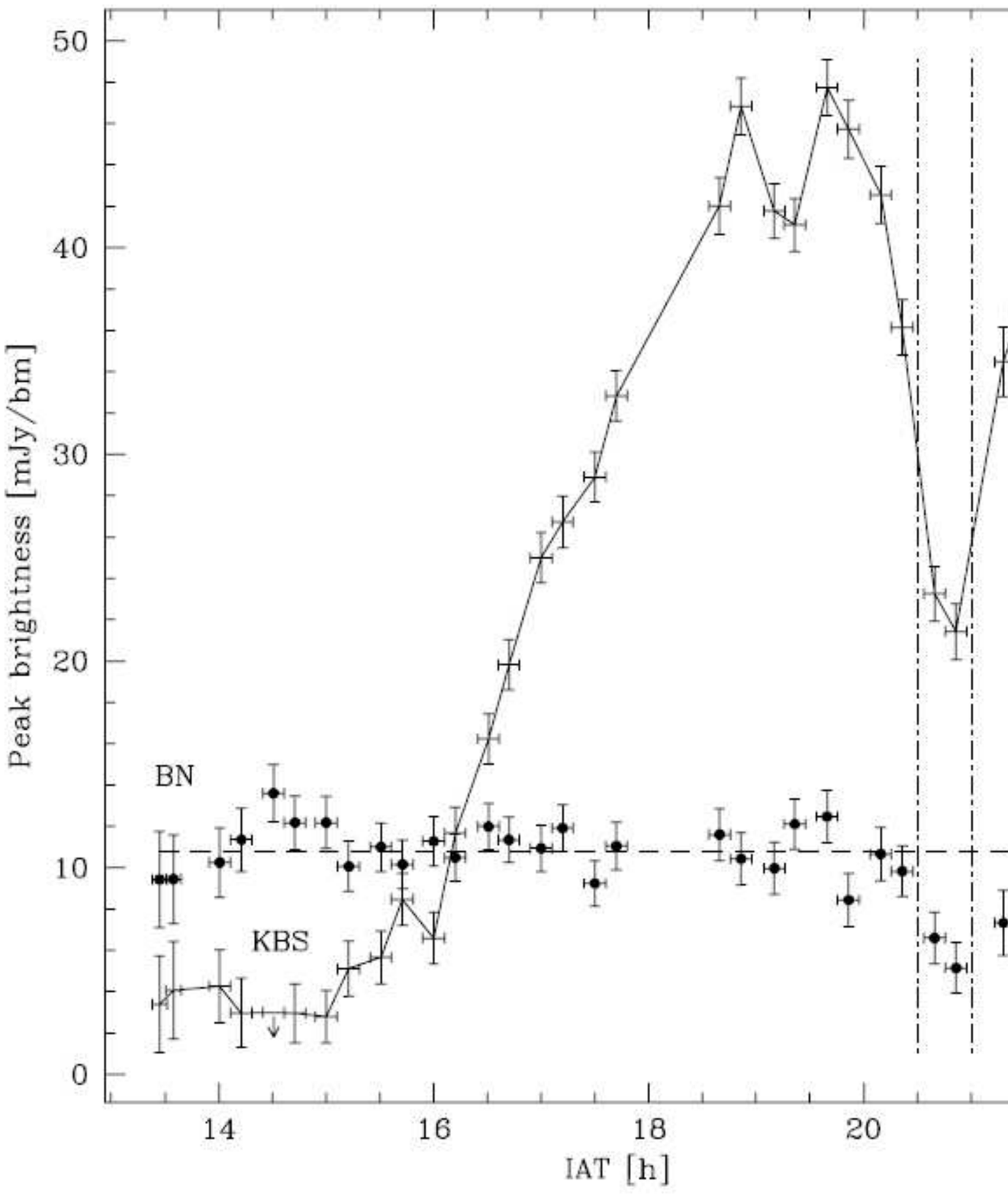}
\includegraphics[scale=0.25]{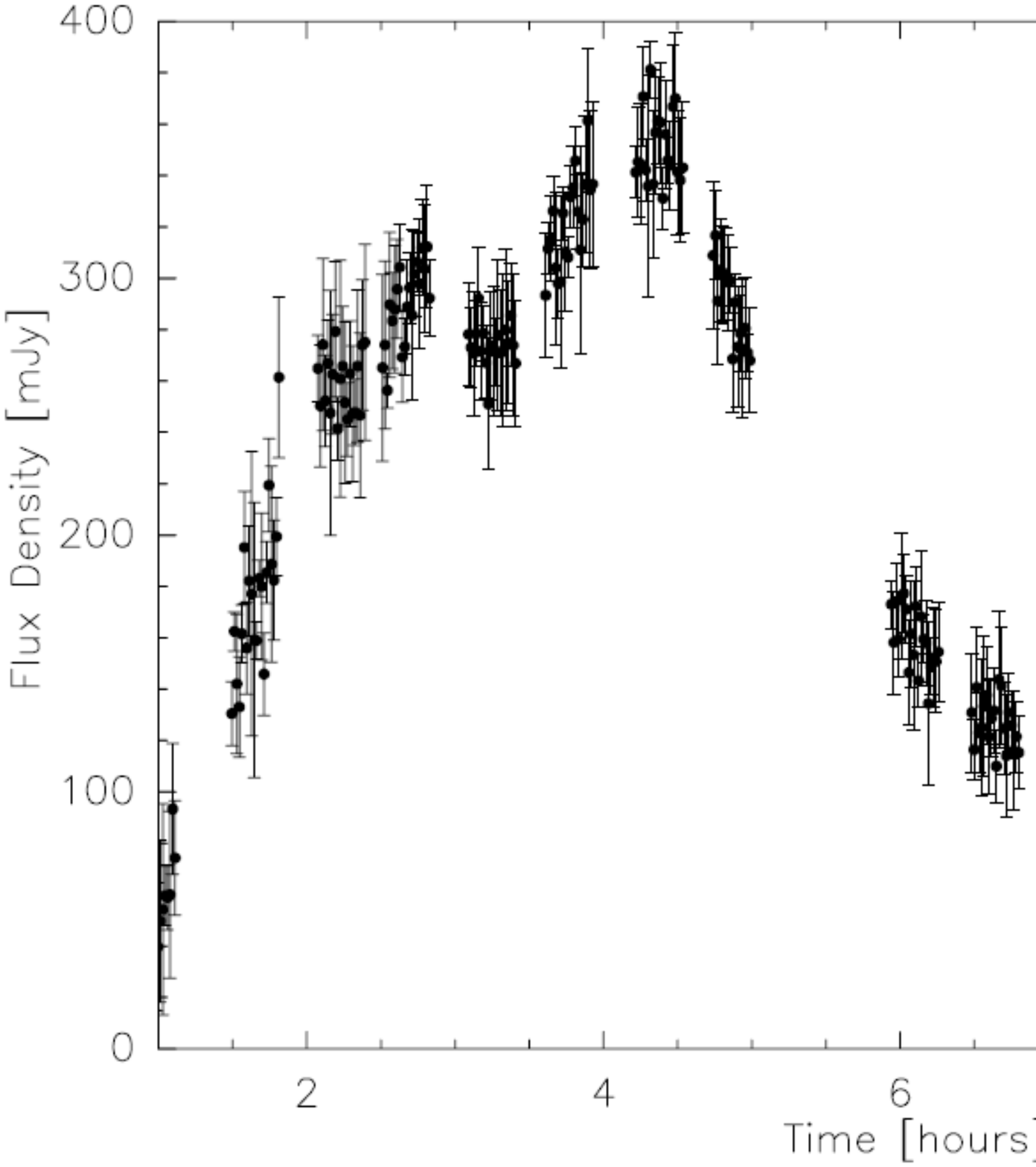}
\caption{\small {\em Left:} Light curves at 22 GHz of the Orion flaring source RBS 
(solid line) and the BN source (circles), detected with VLA (Forbrich et al., 2008). 
{\em Right:} Powerful millimeter flare (90 GHz) towards the pre-main sequence 
binary system V773 Tau A observed in 2003 August 6 with the Plateau de Bure 
Interferometer (Massi, Forbrich, et al. 2006).}
\label{flare}
\end{figure}

Although long-term centimeter variability on timescales of months to years has
been observed in star-forming regions (Felli et al.\ 1993; Zapata et al.\ 2004;
Forbrich et al.\ 2007), it is still not fully clear that these variations are
caused by long-term mechanisms. In the last years it is becoming clear that they
may be the result of a continuous sequence of events occurring on shorter
timescales. Systematic observations looking for short-term variability are
required to answer this question. Recently, Rivilla et al. (2015) have
detected a new flaring centimeter source embedded in the Orion Hot Core region
(Fig.~\ref{OHCE}) using the Q-band (7~mm) of Very Large Array (VLA). The source
remains below the sensitivity of the observation ($\sim$0.4~mJy) at 2009 March 9,
and appeared ten days later with a flux of 8 mJy. A subsequent monitoring of 11
epochs using the Ka band (9~mm) detected this source only once (lower right panel
of Fig.~\ref{OHCE}), confirming that the source is highly variable. This work has
also confirmed that the variability from young stars occurs in timescales down to
hours towards several sources (Fig.~\ref{hours-variability}). Liu et al. (2014)
also detected centimeter variability on hour timescales in the young stellar
cluster R Coronae. However, only a few serendipitously detected impressive flares
with a good flux density curve have been reported so far (e.g. Forbrich et al.,
2008; left panel of Fig.~\ref{flare}).

The low number of observed events have prevented a good undestanding of these
flares. Comparing the centimeter and X-ray populations, Forbrich et al.\ (2013) and
Rivilla et al. (2015) have found that the centimeter detections
correspond to the brighter X-ray stars. This suggests that the observations at
long wavelengths have been strongly limited by sensitivity. The new Karl Janky
Very Large Array (JVLA) and ALMA are capable now to significantly increase the
number of cm/mm detections of flares from young stars.




\subsubsection{A new window on high-energy processes in young stellar objects: millimeter-wavelength flares seen with ALMA}

ALMA will allow us to study these stellar flares in the millimeter range. 
The number of flares detected with millimeter instruments so far is very low (e.g.
Bower et al., 2003; or Massi et al.\ 2006, see Fig.~\ref{massi}). The unprecedented
sensitivity of ALMA will surely provide a large number of detection of flares. But
ALMA observations not only will allow us to merely expand previous centimeter studies
to shorter wavelenghts, but also they will open up a new window of very high-energy
processes in young stellar objects. While the centimeter range is probing
gyrosynchrotron radiation from mildly relativistic electrons, even greater electron
energies (MeV) are accessible in the millimeter wavelength range, where synchrotron
radiation from relativistic electrons can be observed. Millimeter-wavelength
observations thus probe electron energies that are orders of magnitude above those of
stellar X-ray flares. While both gyrosynchrotron and synchrotron radiation can occur
in the same source, it is not clear whether they are correlated, and even the
centimeter range is thus providing different windows on different physical processes
when going from centimeter to millimeter wavelengths. The relativistic electron
population causing the synchrotron emission may be entirely independent of its lower
energy counterpart, as has been observed on the Sun, where millimeter gyrosynchrotron
flares without centimeter counterparts have been observed (e.g., Kundu et al.\ 2000).
With unprecedented sensitivity in the millimeter range, ALMA is thus providing us
with a new window on high-energy processes, providing systematic access to very high
energy electrons for the first time. ALMA will allow us to test acceleration physics
in extremely active stars that may accelerate particles for much longer periods and
to much higher energies than the Sun.


\paragraph{\large ALMA Band 2+3: Nature of the flaring emission from wide band
spectral indeces}

To constrain the nature of the millimeter flaring emission, a good determination
of the spectral indeces (i.e., how the flux density changes with the frequency)
will be crucial. Since these events are highly variable in very short timescales,
simultaneous observations with high sensitivity in time bins of a few minutes and
covering a wide frequency band are required. The ALMA Band 2+3 will suffice these
two conditions, since it will allow us to observe simultaneously several spectral
windows spanning a wide frequency range between 67 and 116 GHz. This will result
in a robust derivation of quasi-instantaneous spectral indeces of the flares, and
thus in a good determination of the nature of the emission.

\paragraph{\large mm-VLBI+ALMA: Revealing the geometry of millimeter flares}

The catalogues of millimeter flaring sources that ALMA will provide in the several
years will allow us to select the best potential sources to study in detail with
millimeter Very long Baseline Interferometry (mm-VLBI) including ALMA. Only such
observations will provide the needed sensitivity and spatial resolution to resolve
the small-scales ($<$0.1 AU; $<<$ 1 mas) where the flaring emission is expected to
be originated. These extremely high resolution images will tell us the geometry of
the flares: do they arise directly from the star or alternatively are originated
in magnetic loops connecting central star and the circumstellar disk (Feigelson
\& Montmerle 1999)? Furthermore, these studies will give us a much improved
understanding of the high-energy irradiation of protoplanetary disks and its
impact on planet formation, including ionization and dissociation. This is key
because the high energy environment not only ionizes and dissociates molecules
over parts of the disk, but also drives magneto-rotational instability inside them
and thus influences the evolution of exoplanet atmospheres.

\section{Extragalactic Science}

\subsection{\underline{Dense gas and CO in galaxies with ALMA Band 2}}

In the context of galaxies in the Local Universe and beyond, there is a 
unique top-level science driver made possible by ALMA Band 2: 
characterizing the cool and dense gas 
in the crucial redshift range ($0.4\la z \la 2$)
where strong evolution is occurring in galaxy populations.

\subsubsection{Low-J CO lines and the redshift desert}

Although the power of ALMA as a ``redshift engine'' has been clearly demonstrated
(e.g., Swinbank et al.\ 2012; Weiss et al.\ 2013; Simpson et al.\ 2014), arguably
ALMA's greatest contribution will be to characterize the cool gas content of
galaxies over the epoch of galaxy formation from $z\ga8$ down to nearby galaxies
in the Local Universe. The epoch of galaxy formation is most commonly traced by
the cosmic star-formation rate density of the Universe (SFRD). The SFRD peaks at
$z\sim1-3$, during a period usually associated with the main epoch of galaxy
assembly.  In this epoch, roughly half of the stars present in galaxies today are
formed (e.g., Shapley et al.\ 2011).  The SFRD then declines dramatically toward lower
redshift, by roughly an order of magnitude, from $z\sim2$ to $z\sim0$, with the
most significant decrease at $z<1$.

One of the most important diagnostics of the epoch of galaxy assembly and the
successive decline of the SFRD are observations of cool gas. Theory predicts and
observations confirm that  the molecular gas fraction increases with lookback
time, by as much as a factor of 7 from $z=0$ to $z\sim2$ (e.g., Schaye et al.\
2010; Daddi et al.\ 2010; Tacconi et al.\ 2010; 2013; Lagos et al.\ 2011; Genzel
et  al.\ 2015). At $z\sim2$, molecular gas can contribute 60$-$70\% to the total
baryonic inventory.  Such a high gas content and the decline in gas mass fraction 
toward lower redshifts is undoubtedly associated with the peak of the SFRD, and
its decrease toward the present day. The epoch from $z<1$ is thus crucial to
understand why galaxies exhausted their fuel for star formation in the latest
stages of the age of the Universe. 

\begin{figure}[!h]
\begin{center}
\vspace{0pt}
\hbox{
\includegraphics[height=0.33\linewidth]{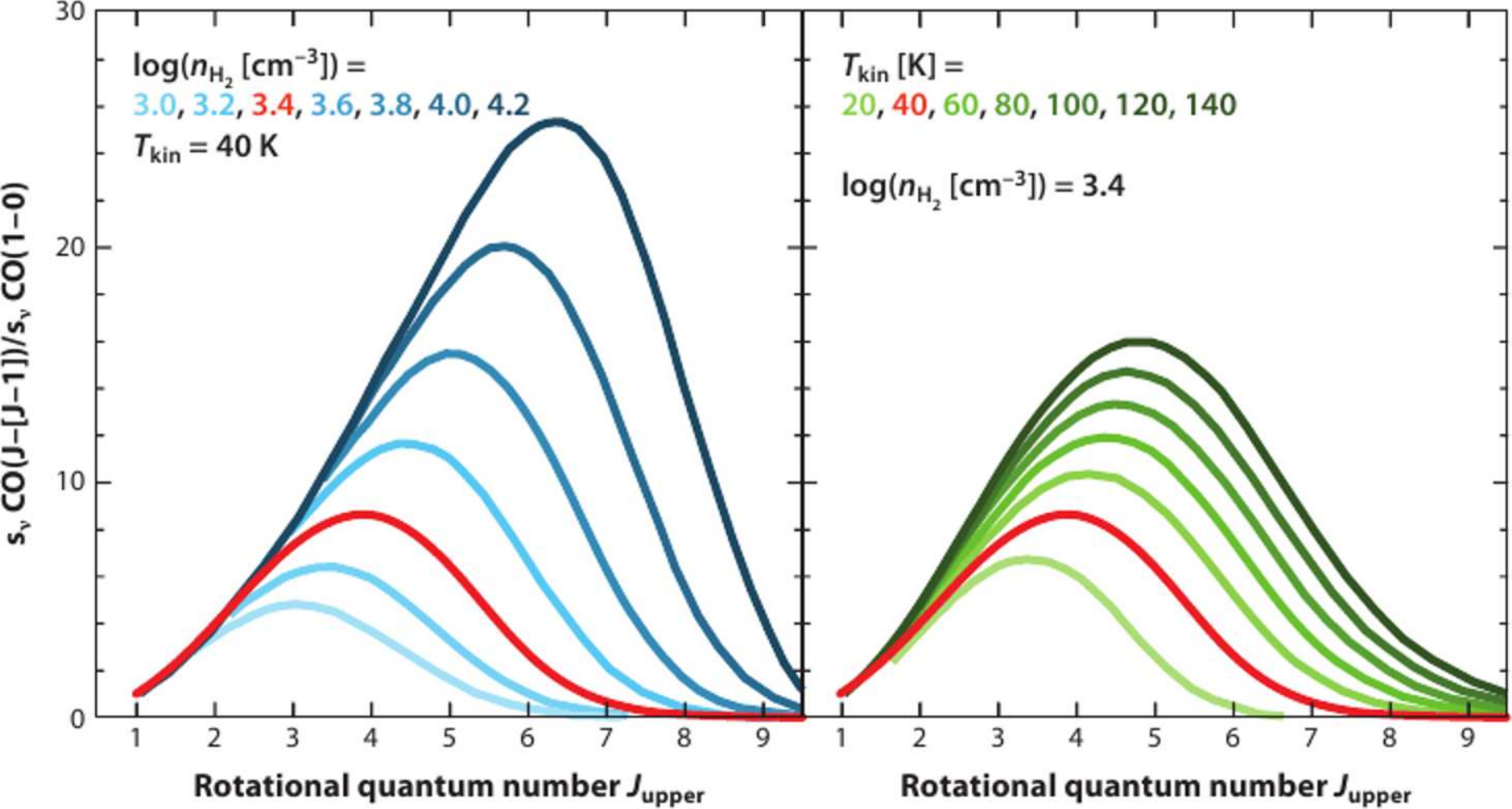}
\hspace{0.2\baselineskip}
\includegraphics[height=0.33\linewidth]{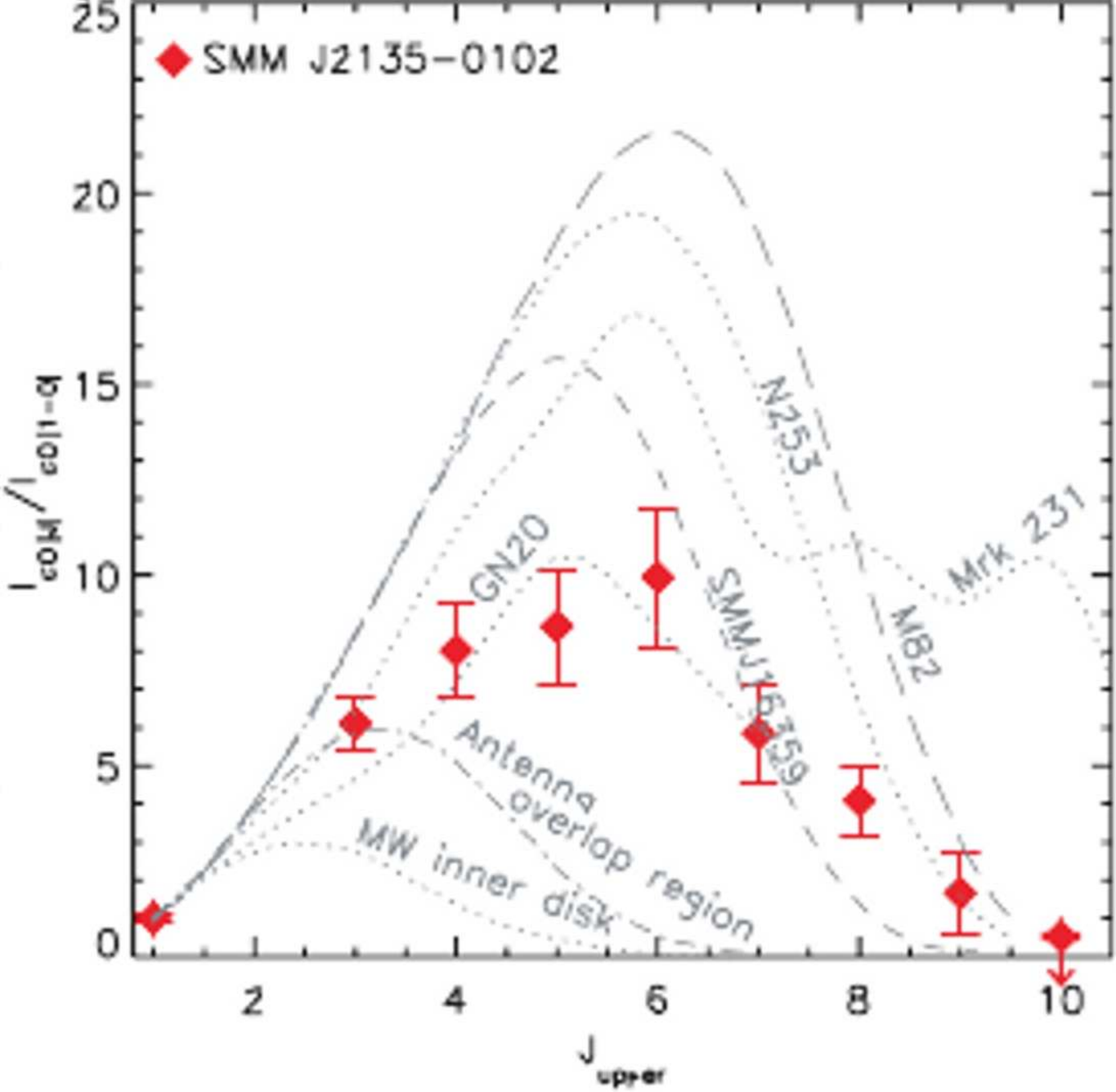}
}
\caption{\it{Left:} Theoretical CO Spectral Line Energy Distributions (SLEDs) 
for different physical conditions, taken from Carilli et al.\ (2013).
\it{Right:} Integrated CO SLED for a sub-millimeter galaxy (SMG),
SMM J2135-0102 at $z=2.3$, taken from Danielson et  al.\ (2011).
The gray curves show observed SLEDs from different galaxies, illustrating
the contrast between starbursts such as NGC\,253 and M\,82, 
and more quiescent systems such as the Milky Way inner disk.
\label{fig:sled}
}
\end{center}
\end{figure}

Recent years have seen the number of high-$z$ galaxies with molecular gas-mass
estimates increase exponentially. However, uncertainties in
two critical factors still hamper the interpretation of observations:
one is the large uncertainty
in \xco, the factor that converts observed CO column densities to \htwo\ 
masses.
This factor depends on physical conditions in the molecular clouds,
including density, temperature, radiation field, and metallicity,
and can vary by a factor of 50 or more over the variety of
galaxy populations that dominate the SFRD at $z\sim0-3$
(see Bolatto et al.\ 2013, and references therein).
Ultimately, assumptions are made based on observations of a single CO
transition, but the more varied galaxy populations and different
physical conditions at high redshift make \xco\ highly uncertain.

The other uncertainty is linked with the necessity, up to now, of
observing high-redshift galaxies in high-\jj\ CO transitions.  Most of
the observations of gas in high-$z$ galaxies have been carried out in
CO~(3--2) or higher-\jj\ lines (e.g.., Daddi et al.\ 2010; Tacconi et al.\ 2010;
2013). However, the estimate of gas masses is ultimately based on the CO~(1--0)
transition, so that ratios of the higher-\jj\ lines relative to
1$\rightarrow$0 must be assumed.  As illustrated in the left panel of
Fig. \ref{fig:sled}, these ratios are critically related to
excitation conditions, and the temperature and density in the molecular clouds.
Physical conditions are known to vary between low-
vs. high-SFR galaxies, and compact vs. extended star-forming regions
(e.g., Weiss et al.\ 2005; Danielson et al.\ 2011), as illustrated in
the right panel of Fig. \ref{fig:sled}. 
{\it Observing only high-\jj\ lines skews mass estimates toward the warm, dense gas
traced by these transitions, and can cause molecular gas masses to be
severely underestimated} (e.g., Narayanan et al.\ 2009; Dannerbauer et al.\ 2009; 
Aravena et al.\ 2010; Ivison et al.\ 2011).

\subsubsection{Band 2 and cool gas mass }

\begin{figure}[!h]
\begin{minipage}[c]{0.55\linewidth}
\vspace{0pt}
\includegraphics[width=\linewidth]{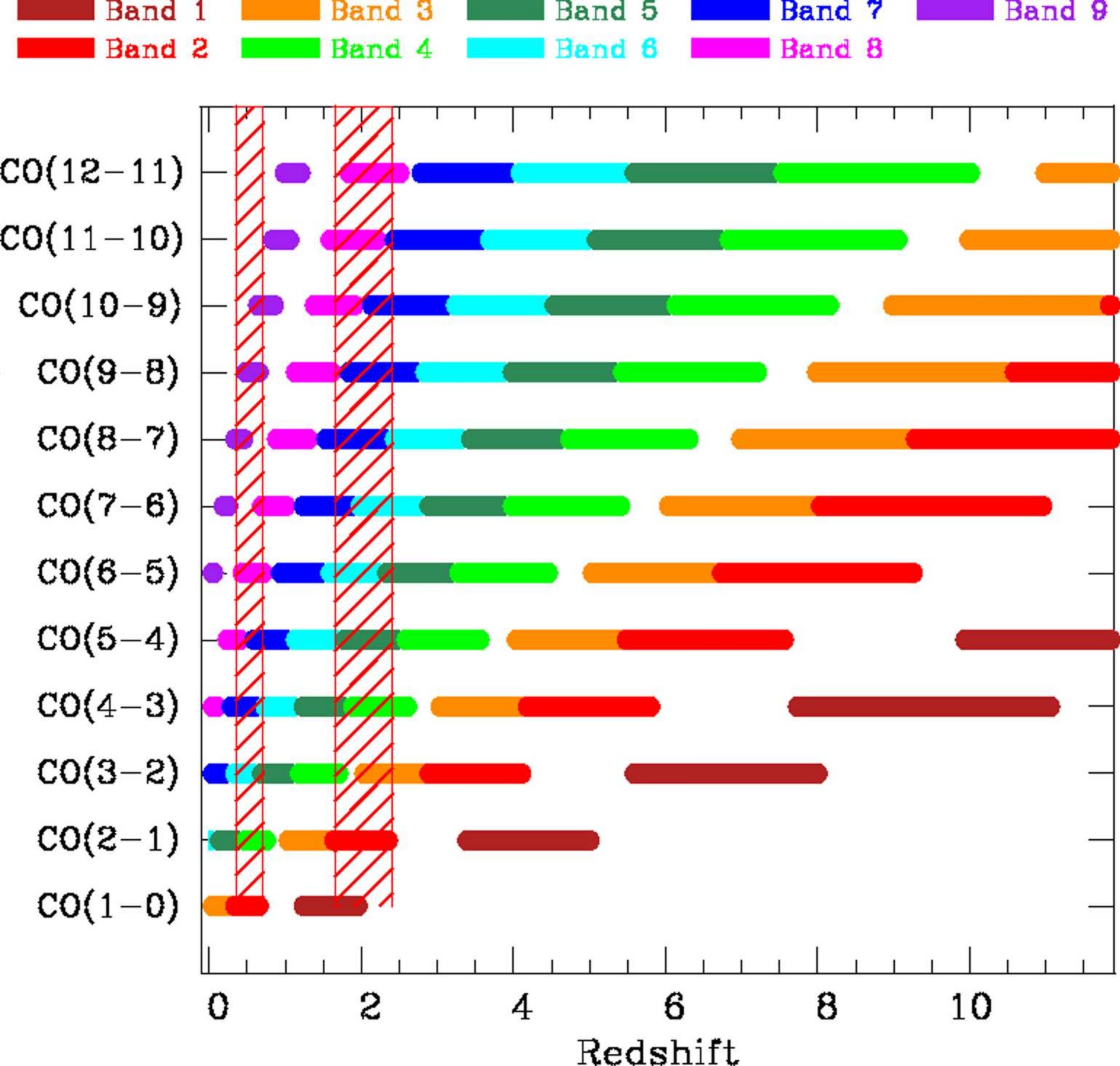}
\end{minipage}
\hspace{0.5\baselineskip}
\begin{minipage}[c]{0.45\linewidth}
\caption{\it CO ladder coverage by the ALMA receivers plotted against redshift. 
The vertical hatched strips show the potential of ALMA with Band 2
to measure multiple transitions in two crucial redshift regions,
the first where the Universe is 3/4 of its present age,
and the second where the cosmic SFRD peaks.
\label{fig:bands_sled}
}
\end{minipage}
\end{figure}

\textit{Band 2 can mitigate, if not completely resolve, these uncertainties.}
CO~(1--0) is unobservable at $z>0.3$ with only Band 3, 
and CO~(2--1) is recovered by Band 3 only at $z\sim$1.0. 
However, Band 2 enables CO~(1--0) measurements 
over this crucial range in redshift, 
from $z\sim0.3$ to $z\sim0.7$ where SFRD falls dramatically and
the Universe is about 3/4 of its present age
($0.3<z<1.0$ corresponds to lookback times from 3.4\,Gyr to 7.7\,Gyr).
Figure \ref{fig:bands_sled} illustrates the CO ladder and
the band coverage as a function of redshift.  From $z=0.29$ to
$z=0.72$, Band 2 combined with the other ALMA bands enables
observations of between seven and nine transitions, including the four with
lowest \jj; Band 2 covers 1$\rightarrow$0, thus making possible a
complete assessment of excitation conditions and the CO spectral line
energy distribution (SLED).
 
No CO lines can be observed in Band 1 until $z\sim1.2$, where the (1--0)
transition enters the band; Band 2 recovers CO~(2--1) at $z\sim1.6$
which remains observable with Band 2 until $z\sim2.4$.  Thus, in the
redshift range where the SFRD is peaking, from $z\sim1.6$ to
$z\sim2.4$, Band 2, together with the other ALMA bands, provides
coverage of between eight and eleven CO transitions; Bands 2+1 measure
the two lowest-\jj\ transitions, indispensable for accurate gas mass
estimates.  As shown in Fig. \ref{fig:sled}, such
band coverage is fundamental because of the shape of the CO cooling
curve, which peaks around 6$\rightarrow$5 in the case of starbursts,
or toward lower \jj\ in the case of more quiescent disks.  The factor to
scale the higher-\jj\ lines to 1$\rightarrow$0 depends on excitation, and
is virtually impossible to estimate without multiple transitions.

Expanding the frequency coverage of ALMA by installing Band 2 (or 2+3) will
allow the almost complete exploration of
the cool gas content in galaxies in this important stage of the age
of the Universe.  There would remain only a gap of $\sim$0.7\,Gyr in
the evolution of the
Universe (between $z\sim0.84$ and $z\sim1$, where CO~(2--1) shifts out of Band 4
and into Band 3; CO~(1--0) is recovered by Band 1 at $z\sim1.2$).
These low-\jj\ transitions are fundamental to accurately estimate the cool
molecular gas mass.
As described below, this epoch in redshift is crucial for observationally
constraining the physical mechanisms which quench star 
formation and cause galaxies to transform from blue star-forming systems to 
``red and dead'' ones.

\subsubsection{Dense-gas diagnostics of physical conditions in galaxies \label{sec:dense}}

\begin{figure}
\includegraphics[width=.45\textwidth]{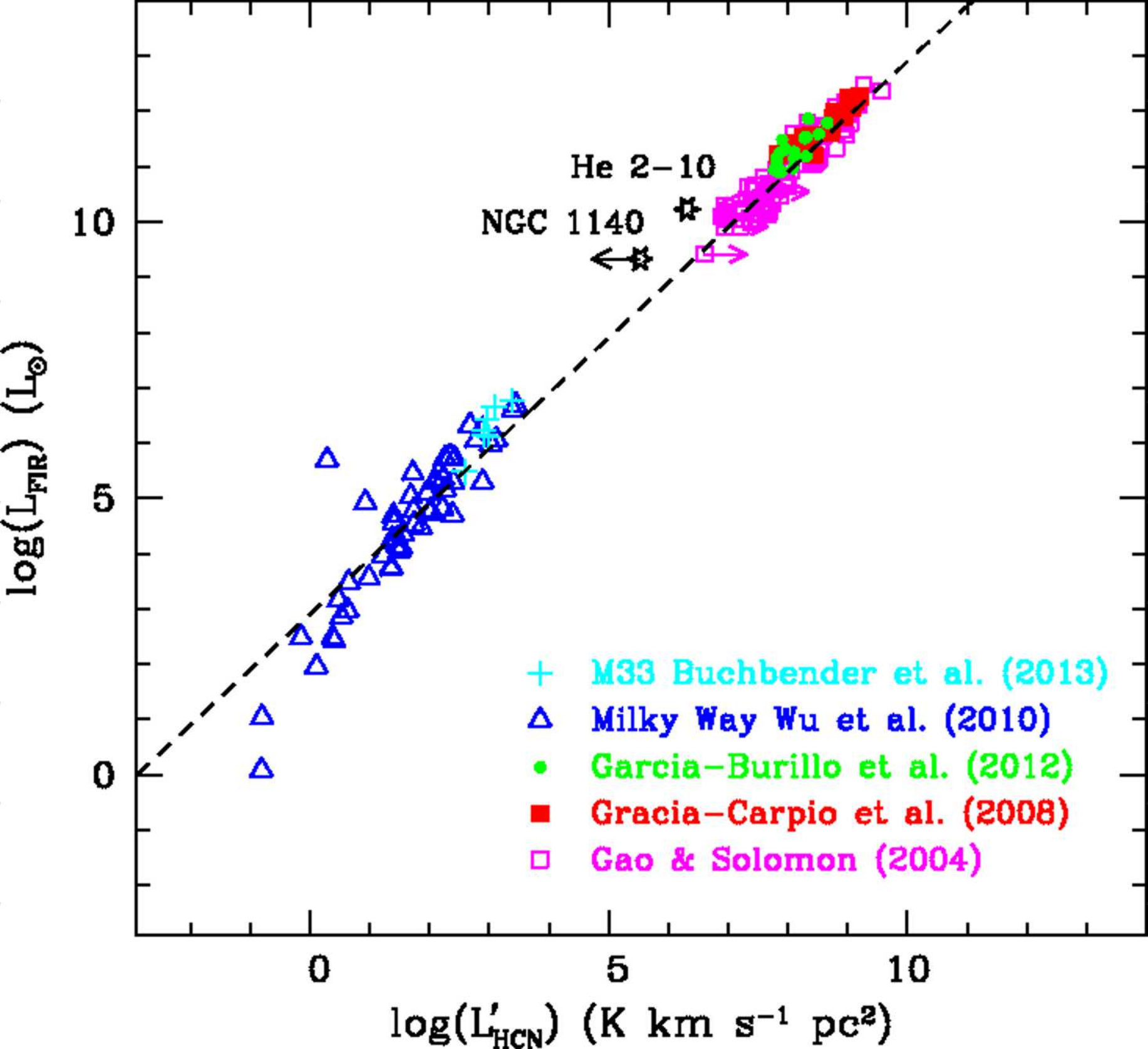}\quad
\includegraphics[width=0.53\textwidth]{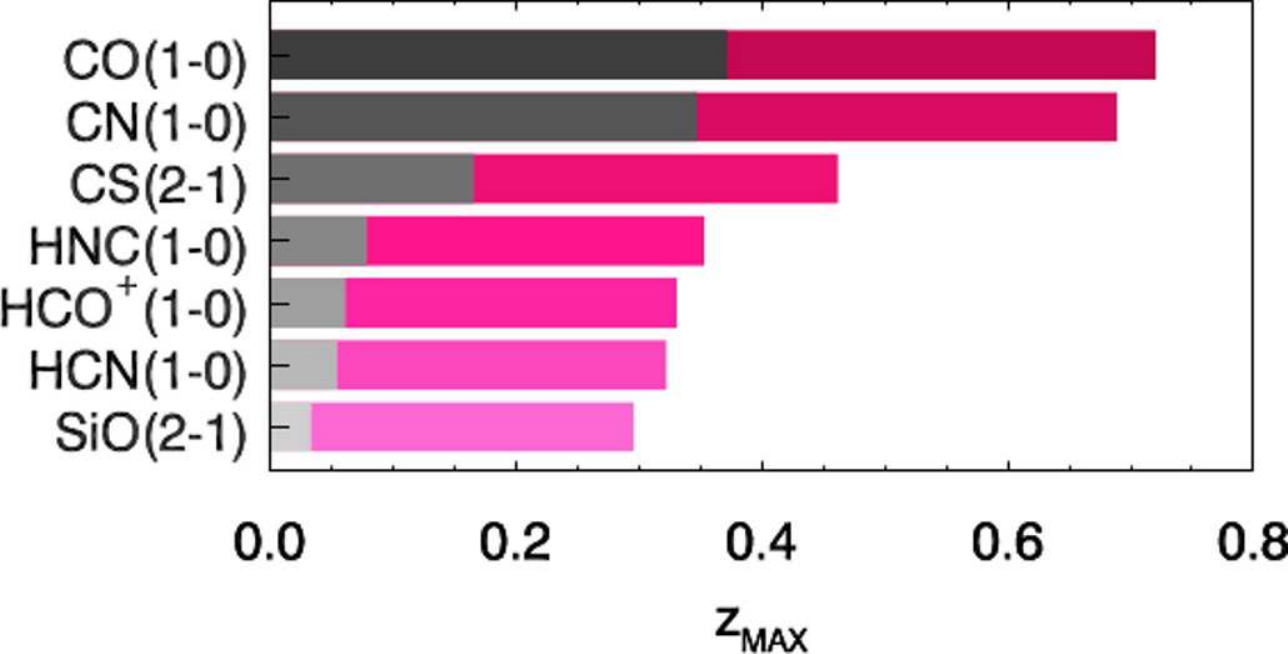}
\caption{\it {Left:} Correlation between the star formation rate (traced by total IR luminosity) and 
HCN luminosity in a sample of nearby galaxies, taken from Hunt et al. (2015, in prep.).
\it{Right:} Molecular transitions of interest for studying dense gas, 
plotted as a function of the maximum redshift at which they are detectable with ALMA.
Grey bars show the current possibilities, and pink bars indicate the improvement that can be achieved 
with Band 2 (Image courtesy of C.\ Cicone; taken from the European 
Science Case for ALMA Band 2: Fuller et al., in preparation).}
\label{fig:HCN}
\end{figure}

Although observations of CO line emission are an effective means
of studying the cold molecular gas reservoir in the interstellar
medium (ISM) of high-redshift galaxies, it is not directly related to star
formation.
The dense molecular gas ($n_{H2} \ga$10$^5$~cm$^{-3}$) more directly associated with
star-formation is best studied through observations of higher dipole
moment molecules with higher critical densities, such as HCN, HNC or HCO$^+$. 
The luminosity in the
HCN~(1--0) line is known to correlate tightly with infrared
(IR) luminosity (equivalently star-formation rate, SFR) in nearby galaxies
(Gao et al.\ 2004; Wu et al.\ 2010; Garcia-Burillo et al.\ 2012).
As shown in Fig.~\ref{fig:HCN}, this linear correlation extends over almost 10 orders of
magnitude, from the IR luminosities of dense
Galactic molecular cloud cores ($L_{IR} \ga 10^{2}~$\,\lsun) (Wu et al.\ 2005; 
2010) 
to luminous IR galaxies ($L_{IR} \ga 10^{11}$\,\lsun) 
(ULIRGs: Gao et al.\ 2004; Garcia-Burillo et al.\ 2012). 
Beyond $L_{IR}\sim10^{11}$\,\lsun, the correlation becomes super-linear,
implying some change in the physical conditions in the
galaxy's ISM (Juneau et al.\ 2009; Garcia-Burillo et al.\ 2012).
As this dense gas in our own Galaxy is
generally associated with sites of ongoing star-formation (indicated
by the IR luminosity), it is believed that the presence of strong
HCN~(1--0) line emission in a high-redshift galaxy may be
interpreted as coming from the dense gas dominating an ongoing starburst
(e.g., Juneau et al.\ 2009).
By comparing the
dense gas masses estimated from HCN~(1--0) with the total
molecular gas masses estimated from low-\jj\ CO line emission (e.g., Aalto et al.\
1995), 
the dense fraction and its variation in galaxies can be estimated over cosmic time. 
Similarly, the dense gas mass can be compared to the SFRs
in order to estimate a true star-formation efficiency.
As the number density of IR-luminous
star-forming galaxies increases with redshift, surveys of 
HCN~(1--0) with Band 2 would provide the only means of
constraining the apparent non-linearity at the highest end of the
$L_{IR}-L^{\prime}_{HCN}$ relationship. 
Molecular line surveys(e.g., Costagliola et al.\ 2011)
with Band 2+3 will be fundamental for efficiently
constraining physical conditions in star-forming galaxies 
up to $z\sim0.5$.

Figure \ref{fig:HCN} also shows two galaxies which deviate from the global trend:
these are dwarf galaxies He2$-$10 (Santangelo et al.\ 2009) and NGC\,1140 (Hunt et
al. 2015, in prep.). In these galaxies, HCN is under-luminous, almost certainly a
result of different physical conditions in their nuclei including low metallicity
(Meijerink et al.\ 2007). In such situations, \hcoplus\ may be a better tracer of dense
gas as it can outshine \hcn\ by factors of a few (e.g., Anderson et al.\ 2014). 
Since dwarf galaxies are the most plentiful galaxies by number in the universe, 
and thought to be the ``building blocks'' for hierarchical galaxy assembly,
probing their dense gas will be important for understanding how galaxies assemble
their stellar mass. ALMA with Band 2 will be sufficiently sensitive to probe the
dense-gas content of such galaxies up to $z\la 0.4$.

\subsubsection{Molecular outflows and AGN feedback}

Since the seminal discovery that active-galactic nuclei (AGN) can drive powerful
outflows that effectively quench star formation (e.g., Di Matteo et al.\ 2005),
AGN feedback has become a cornerstone of galaxy evolution. The most effective mode
of studying AGN feedback is arguably through the molecular component of the ISM
(e.g., Feruglio et al.\ 2010; Aalto et al.\ 2012; Cicone et al.\ 2012, 2014;
Combes et al.\ 2013, 2014; Garcia-Burillo et al.\ 2014).
The massive molecular outflow in the ULIRG, Mrk\,231, the best-studied case so
far, has been detected not only in low-\jj\ CO transitions
(Feruglio et al.\ 2010; Cicone et al.\ 2012), but also with high-density tracers including \hcn,
\hcoplus, and HNC~(1--0) (Aalto et al.\ 2012). Moreover, the high-velocity wings that
characterize the outflow are  more prominent in the higher density gas (traced by
HCN) than in the  low-\jj\ CO lines (Cicone et al.\ 2012). The implication is that not
only very dense ($n_{\rm H2} \ga$10$^4$~cm$^{-3}$)  molecules survive in the outflow,
but also that the high dipole moment molecules may be enhanced in the wind, and
may thus offer the best probe of the energetics of the outflow.

A possible explanation for such enhanced HCN emission in the massive molecular wind
can be the presence of strong shocks. While in Mrk\,231 the hypothesis of strong
shocks is inconsistent with the CO excitation in the outflowing gas (Cicone et al.\
2012), in NGC\,6240, another ULIRG, there is evidence for shocked gas at the confines
of the outflow (Feruglio et al.\ 2013). Moreover, in addition to negative AGN
feedback which quenches star formation,  in some cases star formation may be enhanced
by the compression of the gas(e.g., Zinn et al.\ 2013; Cresci et al.\ 2015). Because
of the increased redshift range for tracers of dense-gas outflows,  ALMA Band 2+3
will provide a key diagnostic into the physical processes behind feedback in
galaxies, exactly over the crucial redshift range where the cosmic SFRD is
declining. 

Other important molecular transitions can also be observed in galaxies up to
$z\sim0.4$ within Band 2(or 2+3), including CS~(2--1) (97.98\,GHz) and SiO~(2--1)
(86.85\,GHz). CS, being particularly resistant to shocks and UV photo-dissociation,
is a very good and unbiased tracer of dense molecular gas, and can be used to
determine the amount of dense gas in the outflows and its relation with the diffuse
component.   Silicon monoxide (SiO) is an unambiguous tracer of strong shocks which 
can modify the chemistry of the local ISM; by destroying dust grains and injecting
silicon and SiO into the gas phase, the SiO abundance is enhanced. This molecule
represents an  independent tool to test the presence of shocks and their relevance in
the feedback process.  The (1--0) transition of CS and SiO are covered by the ALMA
Band 1 and the Q Band of JVLA, but the (2--1) transitions are {\it crucial to study
the excitation of these species} in the outflow. These transitions could also open a
new diagnostic for galactic-scale outflows, given that similar physics seem to be
driving bipolar outflows from stellar-sized accretion  disks (e.g., S\'anchez-Monge
et al.\ 2013).  ALMA Band 2+3 will open a new window on the physical conditions
behind dense gas tracers  in galaxies.

\subsection{\underline{The radio-loud AGN duty cycle: the role of cold gas}}

Early-type galaxies (ETGs) host a wide range of kinematic sub-structures (e.g.\
Emsellem et al.\ 2004), with decoupled prograde stellar disks in the
preferentially low-mass fast-rotating galaxies and
kinematically-misaligned cores in the preferentially massive
slow-rotating spheroids (Emsellem et al.\ 2011).
Why star formation is suppressed in ETGs and how sub-structures form
are both hotly debated issues. The solutions to both problems may well
be connected. On the one hand, active galactic nucleus (AGN) feedback
through radio jets ({\em jet mode}; Heckman \& Best 2014) and
radiation from the accretion disk ({\em radiative mode}) are both
thought to to be essential to the suppression of star formation (see
Fabian 2012 for a review). On the other hand, the interstellar gas
which accretes onto the supermassive black hole (SMBH) and powers the
AGN may also form decoupled stellar components, in a manner depending
on its mass, origin, and ability (or not) to turn into stars. The
SAURON and ATLAS$^{\rm 3D}$ optical integral-field spectroscopic (IFS)
surveys and associated CO follow-ups have clearly demonstrated a
causal link between the two in radio-quiet ETGs (e.g.\ Crocker et al.\
2008, 2009, 2011). How feedback loops can be established and sustained
over many orders of magnitude in spatial scale, and how the gas
migrates from kpc scales to the vicinity of the SMBH, have yet to be 
understood.

\subsubsection{AGN fueling: the role of cold gas}
It is generally accepted (Heckman \& Best 2014) that radiatively
efficient, high-excitation (typically powerful) radio galaxies (HERGs)
are triggered by cold gas transported to the centre through merging or
collisions with gas-rich galaxies. Allen et al.\ (2006) however
suggested that accretion in radiatively inefficient, low-excitation
(typically low-power) radio galaxies (LERGs), that dominate in the
local Universe, may occur directly from the hot phase of the
intergalactic medium. However, radio galaxies (RGs) of all powers have
complex, multi-phase interstellar media, with cold (molecular), cool
(H{\small I}), warm (ionised) and hot (X-ray) components. LERGs often
possess large amounts of dust and molecular gas (e.g.\ de Koff et al.\
2000; Prandoni et al.\ 2007, 2010; Ocana Flaquer et al.\ 2010),
providing compelling evidence that cold gas could also be the fuel
supply. In fact, larger reservoirs of molecular gas are present in
LERGs than in radio-quiet galaxies (Senatore et al.\ 2015).

Several of the LERGs detected in CO have double-horned line profiles
(Prandoni et al.\ 2010; Senatore et al.\ 2015), consistent with
ordered rotation, while HST observations show that dust in these
objects is often confined to disks on small (kpc or sub-kpc) scales.
The presence of nuclear disks of molecular gas (and dust) may
therefore be a common feature in these objects. This is indirectly
supported by the systematic CO~(1--0) line imaging campaign of a large
sample of gas-rich but generally radio-quiet ETGs from ATLAS$^{\rm
  3D}$ (Alatalo et al.\ 2013), where kpc-scale molecular gas disks
were found in $50\%$ of the targets and tend to follow the dust. A
regular disk morphology is preferentially associated with more massive
ellipticals, the typical hosts of RGs. 
Nuclear disks may thus be an essential link in the feeding/feedback cycle. One
possible scenario is that the CO traces the outer parts of a thin disk
that becomes progressively hotter and then ionised at smaller
radii. The molecular emission would then have an inner edge, that
should be seen at high resolution. It is likely that most of the
cold gas in LERGs is in stable orbits (as suggested by Okuda et al.\
2005 for the proto-typical LERG 3C\,31) and that the black holes are
``drip-fed'' by molecular clouds from the inner disk.
Whether the cold gas in LERGs originates externally (accretion,
mergers) or internally (stellar mass loss, cooling of an initially hot
gas phase) remains an open question. Comparing the kinematics of the
molecular and ionised gas with that of the stars can differentiate
between these alternatives (see Davis et al.\ 2011, 2013a).

\subsubsection{Jet-induced Feedback}
Radio jets can transport energy efficiently to large distances, they
couple efficiently to the interstellar medium (ISM), and they produce
fast outflows emanating from the central regions (as required from
feedback models; Wagner, Bicknell \& Umemura 2012). Several outflows
are known to be driven by low-power radio jets, including cases of
massive neutral and molecular outflows (e.g.\ Alatalo et al.\ 2011;
Combes et al.\ 2013; Morganti et al.\ 2013; Santoro et al.\
2015). This suggests that even jets with low kinetic power can drive
outflows, and that the jet-ISM interaction may be a relatively common
phenomenon associated with the active phase of a galaxy. It is however
surprising that despite the violent interaction, these off-nuclear
outflows still have a component of relatively cold gas ($<1000$~K), as
detected in H{\small I} and CO (see also discussion in previous section). 
One possibility is that the radio jet
interacts with the gas close to the nucleus and decelerates from
relativistic speeds on kpc scales (Laing \& Bridle 2002), entraining
some of the denser surrounding material along with it. However,
whether this is a common property remains an open question. To probe
the above scenarios, we need sub-kpc/kpc imaging of CO or other
molecular lines, but only a few of the most spectacular examples have
been studied in detail so far. Understanding the incidence,
characteristics, and effects of such interactions in a broader
statistical manner is crucial to assess the role of jet-induced
feedback in galaxy evolution, as well as the jet-ISM interaction physics.

\begin{figure}[t]
 
\resizebox{16.5cm}{!}{\includegraphics[scale=0.42]{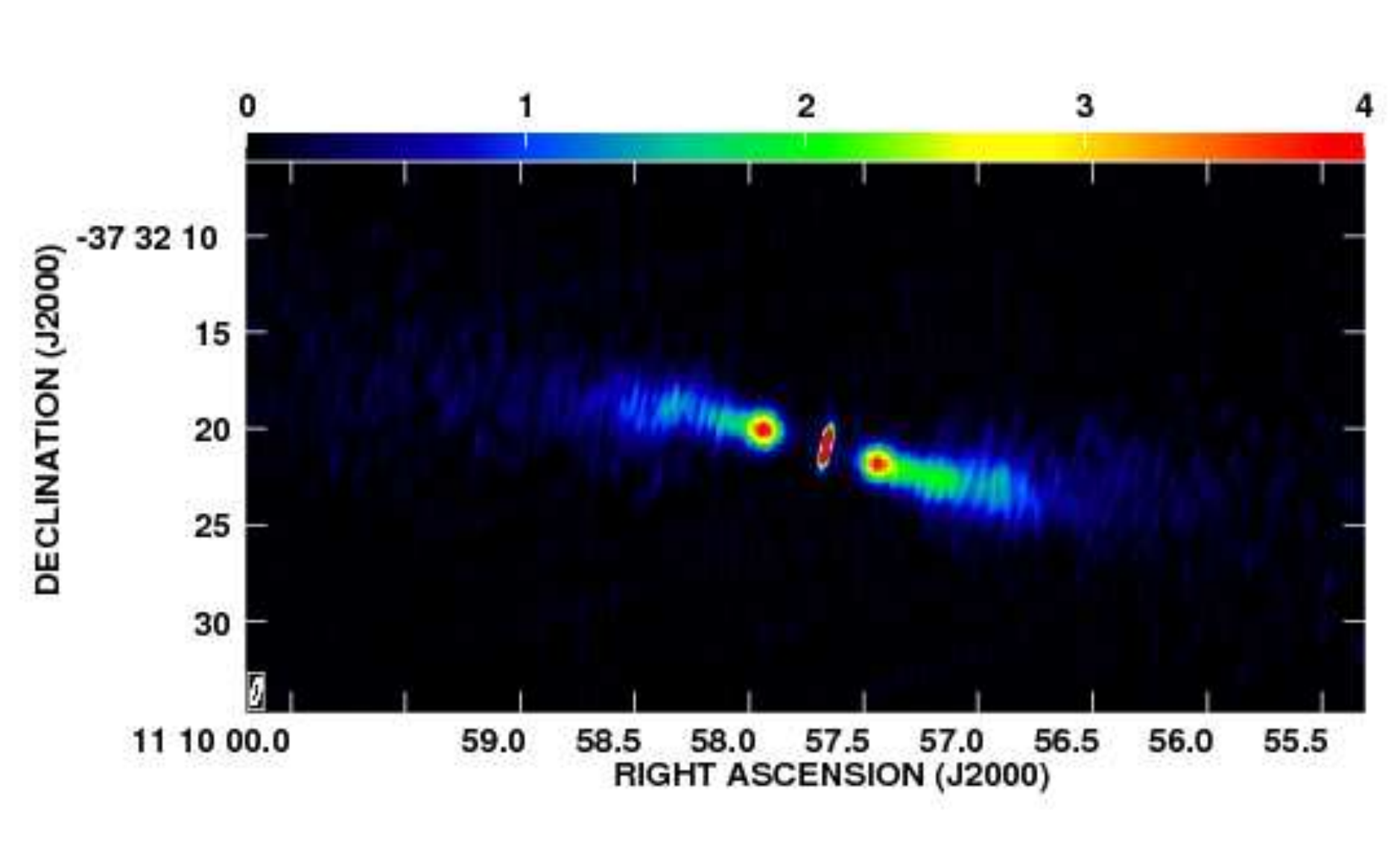}\includegraphics[scale=0.33]{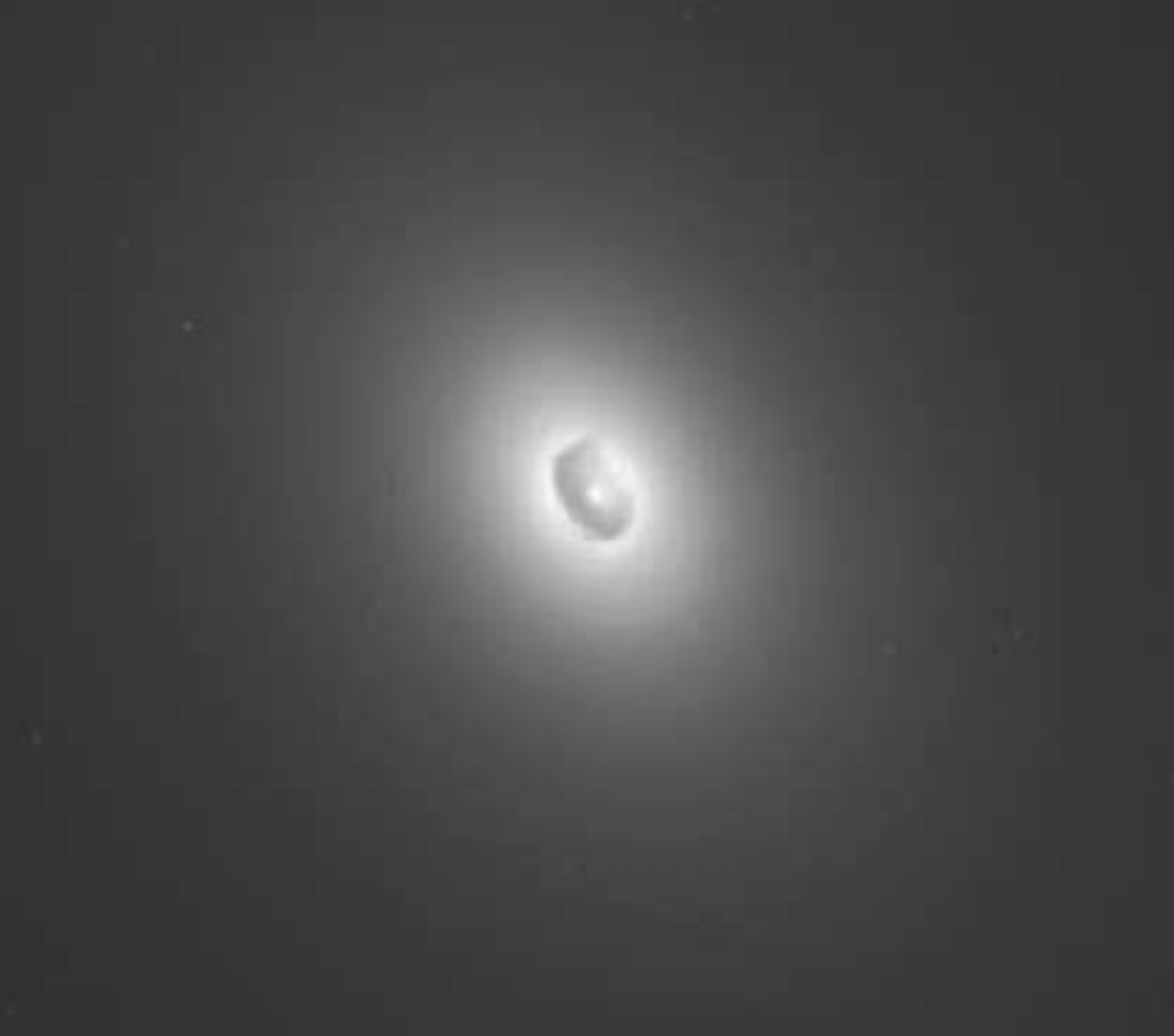}\includegraphics[scale=0.57]{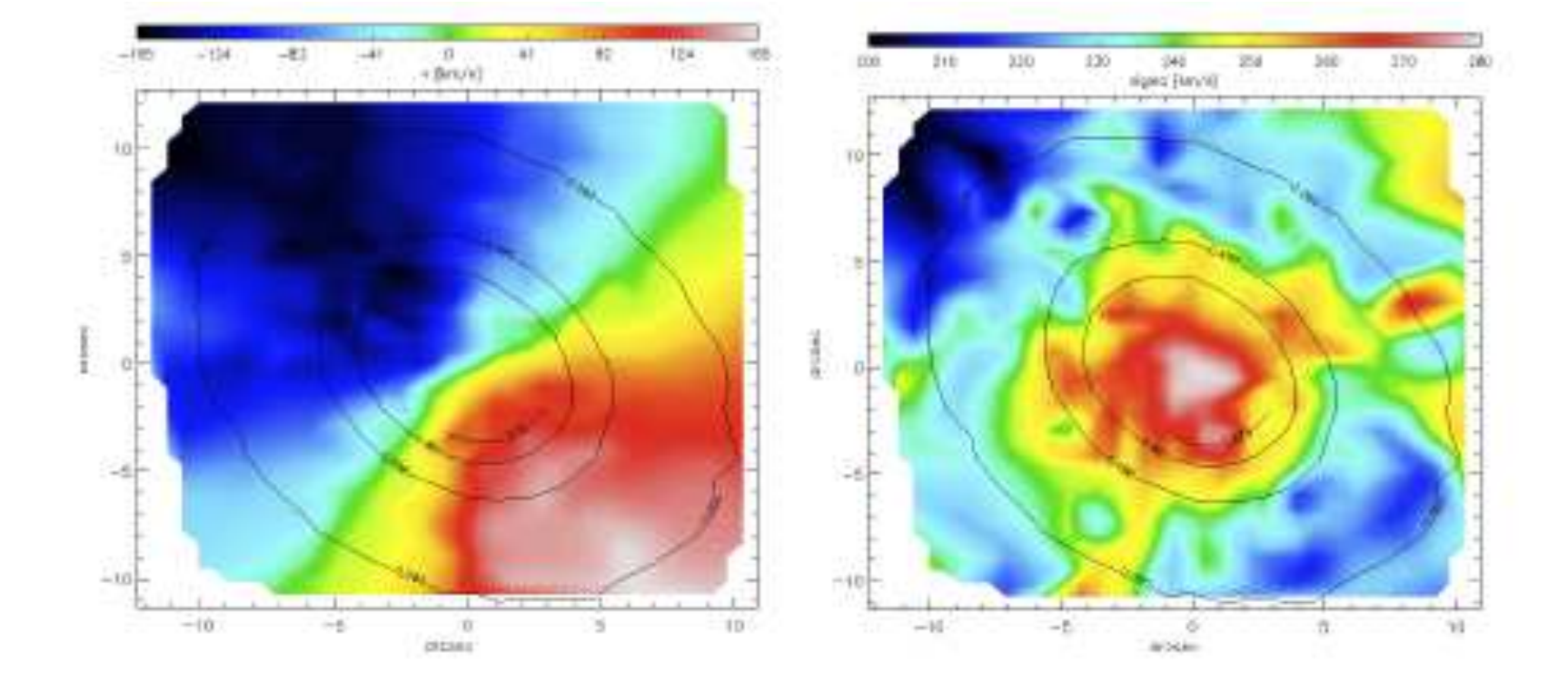}}
  \caption{\it Observations of NGC~3557. {\bf Extreme-left:}
    5~GHz VLA radio continuum image, showing the inner part of the
    radio galaxy ($60^{\prime\prime}\times30^{\prime\prime}$). {\bf
      Centre-left:} HST image of the central regions, revealing a
    regular nuclear dust disk (Lauer et al.\ 2005). {\bf
      Centre-right:} VLT/VIMOS IFS mean stellar velocity map of the
    central regions ($27^{\prime\prime}\times27^{\prime\prime}$) with
    overlaid isophotes, revealing ordered rotation. {\bf
      Extreme-right:} Associated stellar velocity dispersion map.}
\label{fig-prandoni}
\end{figure}

\subsubsection{Justification for ALMA Band 2+3}
A systematic study of the various gas phases (ionised and molecular)
and of the stellar and dust components in the cores of representative samples of
massive, radio-loud ETGs (in which jets are currently active) 
would enable a better understanding of the feeding of AGNs, and would
isolate the role played by jet-induced feedback  in the overall formation and
evolution of ETGs. 

Exploratory studies in the local Universe ($z<0.03$) are ongoing with VLT-VIMOS (eg. Prandoni et al.\ 2010; Senatore et al.\ 2015, see Fig.~\ref{fig-prandoni})
with the aim of identifying kinematical signatures of feeding/feedback loops that can be causally
related to the presence of radio jets. For such studies ALMA can play a crucial role, by imaging the cold (molecular) gas
component in CO on kpc and sub-kpc scales, using Bands 3 and 6 to probe the CO~(1--0) 
and (2--1) transitions respectively. The ATLAS$^{\rm 3D}$ IFS survey (Cappellari et al.\ 2011), for which
molecular-line imaging is available (Alatalo et al.\ 2013) provides an excellent control sample of predominantly
radio-quietETGs. 

With the advent of new-generation integral-field spectrometres, spatially-resolved
kinematic studies of the stellar and  ionized gas galaxy components can be pushed to
significantly higher redshifts. For instance the second-generation {\it VLT Multi
Unit Spectroscopic Explorer (MUSE)} in  {\it Narrow Field Mode} allows to image
galaxies on $< 300$ pc  scales at $z < 0.7$.  This will allow to identify possible
trends with galaxy evolution, using $z\sim 0$ samples as a local Universe
constraint.  The combination of ALMA Band 2+3 would nicely complement MUSE,
allowing to image the CO~(1--0) molecular gas components in galaxies on
comparable scales up to similar redshifts.  Millimeter-VLBI, including ALMA will
allow to probe the innermost regions, down to pc scales (see Sect. 3.4 for more
details).

\subsection{\underline{Evolutionary history of galaxy environments}}

Galaxies occur in a range of environments, from close-pairs to clusters which are
the largest collapsed structures in the Universe with total masses up to
$10^{15}M_{\odot}$ (e.g., Arnaud et al.\ 2009).  One of the most important issues
is the spatial and temporal evolution of star formation activity within these
environments.  A key requirement for star formation is the presence of a reservoir
of dense, cold gas that can be efficiently converted into stars.  This is
especially crucial for galaxies in rich clusters, because they are expected to be
a affected by mechanisms able to remove cold gas from the haloes and disks of
infalling galaxies (e.g., ram pressure stripping, Gunn \& Gott 1972) or to prevent
further cooling of gas within galaxies' dark matter haloes (starvation or
strangulation, e.g., Larson et al.\ 1980; Bekki et al.\ 2002).  This environmental
dependence profoundly influences the evolutionary histories of galaxy clusters and
the star formation-galaxy density relation represents a well-established
observational hallmark of how galaxies evolve as a function of environment (e.g.,
Geach et al.\ 2009).

Studies of star formation activity based on multi-wavelength tracers have shown a
gradual truncation, from $z = 0$ to $z \sim 1$, in the cores of rich clusters
(e.g., Hashimoto et al.\ 1998; Ellingson et al.\ 2001; G{\'o}mez et al. 2003;
Patel et al.\ 2009). Determining the nature and modes of star formation requires a
robust understanding of the relationship between the gas content of a galaxy and
its star formation rate.  Remarkable progress has been made in understanding the
conversion mechanisms in field galaxies  (e.g., Wong \& Blitz 2002; Bigiel et al.\
2008; Daddi et al.\ 2010; Genzel et al.\ 2010; Tacconi et al.\ 2010; Combes et
al.\ 2011; Casasola et al.\ 2015), but the cold  and dense gas fueling the star
formation has been difficult to investigate in clusters (e.g., Wagg et al.\ 2012), 
despite their exceptional opportunities for cosmological studies. 

\begin{figure*}
\centering
\includegraphics[width=0.4\textwidth]{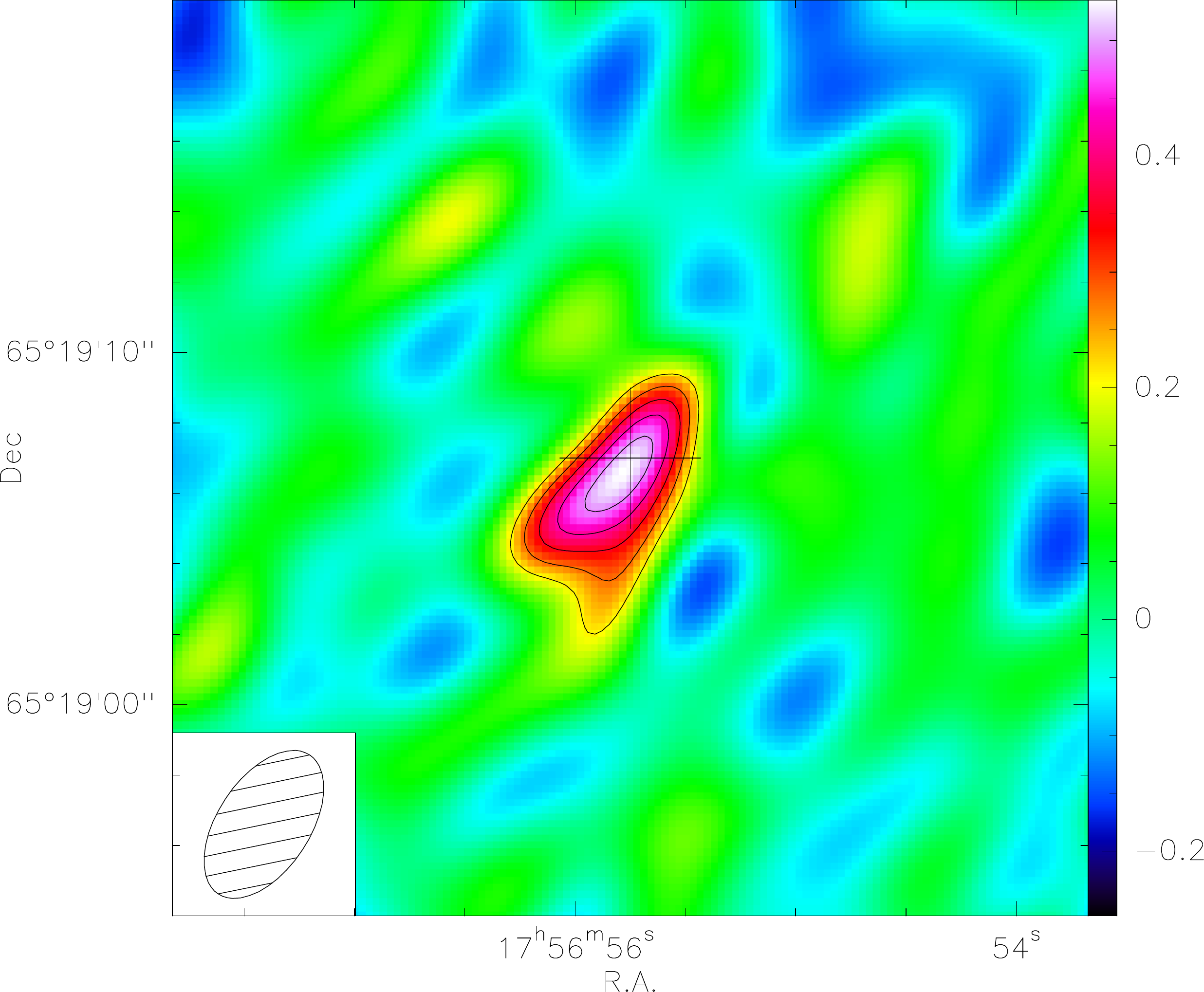}
\hspace*{1cm}
\includegraphics[width=0.5\textwidth]{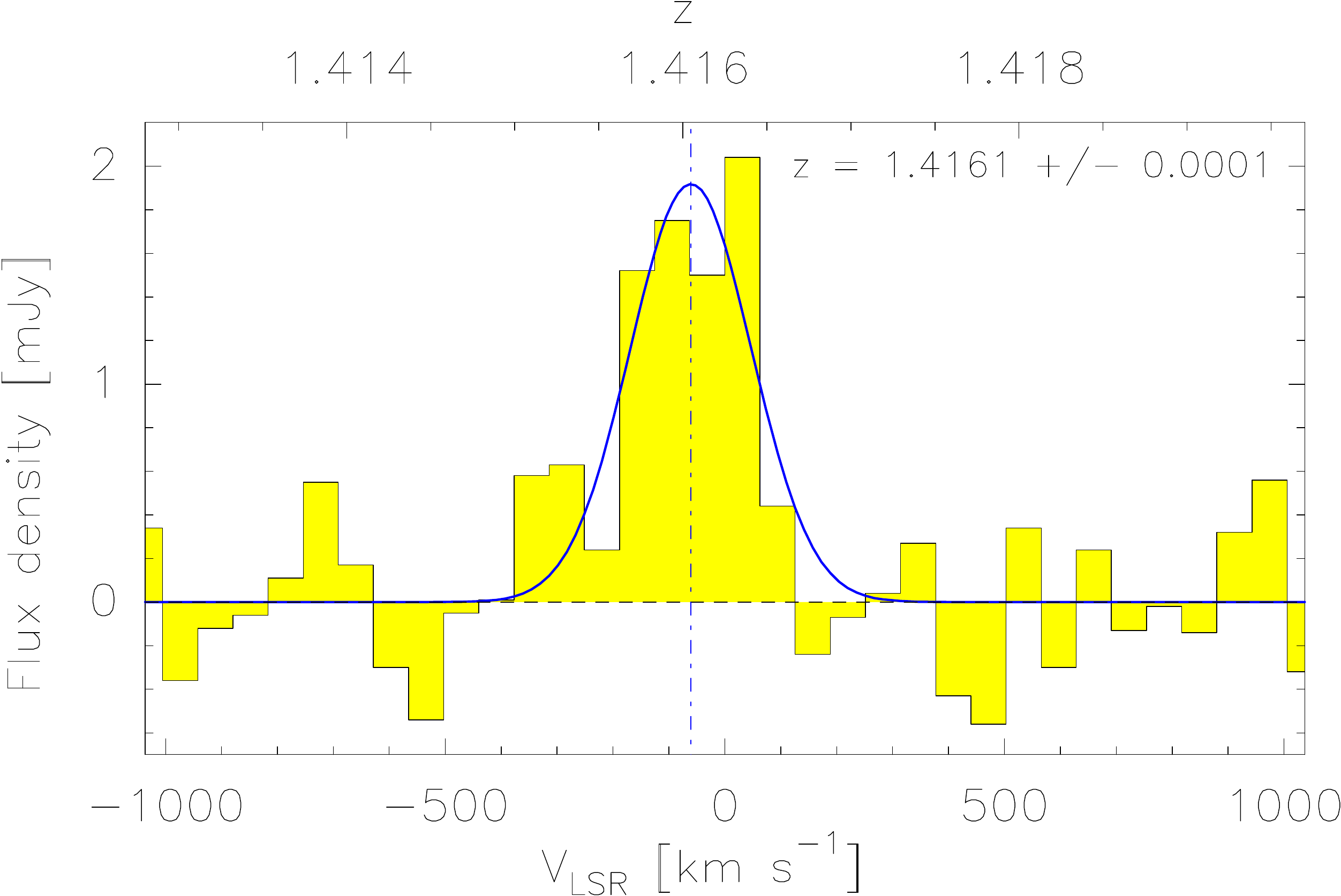}
\caption{
\textit{Left Panel:} CO~(2--1) intensity map obtained with the IRAM PdBI toward AGN.1317. 
The color wedge of the intensity map is in Jy~beam$^{-1}$~km\,s$^{-1}$.
The black cross marks the coordinates of the phase tracking centre of observations.
The {\it rms} noise level is $\sigma = 0.07\,{\rm Jy\,beam^{-1}\,km\,s^{-1}}$ 
and contour levels run from 3$\sigma$ to 7$\sigma$ with 1$\sigma$ spacing.
In this map a velocity range of $\sim$360~km~s$^{-1}$ is used.
The beam of $4.73^{\prime\prime} \times 2.33^{\prime\prime}$ (PA = 147$^{\circ}$)
is plotted at lower left.
\textit{Right Panel:} CO~(2--1) integrated spectrum obtained with the IRAM PdBI toward AGN.1317
centred on the optical velocity of AGN.1317, $ {\rm V_{opt}} = 4.25 \times 10^{5}$~km~s$^{-1}$ ($z = 1.4162$, Galametz et al. 2010)
Figures from Casasola et al. (2013).  
}
\label{fig:map}
\end{figure*}

\begin{figure}[tb]
\centering
\includegraphics[width=0.55\textwidth]{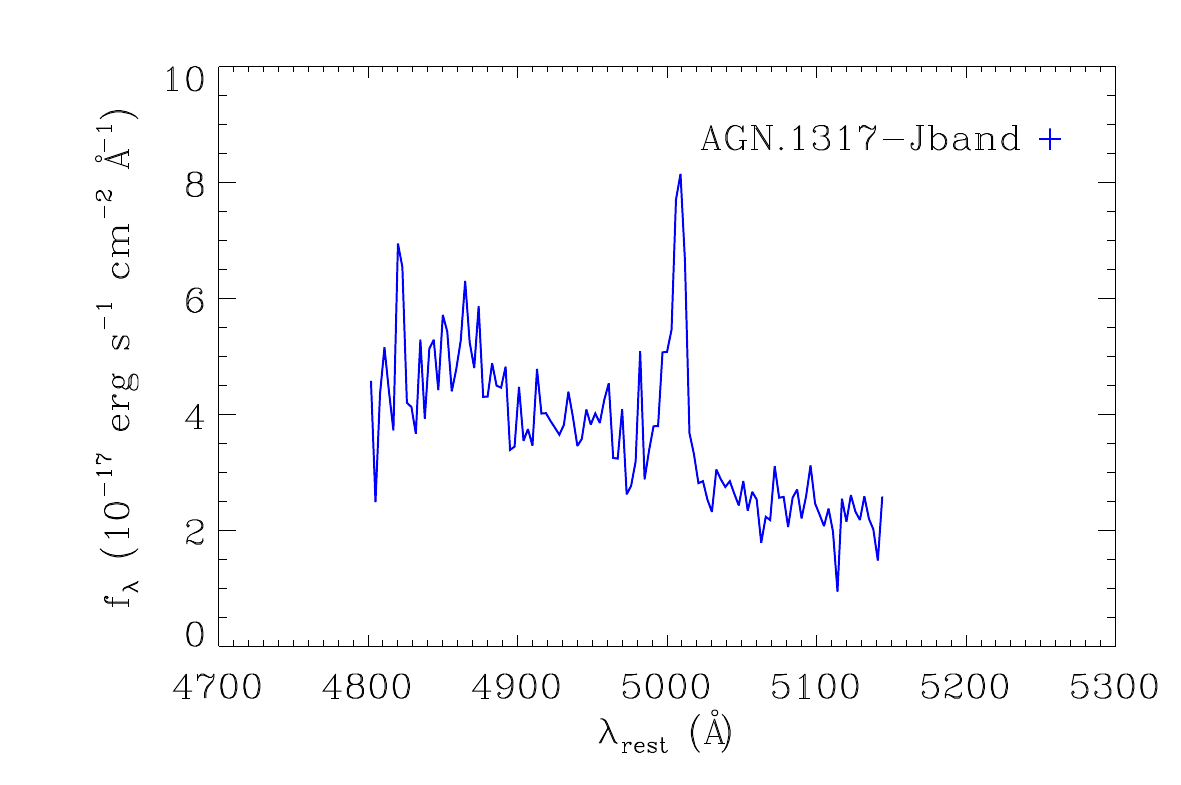}
\caption{\it 
J-band spectrum obtained with the LBT toward AGN.1317.
It shows the detection of {\sc [O~iii]} at 5700~\AA\, and H$\beta$ at 4861~\AA.
Figure from Casasola et al.\ (in preparation).}
\label{fig:lbt}
\end{figure}

We detected, for the first time, CO~(2--1) emission with IRAM PdBI in
AGN.1317, a source belonging  to the galaxy cluster associated with the radio
galaxy 7C~1756+6520 at $z \sim 1.4$ and located nearby the centre of the cluster 
(see Fig.~\ref{fig:map}). Such CO detections in high-$z$ clusters are rare. From
the CO~(2--1) line luminosity, we measured  H$_{2}$ mass of $1.1 \times
10^{10} M_{\odot}$, comparable to that in massive sub-millimeter galaxies. The
optical images of the cluster show that  AGN.1317 is an isolated object, and
thus   its gas mass might be an intrinsic characteristic associated with its early
evolutionary phase,   when most of its baryonic mass was in gas phase, than due to
a merger episode with nearby companions. The H${\alpha}$-derived star formation
rate (SFR) is $\sim 65\,M _{\odot}\,\rm{yr}^{-1}$, and so AGN.1317 would exhaust
its reservoir  of cold gas in $\sim$0.2--1.0~Gyr. This relatively high molecular
gas content and SFR for an AGN near the center of a young cluster is compatible
with  models of evolution of galaxy clusters that predict high SFR at the epoch of
their formation,  i.e. $z \sim 1.5$. Following the predictions of those models, 
at $z \sim 1$, the SFR would  rapidly be quenched  through environmental  effects
starting from the innermost regions of clusters.  Very recently, we obtained
near-infrared observations at the Large Binocular Telescope (LBT) to map the whole
AGN population associated with the radio galaxy 7C~1756+6520. We detected
H$\alpha$, H$\beta$,  {\sc [O~iii]}, {\sc [N~ii]}, and Fe~{\sc ii} line emissions
in four AGN and defined other new eleven sources  with redshift consistent with
that of the cluster (Fig.~\ref{fig:lbt}, Casasola et al. in prep.). Based on these
LBT results we are selecting single cluster galaxies for observations of
CO~(2--1) at the IRAM PdBI   and ­of CO~(1--0) at the  JVLA and to
deeply investigate the nature and modes of star formation in this primordial
cluster.

So far, cold gas has been detected in cluster galaxies only up to $z \sim 0.5$,
especially in the outskirts of rich clusters (e.g., Geach et al.\ 2009; Jablonka et
al.\ 2013), while the range $z \sim 0.5 - 1$ is yet unexplored. For the first time,
ALMA Band 2 will offer the opportunity to study the CO~(1--0) line in group
and cluster galaxies at $z \sim 0.3 - 0.7$, in a redshift range where the gas-star
conversion in clusters is completely unexplored.  Although similar redshift ranges
are observable in other ALMA bands (i.e., Band 6 for CO~(3--2) and  Band 4
for CO~(2--1)), Band 2 enables accurate estimates of molecular gas content
without  ambiguity on the H$_{2}$--CO conversion factor. In addition, Band 2
offers a field-of-view $\sim$2--3 times larger (depending on the redshift range)
than other ALMA bands at higher frequencies; clusters at $z > 1$ are generally
$<1{^\prime}$ in size (e.g., Wagg et al. 2012; Casasola et al. 2013)  so that the
Band 2  field-of-view of $70-90^{\prime\prime}$ gives coverage of the entire
cluster and resolves individual cluster members with only a single pointing.
Observations of cluster members covering a wider radial range can help distinguish
the various physical processes expected to play a role (e.g., ram-pressure
stripping, strangulation, tidal interactions, and mergers) because these processes
peak in effectiveness at different clustercentric radii (Moran et al.\ 2007; De
Lucia et al.\ 2010). How, when, and where such mechanisms affect the evolution of
galaxies has yet to be explored.

\subsection{\underline{Observations of deuterated molecules in nearby galaxies}}

In the 40 years since the first molecular detection in the extragalactic
interstellar medium, the number of species identified in external galaxies is now
more than 50.

Many of these species have been observed in nearby starburst galaxies, an obvious
target of  mm/submm spectral surveys (Martin et al.\ 2006, 2011; Aladro et al.\
2011, 2012; Requena-Torres et al.\ 2011), due to their molecular brightness.

\begin{figure}[h!]
    \centering
    \includegraphics[angle=90,width=0.9\textwidth]{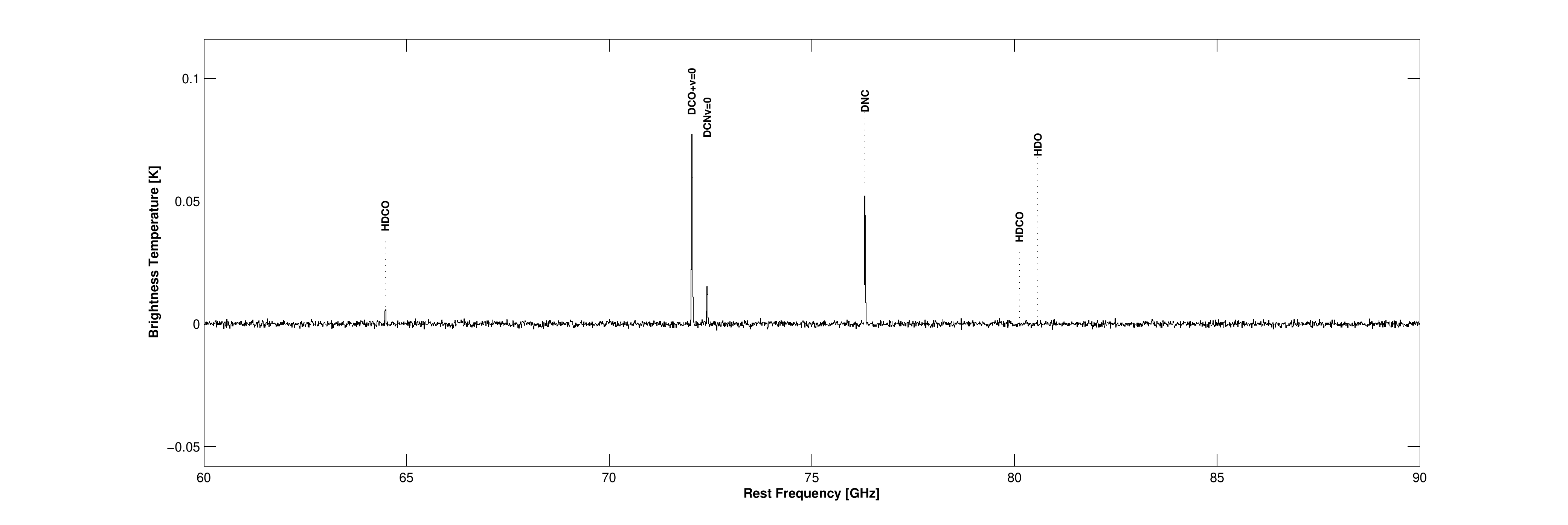}
    \caption{\it The simulation is based on a on time of 6h;  rms 0.00072 K;  
    resolution 4";  full ALMA freq: 90 GHz;  Spectral resolution 50 km/s;  
    Line width=100~km/s; N(H$_2$)=1$\times$10$^{23}$~cm$^{-2}$;  N(Deut.
    Mol)=1$\times$10$^{13}$~cm$^{-2}$;  T$_k$=10 K}
\label{paladino}    
\end{figure}

ALMA's sensitivity will certainly increase the number of detections in "normal"
starforming galaxies,  so far mainly observed in the two most abundant molecules,
tracing star formation: CO and HCN.  The study of chemical composition and physics
of the interstellar gas in star-forming regions is extremely important in order to
understand the complex processes of star formation. Observations of molecular
lines of many species provide comparative molecular abundances and inferences on
the evolution of physical conditions and atomic abundances in galaxies.  CO is
widely used to map the distribution of molecular hydrogen, but in dense star
forming regions it can be highly depleted or photodissociated: in these cases
observations of other molecular tracers become extremely important.  In highly 
embedded star forming regions, where CO is depleted, the chemistry is driven by
deuterated species. In our Galaxy, in such regions, a large deuterium
fractionation is observed, well above the elemental abundance ratio D/H of
$1.5\times10^{-5}$. Deuteration in extragalactic star-forming environments, where the
physical conditions can be drastically different from what we see in our own
Galaxy, is almost an unexplored territory. Observations of deuterated molecules in
external galaxies will help determining (i) the D/H ratio, which depends on the
degree of processing of the gas, and hence can give us clue on the evolution of
the gas in a galaxy as well as be used as tracer of nucleosynthesis in the Big
Bang, and (ii) the depletion in dense gas in external galaxies which will give us
clues on the star formation process.

\subsubsection{Justification for ALMA Band 2+3}

ALMA Band 2 will make accessible a range of frequencies where no interferometric
capabilities currently exist. Some deuterated molecular transitions are uniquely
observable in ALMA Band 2, such as: DCO$^+$~(1--0), DCN~(1--0), DNC~(1--0),
N$_2$D$^+$~(1--0).  Astrochemical models (Bayet et al. 2010) have explored a large
parameter space of physical conditions covering different extragalactic
environments and provided a guide to observations of deuterated molecules arising
from the dense gas in external galaxies. Although qualitative in nature, the
models led to some important conclusions: (i) HDO and DCN are abundant, regardless
of the extragalactic environment; and (ii) DCO$^+$ is a tracer of cosmic rays
enhanced gas. Clearly, observations of the fundamental transitions, in Band 2 give
the best chance of optimizing the detection of these species.  A tentative
detection of DCN~(2--1) line has been reported by Martin et al.\ (2006) in the
nucleus of NGC253. This detection, if confirmed, would be in agreement with the
prediction of a cosmic ray enhanced environment in that region. NGC253 would be
the ideal target for an observational campaigns aiming at validating the model
predictions.  Figure~\ref{paladino} shows an example of a simulated ALMA spectrum in 
Band 2 containing the selected deuterated species. 

\subsection{{\underline{Imaging super-massive black holes and origin of AGN jets 
with  mm-VLBI}}}

\subsubsection{mm-VLBI with ALMA}

The huge potential of  Very Long Interferometer (VLBI)  technique at submm/mm
wavelengths is that it enables the highest resolution imaging currently possible
at any wavelength in astronomy, achieving about 20 microarcsecs for a 9000~km
baseline at 1.3~mm.  The use of ALMA as a phased array to form part of a global
VLBI network will offer unprecedented sensitivity at very high angular resolution,
making one order of magnitude weaker sources accessible to $\mu$as-resolution
studies and greatly improving image fidelity for the brighter objects (Fish et
al.\ 2013, Tilanus et al.\ 2014) . Existing activities at 3~mm (the Very Long Baseline
Array and the Global Millimeter VLBI Array) and at 1~mm (the Event Horizon
Telescope project) have laid the ground work for future coordinated observations.
An international consortium is presently constructing a beamformer for the ALMA
that will be available as a facility instrument providing substantial improvements
to existing VLBI arrays at 7 and 3~mm.  The wide mm-VLBI science achievable by a
beamformer for ALMA has recently been discussed by Fish et al.\ (2013) and Tilanus
et al.\ (2014). In the following, we discuss the AGN science cases that will be
possible thanks to ALMA Band 2 or Band 2+3 observations in conjunction with other
millimetre telescopes.

\subsubsection{Event horizon scale imaging} 

The inclusion of the phased ALMA in global mm-VLBI observations will open new
frontiers, offering the very high sensitivity and high angular resolution
necessary to image the emission around nearby super massive black holes, even on
event horizon scales. This will provide tests not only of the black hole paradigm,
Einstein's General Theory of Relativity (GR) in its strong-field limit, but also
of e.g., the ``cosmic censorship" conjecture and the "no-hair" theorem (Fish et al.\
2013, Tilanus et al.\ 2014). Because of their vicinity, the principal targets for
this kind of analysis are SgrA$^*$ and M87.  Each mm-VLBI frequency penetrates a
different depth into the black hole environment. Full frequency coverage,
including Band 2 or 2+3, is needed to distinguish between changes in spectral index
due to scattering or due to intrinsic changes in emitting material and
relativistic effects. While for SgrA$^*$ the scattering in Band 2 could be a critical
issue, for M87, which is actively accreting material and emitting jets on
kilo-parsec scale,  Band 2 or 2+3 observations would probe (European Science Case for
ALMA Band 2: Fuller et al., in
preparation) the material at around 12 Schwarzchild radii where the spectral
energy distribution (SED) deviates from a power law and variability is
increasingly rapidly (Hada et al.\ 2012). 

\subsubsection{Jet launching region study} 

As well explained in Fish et al.\ (2013) and Tilanus et al.\ (2014),  the mm-VLBI
is also the only method to directly image the AGN jets down to their formation
region and study the acceleration/collimation zone in the optically thin regime,
i.e.\  on Schwarzschild radius scales where magneto-hydrodynamic (MHD)
instabilities form.  The very high sensitivity feasible thanks to the inclusion of
the phased ALMA in mm-VLBI observations  will provide a great chance to explore
for the first time these jet regions in fainter AGN. Intriguing targets  (see
Tilanus et al.\ 2014) are  for example the LLAGN,  where cm-wavelength VLBI has
already provided interesting clues;  blazars,  where the high-energy emission
sites and mechanisms are uncertain; and sources with double-sided jets  for which
mm-VLBI observations can provide a direct determination of the location of the
central engine relative to the jet and counter-jet features. Band 2 VLBI could
provide an ideal balance of resolution and flux density necessary to investigate
the launching jet region in these nearby (week) AGN. This is the case of the
compact active galaxies studied by Liuzzo et al. 2013 with observed small-scale
nucleus radio structures posing questions such as the origins of FRI and the
evolution of BL Lac objects  but with SEDs suggesting significant features in the
frequency range covered by Band 2 or 2+3. 

\subsubsection{Jet polarization analysis}

Not only the mm-VLBI  continuum analysis will be essential but also the
polarization study is fundamental in the understanding of the jet launching
mechanisms. Polarimetry with mm-VLBI is indeed a unique tool to image at
Schwarzchild radii the intrinsic magnetic field configuration, helping to
discriminate whether the jets are launched by MHD effects (Tilanus et al.\ 2014).
Observations of changes in jet position angle with frequency could reveal a
helical nature, or  time-change in orientation of the mm-wavelength jet could be
related to precession of the relativistic jet. Good frequency coverage, including
Band 2 or 2+3, is therefore essential (European Science Case for
ALMA Band 2: Fuller et al., in preparation) to trace the spectral
curvature, identify the jet flaring region and avoid ambiguities in Faraday
rotation analysis due to the polarization angle changing very rapidly with
frequency (Porth et al.\ 2011).  Recently, it has been also suggested that the Band
2 or 2+3 emitting region is complex, with a three-dimensional structure. The possible
comparison of Band 2 or 2+3 polarimetric observations with space-VLBI (e.g.,
RADIOASTRON) ones at 22~GHz  will allow to trace the 3D-field  with an angular
resolutions better than 50 microarcseconds and to determine the Rotation Measure
(Tilanus et al.\ 2014).

Finally, mm-VLBI polarimetric observations could trace (European Science Case for
ALMA Band 2: Fuller et al., in
preparation)  the evolution of the magnetic field after flares in blazars and give
important hints on the origin of  their high energy emission  (e.g., Orienti et
al.\ 2013). For example, the possible change of the mm-VLBI polarization
properties close in time with some high energy flares could be related either to
the propagation of a shock along the jet that orders the magnetic field, or a
change of the opacity regime. 

\subsection{\underline{ALMA Band 2 Science case: The Sunyaev-Zel'dovich effect}}

\subsubsection{Sunyaev-Zel'dovich effect in Galaxy clusters}

The Sunyaev-Zel'dovich effect (SZE) is the spectral distortion of the Cosmic
Microwave Background (CMB) due to inverse Compton scattering of CMB photons off a
cloud of electrons (Birkinshaw 1999; Carlstrom et al.\ 2002). The increase of the
photon energy implies a decrease of the CMB brightness at frequencies below 218
GHz and an increase at higher frequencies. The minimum of the CMB brightness
(i.e., the maximum of the distortion) is reached at about 128 GHz. The amplitude of the
distortion depends only on the properties of the electron cloud: in the case of
thermal electrons it is proportional to the integral of the electron pressure
(i.e. $T_{SZ}\propto T_e n_e$, where $n_e$ is the electron density and $T_e$ is
the electron temperature) along the line of sight, and is independent of
distance. 

Galaxy clusters are massive structures permeated by hot dense ionized gas that
preserves the same cosmic baryonic fraction of the epoch of the cluster
virialization, so they are the ideal targets for SZE observations and to infer
pieces of information about cosmology. The absolute X-ray luminosity from the hot
cluster gas has a different dependence on gas properties and is dependent on the
cluster redshift. For virialized clusters, where the distribution of density and
temperature are easy to model with regular forms (e.g., isothermal and spherically
symmetric), the different distance dependence provides a powerful method to
extract cosmological parameters by combining the two signals. The comparison of
the two signals is also a particularly effective tool to estimate the gas mass but
the different dependence on density makes X-ray more susceptible to the gas
clumping. The SZE signal is a better tracer of less dense hotter region. 

Current theories of structure formation predict that clusters form hierarchically
via merger of smaller structures (Borgani et al.\ 2001). X-ray observations
(Jeltema et al.\  2005; Maughan et al.\ 2008) of high redshift ($z>0.5$) clusters
revealed that they are more morphologically complex, less virialized and
dynamically more active than low-redshift clusters. Studies of $z > 0.8$ clusters
show clumpy and elongated structures suggesting that they are close to the epoch
of cluster formation (Rosati et al.\ 2004). The study of high redshift cluster SZE
can provide constraints on the theories of cluster formation and evolution.
However,for these objects the temperature and density distributions might be so
complex that the comparison of X-ray and SZE signals is seriously compromised, and
the information about cosmology or cluster evolution will be misleading. Sub-arcmin resolution centimetric wavelength
observations with ATCA and EVLA (Massardi et al.\ 2010; Malu et al.\ 2010; Mason
et al.\ 2010) of 3 $z>0.8$ clusters (Cl0152-1357, the Bullet cluster and Cl
1357-1145 respectively) confirmed that the structure are morphologically complex
and that there is a displacement of the peak of SZ with respect to the X-ray
emission peak, most likely under the effect of merging events, and strongly
indicating the need for sub arcmin resolution observations to properly compare the
signals and characterize the cluster structure if far from virialization. These conclusions
were confirmed by single dish observations at 90~GHz (Korngut et al. 2011).

Furthermore, by combining X-ray emission with sub-arcmin scale SZE can probe
turbolence and entropy excesses investigating the intra-cluster medium (ICM) pressure
fluctuations, the dynamics of shocks and blasts that enhance the SZ signal within the
ICM, the temperature and density profile of the ICM. This will also help to
understand the cooling flows mechanisms and the AGN feedback within the ICM. Finally,
ALMA will be crucial to follow-up of galaxy clusters detected in large area surveys
(e.g. PCSZ, Planck Collaboration et al. 2014)

\paragraph{\large Justification for ALMA Band 2+3}

The SZE signal is also typically contaminated by radio source emission which is
usually associated to early-type galaxies which preferentially reside in galaxy
clusters. In mm bands, the radio component decreases and the contribution due to
dust emission from star-forming galaxies in the cluster increases with frequency.
The minimum of the contribution is in the so-called ``cosmological window'' at
about 70 GHz. Observations in band 2+3 earn in resolution with respect to centimetric
observations so far available without losing in sensitivity as the absolute
value of the SZE increase by almost an order of magnitude. Point sources can be also directly observed (and removed on the uv
plane) by exploiting the longer baselines of the main ALMA array, and the overall
SZ signal can be measured by mean of the total power antennas.

\subsubsection{Sunyaev-Zel'dovich in early stages of galaxy formation}

Since SZ is independent of distance it is a powerful tool of investigation of high
redshift structures permeated by hot electron clouds like galaxy clusters and star
forming galaxies. In the standard galaxy formation scenario plasma clouds  with a
high thermal energy content must exist at high redshifts since the protogalactic
gas is shock heated to the virial temperature, and extensive  cooling, leading to
efficient star formation, must await the collapse of massive haloes (downsizing,
see Granato et al. 2004; Massardi et al.\ 2008). Model of emission from the early
stages of galaxy formation pointed out that thermal SZE from the virializing gas
is their most significant signal of emission in the radio-mm band. The signal is
of the order of 1~$\mu$K at 3--4~mm wavelength on 10~arcsec angular scales

\paragraph{\large Justification for ALMA Band 2+3}

Model-based estimations indicate that we will need about 30~h on source to survey
1sqdeg down to about $10^{-4}$ Jy, where we can detect up to 100 proto-galaxies
with ALMA Band 2+3 (70--100 GHz). Band 2 should be preferred for the lower
level of dust contamination. Radio luminosity from star formation in the core
regions steeply increases with frequency.  Thermal dust emission from the core
region increases with frequency and overwhelms SZ signal at 100~GHz. In the range
35--80 GHz observations of thermal SZ is possible for most massive haloes. Arcsec
resolution images may reconstruct the uncontaminated SZ signal by distinguishing
contaminating emissions on the scale of the stellar distributions ($<$1 arcsec at
$z>1$) and the SZ effects on the scale of the dark matter halo (typically ten
times larger) and by detecting and subtracting out the confusion effect by radio
sources.

\noindent

{\small
\section*{References}

Aalto, S., Booth, R.~S., Black, J.~H., \& Johansson, L.~E.~B.\ 1995, \aap, 300,
369  \\
Aalto, S., Garcia-Burillo, S., Muller, S., et al.\ 2012, \aap, 537, AA44  \\
Aladro, R. et al., 2011, A\&A, 535A, 84A\\
Aladro, R. et al., 2012, A\&A, 549A, 39A\\
Alatalo, K., Blitz, L., Young, L.M., et al.\ 2011, ApJ, 735, 88 \\
Alatalo, K., Davis, T.A., Bureau, M., et al.\ 2013, MNRAS, 432, 1796 \\
Allen, S.~W., Dunn, R.~J.~H., Fabian, A.~C., Taylor, G.~B., Reynolds, C.~S. 2006, MNRAS, 372, 21 \\
Anderson, C.~N.,  Meier, D.~S., Ott, J., et al.\ 2014, \apj, 793, 37 \\
Aravena, M., Carilli, C., Daddi, E., et al.\ 2010, \apj, 718, 177 \\
Arce, H.G., Santiago-Garc\'{\i}a, J., J\"orgensen, J.K., Tafalla, M., Bachiller,
R. 2008, A\&A 681, L21  \\
Armijos-Abenda\~no, J., Mart\'in-Pintado, J., Requena-Torres, M. A., et al.\ 2015, MNRAS, 446, 3842 \\
Arnaud, M.\ 2009, \aap, 500, 103 \\
Arzner K., G\"udel M., Briggs K., Telleschi A., Audard M., 2007,
A\&A, 468, 477 \\
Bachiller, R., P\'erez Guti\'errez, M., Kumar, M.S.N., Tafalla, M. 2001, 
A\&A 372, 899 \\ 
Bacmann, A., Taquet, V., Faure, A., Kahane, C., Ceccarelli, C. 2012, 
A\&A 541, L12 \\
Bartkiewicz A., Szymczak M., van Langevelde H.~J., et al. 2009, A\&A, 502, 155\\
Bayet, E. et al., 2010, ApJ, 725, 214\\
Bekki, K., Couch, W.~J., \& Shioya, Y.\ 2002, \apj, 577, 651 \\
Beltr\'an, M.\ T., Codella, C., Viti, S., Neri R., Cesaroni, R. 2009, A\&A, 690,
L93 \\
Beltr{\'a}n M.~T., S{\'a}nchez-Monge {\'A}., Cesaroni R., et al. 2014, A\&A,
571, A52 \\
Benedettini, M., Viti, S., Codella, C., et al. 2013, MNRAS 436, 179 \\ 
Bergin, E.~A., Aikawa, Y., Blake, G.~A., \& van Dishoeck, E.~F.\ 2007,
  Protostars and Planets V, 751 \\
Bergin, E.~A., Cleeves, L.~I., Gorti, U., et~al.\ 2013, Nature, 493, 644 \\
Bigiel, F., Leroy, A., Walter, F., et al.\ 2008, AJ, 136, 2846 \\
Birkinshaw, M.\ 1999, Physics Reports, 310, 97 \\
Bisschop, S.E., J\"orgensen, J.K., Bourke, T.L., Bottinelli, S., van Dishoeck, E.F. 2008, A\&A 488, 959 \\
Bolatto, A.~D., Wolfire, M., \& Leroy, A.~K.\ 2013, \araa, 51, 207 \\
Bower et al., 2003, ApJ 598, 1140 \\
Borgani, S., \& Guzzo, L.\ 2001, \nat, 409, 39 \\
Bottinelli, S., Ceccarelli, C., Williams, J.~P., \& Lefloch, B.\ 2007,
  A\&A, 463, 601 \\
Buchbender, C., Kramer, C., Gonzalez-Garcia, M., et al.\ 2013, \aap, 549, AA17 \\
Calcutt, H., Viti S., Codella C., Betr\'an M., Fontani F., 2014, MNRAS, 433, 3157
\\
Cappellari, M., Emsellem, E., Krajnovi{\'c}, D., et al.\ 2011, MNRAS, 413, 813  \\
Carilli, C.~L., \& Walter, F.\ 2013, \araa, 51, 105 \\
Carlstrom, J.~E., Holder, G.~P., \& Reese, E.~D.\ 2002, \araa, 40, 643 \\
Carr, J.~S. \& Najita, J.~R.\ 2008, Science, 319, 1504 \\
Casasola, V., Magrini, L., Combes, F., et al.\ 2013, \aap, 558, A60  \\
Casasola, V., Hunt, L., Combes, F., \& Garcia-Burillo, S.\ 2015, A\&A, 577, A135 \\
Caselli, P. \& Ceccarelli, C.~2012, A\&ARv, 20, 56 \\
Caselli, P., Keto, E., Bergin, E.A., et al.\ 2012, ApJ 759, L37 \\
Cazaux, S., Tielens, A.~G.~G.~M., Ceccarelli, C., et~al.\ 2003, ApJL,
  593, L51 \\
Ceccarelli, C., Caselli, P., Herbst, E., Tielens, A.G.G.M., Caux, E. 2007, Protostars and Planets V, University of Arizona Press, Tucson, 47 \\
Cernicharo, J.\ et al.\  2012, ApJ, 759, L43 \\
Cesaroni, R. 2008, Ap\&SS, 313, 23 \\
Cesaroni, R., Galli, D., Lodato, G., Walmsley, M., Zhang, Q. 2007,
Protostars and Planets V, 197 \\
Cesaroni, R., Hofner, Walmsley, C.\ M., \& Churchwell, E.\ 1998, A\&A, 331, 709 \\
Chapillon, E., Dutrey, A., Guilloteau, S., et~al.\ 2012b,
  ApJ, 756, 58 \\
Chapillon, E., Guilloteau, S., Dutrey, A., Pi{\'e}tu, V., \& Gu{\'e}lin, M.\ 2012a, A\&A, 537, A60 \\
Cicone, C., Feruglio, C., Maiolino, R., et al.\ 2012, \aap, 543, AA99 \\
Cicone, C., Maiolino, R., Sturm, E., et al.\ 2014, \aap, 562, AA21 \\
Codella, C., Benedettini, M., Beltr\'an, M.T., et al. 2009, A\&A 507, L25 \\
Codella, C., Fontani, F., Ceccarelli, C., et al. 2015, ApJ 449, L11  \\
Collins, P. \& Ferrier, R., 1995, Monosaccharides (New York: Wiley) \\
Combes, F., Garc{\'{\i}}a-Burillo, S., Braine, J., et al.\ 2011, \aap, 528, A124 
\\
Combes, F., Garc{\'{\i}}a-Burillo, S., Casasola, V., et al.\ 2013, \aap, 558,
AA124 \\
Combes, F., Garc{\'{\i}}a-Burillo, S., Casasola, V., et al.\ 2014, \aap, 565, AA97
Costagliola, F., Aalto, S., Rodriguez, M.~I., et al.\ 2011, \aap, 528, AA30 \\
Coutens, A., Persson, M. V., J{\o}rgensen, J. K. et al.\ 2015,  A\&A, 576, 5 \\
Coutens, A., Vastel, C., Cabrit, S., et~al.\ 2013, A\&A, 560, A39 \\
Coutens, A., Vastel, C., Caux, E., et~al.\ 2012, A\&A, 539, A132 \\
Cragg D.~M., Sobolev A.~M., and Godfrey P.~D. 2005, MNRAS, 260, 533\\
Crapsi, A., Caselli, P., Walmsley, C.M., et al. 2005, ApJ, 619, 379 \\
Cresci, G., Mainieri,  V., Brusa, M., et al.\ 2015, \apj, 799, 82 \\
Crocker, A.F., et al.\ 2008, MNRAS, 368, 1811\\ 
Crocker, A.F., et al.\ 2009, MNRAS, 393, 1255 \\
Crocker ,A.F., et al.\ 2011, MNRAS, 410, 1197\\
Daddi, E., Bournaud, F.,  Walter, F., et al.\ 2010, \apj, 713, 686 \\
Daddi, E., Elbaz, D., Walter, F., et al.\ 2010, \apjl, 714, L118 \\
Danielson, A.~L.~R., Swinbank, A.~M., Smail, I., et al.\ 2011, \mnras, 410, 1687
\\
Dannerbauer, H., Daddi, E., Riechers, D.~A., et al.\ 2009, \apjl, 698, L178 \\
Dauphas, N., Robert, F., \& Marty, B.\ 2000, Icarus, 148, 508 \\
Davis, T.A., Alatalo, K., Sarzi, M., et al.\ 2011, MNRAS, 417, 882\\
Davis, T.A., Alatalo, K., Bureau, M., et al.\ 2013a, MNRAS, 429, 534\\
Davis, T.A., et al., 2013b, Nature, 494, 328\\
De Lucia, G., Boylan-Kolchin, M., Benson, A.~J., Fontanot, F., 
\& Monaco, P.\ 2010, \mnras, 406, 1533 \\
Di Francesco et al., 2008, ApJS, 175, 277 \\
Di Matteo, T., Springel, V., \& Hernquist, L.\ 2005, \nat, 433, 604 \\
Drake S. A., Linsky J. L., 1989, AJ, 98, 1831 \\
Dullemond, C.~P., Hollenbach, D., Kamp, I., \& D'Alessio, P.\  2007, Protostars and Planets V, 555 \\
Dutrey, A., Guilloteau, S., \& Guelin, M.\ 1997, A\&A, 317, L5 \\
Dutrey, A., Semenov, D., Chapillon, E., et~al.\ 2014, Protostars and Planets VI, 317 \\
Dutrey, A., Wakelam, V., Boehler, Y., et~al.\ 2011, A\&A, 535, A104 \\
Ellingson, E., Lin, H., Yee, H.~K.~C., \& Carlberg, R.~G.\ 2001, \apj, 547, 609 \\
Elsila, J.E., Glavin, D.P., Dworkin, J.P. 2009, M\&PS 44, 1323 \\
Emsellen E., Cappellari, M., Peletier, R.F., et al.\ 2004, MNRAS, 352, 721\\
Emsellem, E., Cappellari, M., Krajnovi{\'c}, D., et al.\ 2011, MNRAS, 414, 888 \\
Fabian, A.C. 2012, ARA\&A, 50, 455 \\
Fedoseev, G. et al., 2015, MNRAS, 448, 1288 \\
Feigelson \& Montmerle 1999 ARA\&A 37, 363 \\
Feruglio, C., Fiore, F., Maiolino, R., et al.\ 2013, \aap, 549, AA51 \\
Feruglio, C., Maiolino, R., Piconcelli, E., et al.\ 2010, \aap, 518, LL155 \\
Fish, V., Alef, W., Anderson, J., et al. 2013, arXiv:1309.3519 \\
Fontani, F., Busquet, G., Palau, A., Caselli, P. et al.~2015, A\&A, 575, 87 \\
Fontani, F., Codella, C., Ceccarelli, C., Lefloch, B., Viti, S., Benedettini, M. 2014, ApJ 788, L43 \\
Forbrich J., Preibisch T., Menten K. M.\ et al.\  2007, A\&A, 464, 1003 \\
Forbrich et al., 2008, A\&A, 477, 267 \\
Forbrich \& Wolk 2013 A\&A 551, 56 \\
Fuente, A, Cernicharo, J., Ag{\'u}ndez, M., et~al.\ 2010, A\&A, 524,
  A19 \\
Fuente, A, Cernicharo, J., Caselli, P., et al., 2014, 568, 65 \\
Fuller, G., et al., in preparation, "The Science Case for ALMA Band 2" \\
Galametz, A., Stern, D., Stanford, S.~A., et al.\ 2010, \aap, 516, A101 \\
Gao, Y., \& Solomon, P.~M.\ 2004, \apj, 606, 271 \\
Garc{\'{\i}}a-Burillo, S., Usero, A., Alonso-Herrero, A., et al.\ 2012, \aap, 539,
AA8 \\
Garc{\'{\i}}a-Burillo, S., Combes, F., Usero, A., et al.\ 2014, \aap, 567, AA125
\\
Geach, J.~E., Smail, I., Coppin, K., et al.\ 2009, \mnras, 395, L62 \\
Genzel, R., Tacconi, L.~J., Gracia-Carpio, J., et al.\ 2010, \mnras, 407, 2091 \\
Genzel, R., Tacconi, L.~J., Lutz, D., et al.\ 2015, \apj, 800, 20 \\
Gerin, M.,  Goicoechea, J.\ R., Pety, J., Hily-Blant, P. 2009, A\&A, 494, 977 \\
Getman K. V., Feigelson E. D., Micela G.\ et al.\ 2008, ApJ, 688, 437 \\
Glavin, D.P., Dworkin, J.P., Aubrey, A., et al. 2006, M\&PS 41, 889 \\
Glover, S.~C.~O., \& Mac Low, M.-M.\ 2011, \mnras, 412, 337 \\
Glover, S.~C.~O., Federrath, C., Mac Low, M.-M., \& Klessen, R.~S.\ 2010, \mnras,
404, 2 \\
G{\'o}mez, P.~L., Nichol, R.~C., Miller, C.~J., et al.\ 2003, \apj, 584, 210 \\
Granato, G.~L., De Zotti, G., Silva, L., Bressan, A., \& Danese, L.\ 2004, \apj, 600, 580 \\
G\"udel 2002 ARA\&A 40, 217  \\
Gueth F., Guilloteau S., \& Bachiller R. 1996, A\&A 307, 891 \\
Gueth F., Guilloteau S., \& Bachiller R. 1998, A\&A 333, 287 \\
Guilloteau, S., Di Folco, E., Dutrey, A., et~al.\ 2013, A\&A, 549, A92 \\
Gunn, J.~E., \& Gott, J.~R., III 1972, \apj, 176, 1 \\
Hada, K., Kino, M., Nagai, H., et al. 2012, ApJ, 760, 52\\
Halfen, D. T., Apponi, A. J., Woolf, N., Polt, R.; Ziurys, L. M., 2006, ApJ, 639,
237 \\
Hartogh, P., Lis, D.~C., Bockel{\'e}e-Morvan, D., et~al.\ 2011, Nature,
  478, 218 \\
Hashimoto, Y., Oemler, A., Jr., Lin, H., \& Tucker, D.~L.\ 1998, \apj, 499, 589 \\
Heckman, T.M., Best, P.N. 2014, ARAA, 52, 589\\
Hogerheijde, M.~R., Bergin, E.~A., Brinch, C., et~al.\ 2011, Science,
  334, 338  \\
Hollis, J.M., Lovas, F.J., \& Jewell, P.R. 2000, ApJ, 540, L107 \\
Hollis, J.\ M. et al., 2002, ApJ, 571, 59 \\
Holtom, P.D., Bennett, C.J., Osamura, Y., Mason, N.J., Kaiser, R.I. 2005,
 ApJ 626, 940 \\
Ivison, R.~J., Papadopoulos, P.~P., Smail, I., et al.\ 2011, \mnras, 412, 1913 \\
Jablonka, P., Combes, F., Rines, K., Finn, R., \& Welch, T.\ 2013, \aap, 557, A103
\\
Jeltema, T.~E., Canizares, C.~R., Bautz, M.~W., \& Buote, D.~A.\ 2005, \apj, 624, 606 \\
Jim\'enez-Serra, I., Testi, L., Caselli, P., Viti, S. 2014, ApJ 787, L83 \\
J\"orgensen, J.K., Favre, C. Bisschop, S.E., Bourke, T.L., van Dishoeck, E.F., Schmalzl, M. 2012, ApJ 757, L4 \\
J\"orgensen, J.K., \& van Dishoeck, E.~F.\ 2010, ApJL, 725, L172 \\
Juneau, S., Narayanan, D.~T., Moustakas, J., et al.\ 2009, \apj, 707, 1217 \\
Keto, E. 2002, ApJ, 580, 980 \\
de Koff, S, Best, P, Baum, S. A., et al\ 2000, ApJS, 129, 33\\
Korngut P.~M., et al., 2011, ApJ, 734, 10 \\
Kuiper R., Klahr H., Beuther H., et al.\ 2010, ApJ, 722, 1556\\ 
Kundu et al., 2000, ApJ 545, 1084 \\
Kurtz, S. 2005, IAUS 227, 111 \\
Lagos, C.~D.~P., Baugh, C.~M., Lacey, C.~G., et al.\ 2011, \mnras, 418, 1649 \\
Laing, R.~A., Bridle, A.H. 2002, MNRAS, 336, 328\\
Larson, R.~B., Tinsley,  B.~M., \& Caldwell, C.~N.\ 1980, \apj, 237, 692 \\
Lauer, T.~R., Faber, S.~M., Gebhardt, K., et al.\ 2005, AJ, 129, 2138  \\
Lee, C.-F., Hirano, N., Zhang, Q., et~al.\ 2014, ApJ, 786, 114 \\
Liu H. B., Galv\'an-Madrid R., Forbrich J.\ et al.\ 2014, ApJ, 780, 155 \\
Liuzzo, E., Buttiglione, S., Giovannini, G., et al.  2013, A\&A, 550, AA76 \\
L\'opez-Sepulcre, A., Mendoza, E., Lefloch, B., et al. 2015, MNRAS, in press \\
Malu, S.~S., Subrahmanyan, R., Wieringa, M., \& Narasimha, D.\ 2010, arXiv:1005.1394 \\
Maret S., Hily-Blant P., Pety J., et al., A\&A 526, A47 \\
Martin, S. et al., 2006,  ApJS, 164, 450\\
Martin, S. et al., 2011, IAUS, 280,  351\\
Mason, B.~S., Dicker, S.,Korngut, P., et al.\ 2010, Bulletin of the American Astronomical 
Society, 42, \#364.06 \\
Massardi, M.,  Ekers, R.~D., Ellis, S.~C., \& Maughan, B.\ 2010, \apjl, 718, L23 \\
Massi et al., 2006, A\&A, 453, 959  \\
Mathews, G.~S., Klaassen, P.~D., Juh{\'a}sz, A., et~al.\\ 2013, A\&A,
  557, A132 \\
Matsui, T. \& Abe, Y.\ 1986, Nature, 322, 526 \\
Maughan, B.~J., Jones, C., Forman, W., \& Van Speybroeck, L.\ 2008, \apjs, 174, 117 \\
Maury, A. J., et al., 2014, A\&A, 563, 2 \\
Meijerink, R., Spaans, M., \& Israel, F.~P.\ 2007, \aap, 461, 793 \\
Mendoza, E., Lefloch, B., L\'opez-Sepulcre, A., et al. 2014, MNRAS 445, 151 \\
Menten K.~M. 1991, ApJ, 380, L75\\
Minier V., Ellingsen S.~P., Norris R.~P., et al. 2003, A\&A, 403, 1095\\
Modica, P. \& Palumbo, M.~E.\ 2010, A\&A, 519, A22 \\
Moran, S.~M., Ellis, R.~S., Treu, T., et al.\ 2007, \apj, 671, 1503  \\
Morbidelli, A., Chambers, J., Lunine, J.~I., et~al.\ 2000, Meteoritics
  and Planetary Science, 35, 13 \\
Morganti, R., et al.\ 2013, Science, 341, 1082\\
Moscadelli L., Cesaroni R., Rioja M.~J., et al. 2011, A\&A. 526, A66\\
Mumma, M.~J. \& Charnley, S.~B.\ 2011, ARA\&A, 49, 471 \\
Mu{\~n}oz Caro, G.~M., Meierhenrich, U.~J., Schutte, W.~A., et~al.\
  2002, Nature, 416, 403 \\
Narayanan, D., Cox, T.~J., Hayward, C.~C., Younger, J.~D., 
\& Hernquist, L.\ 2009, \mnras, 400, 1919 \\
Narayanan, D., Krumholz, M.~R., Ostriker, E.~C., \& Hernquist, L.\ 2012, \mnras,
421, 3127 \\
Neufeld, D.A., Nisini, B., Giannini, T., et al. 2009, ApJ 706, 170 \\
{\"O}berg, K.~I., Guzm{\'a}n, V.~V., Furuya, K., et~al. 2015, Nature,
  520, 198 \\
{\"O}berg, K.~I., Qi, C., Fogel, J.~K.~J., et~al.\ 2010, ApJ, 720, 480 \\
{\"O}berg, K.~I., Qi, C., Fogel, J.~K.~J., et~al.\ 2011, ApJ, 734, 98 \\
Oca{\~n}a Flaquer, B., Leon, S., Combes, F., \& Lim, J.\ 2010, A\&A, 518, A9 \\
Okuda, T., Kohno, K., Iguchi, S., \& Nakanishi, K.\ 2005, ApJ, 620, 673 \\
Orienti, M., Koyama, S., D'Ammando, F., et al. 2013, MNRAS, 428, 2418\\
Palla F., Stahler S.~W., 1993, ApJ, 418, 414\\
Patel, S.~G., Holden,  B.~P., Kelson, D.~D., Illingworth, G.~D., 
\& Franx, M.\ 2009, \apjl, 705, L67 \\
Persson, M.~V., J{\o}rgensen, J.~K., van Dishoeck, E.~F., \& Harsono,
  D.\ 2014, A\&A, 563, A74 \\
Pizzarello, S., Krishnamurthy, R.~V., Epstein, S., \& Cronin, J.~R.\
  1991, GeCoA, 55, 905 \\
Planck Collaboration, et al., 2014, A\&A, 571, A29 \\
Podio, L., Codella, C., Gueth, F., et~al.\ 2015, A\&A, in press, arXiv:1505.05919 \\
Podio, L., Kamp, I., Codella, C., et~al.\ 2013, ApJL, 766, L5  \\
Podio, L., Kamp, I., Codella, C., et~al.\ 2014, ApJL, 783, L26 \\
Pontoppidan, K.~M., Salyk, C., Bergin, E.~A., et~al.\ 2014, Protostars
  and Planets VI, 363 \\
Pontoppidan, K.~M., Salyk, C., Blake, G.~A., et~al.\ 2010, ApJ, 720,
  887 \\  
Porth, O., Fendt, C., Meliani, Z., \& Vaidya, B. 2011, ApJ, 737, 42\\
Prandoni, I., Laing, R.~A., Parma, P., et al.\ 2007, NewAR, 51, 43  \\
Prandoni, I., Laing, R.~A., de Ruiter, H.~R., \& Parma, P.\ 2010, A\&A, 523, A38\\
Requena-Torres, M.\ A., et al., 2011, EAS, 52, 299 \\
Qi, C., {\"O}berg, K.~I., Wilner, D.~J., et~al. 2013a,
  Science, 341, 630 \\
Qi, C., {\"O}berg, K.~I., Wilner, D.~J., \& Rosenfeld, K.~A.\
  2013b, ApJL, 765, L14  \\
Rivilla, V.\ M. et al.\  2015, ApJ, 808, 146  \\
Robert, F., Gautier, D., \& Dubrulle, B.\ 2000, Space Sci Rev, 92, 201 \\
Rosati, P., et al.\ 2004, AJ, 127, 230 \\
Rygl K.~L.~J, Brunthaler A., Reid M.~J., et al. 2010, A\&A, 511, A2\\
Sakai, N., Sakai, T., Hirota, T., et~al.\ 2014, Nature, 507, 78 \\
S\'anchez-Monge {\'A}., Cesaroni R., Beltr{\'a}n M.T., et al. 2013, A\&A,
552, L10 \\
S\'anchez-Monge {\'A}., Beltr{\'a}n M.T., Cesaroni R., et al.\ 2014, A\&A,
569, A11 \\
S{\'a}nchez-Monge, {\'A}., L{\'o}pez-Sepulcre, A., Cesaroni, R., et al.\ 2013,
\aap, 557, AA94 \\
Sanna A., Moscadelli L., Cesaroni R., et al. 2010, A\&A, 517, A78\\
Sanna A., Cesaroni R., Moscadelli L., et al. 2014, A\&A, 565, A34\\
Santangelo, G., Testi, L., Gregorini, L., et al.\ 2009, \aap, 501, 495 \\
Santoro, F., et al., 2015, A\&A, 574, 89\\
Schaye, J., Dalla Vecchia, C., Booth, C.~M., et al.\ 2010, \mnras, 402, 1536 \\
Senatore, F., et al.\ 2015, A\&A, in press \\
Shapley, A.~E.\ 2011, \araa, 49, 525 \\
Simpson, J.~M., Swinbank, A.~M., Smail, I., et al.\ 2014, \apj, 788, 125\\
Snyder, L.\ E., Hollis, J.\ M., Ulich, B.\ L.\ 1976, ApJ, 208, L91 \\
Sobolev A.~M., Deguchi S., A\&A, 291, 569\\
Swinbank, A.~M., Karim, A., Smail, I., et al.\ 2012, \mnras, 427, 1066 \\
Tacconi, L.~J., Genzel, R., Neri, R., et al.\ 2010, \nat, 463, 781 \\
Tacconi, L.~J., Neri, R., Genzel, R., et al.\ 2013, \apj, 768, 74 \\
Taquet, V., L\'opez-Sepulcre, A., Ceccarelli, C., Neri, R., Kahane, C., 
Charnley, S.B., 2015, ApJ, in press \\
Testi, L., Birnstiel, T., Ricci, L., et~al.\ 2014, Protostars and
  Planets VI, 339 \\
Thi, W.-F., M{\'e}nard, F., Meeus, G., et~al.\ 2011, A\&A, 530, L2 \\ 
Thi, W.-F., van Zadelhoff, G.-J., \& van Dishoeck, E.~F.\ 2004, A\&A, 425,
  955 \\
Tilanus, R.~P.~J., Krichbaum, T.~P., Zensus, J.~A., et al. 2014, arXiv:1406.4650
\\
van der Marel, N., van Dishoeck, E.~F., Bruderer, S., \& van Kempen,
  T.~A.\ 2014, A\&A, 563, A113 \\
van Dishoeck, E.~F., Bergin, E.~A., Lis, D.~C., \& Lunine, J.~I.\ 2014,
  Protostars and Planets VI, 835 \\  
Wagg, J., Pope, A., Alberts, S., et al.\ 2012a, \apj, 752, 91 \\
Wagner, A.~Y., et al.\ 2012, ApJ, 757, 136\\
Walsh, C., Millar, T.~J., Nomura, H., et~al.\ 2014, A\&A, 563, A33 \\
Weber, A. L., 1998, Orig. Life Evol. Biosph, 28, 259 \\
Wei{\ss}, A., De Breuck, C., Marrone, D.~P., et al.\ 2013, \apj, 767, 88 \\
Wei{\ss}, A., Walter, F., \& Scoville, N.~Z.\ 2005, \aap, 438, 533 \\
Wolk et al. 2005 ApJS 160, 423 \\
Wong, T., \& Blitz, L.\ 2002, \apj, 569, 157 \\
Woods, P.\ M., Kelly, G, Viti, S.\ et al.\  2012, ApJ, 750, 19 \\ 
Woods, P.\ M.\ et al.\ 2013, ApJ, 777, 90 \\
Woon et al.\ 2002, ApJ, 569, 541 \\
Wu, J., Evans, N.~J., II, Gao, Y., et al.\ 2005, \apjl, 635, L173 \\
Wu, J., Evans, N.~J., II, Shirley, Y.~L., \& Knez, C.\ 2010, \apjs, 188, 313 \\
Zapata et al., 2004, AJ, 127, 2252 \\
Zinn, P.-C., Middelberg, E., Norris, R.~P., \& Dettmar, R.-J.\ 2013, \apj, 774, 66
\\

\end{document}